\documentclass[aps,prd,showpacs,preprintnumbers,notitlepage,nofootinbib,11pt]{revtex4-1}

\usepackage{graphicx}
\usepackage{bm}
\usepackage{enumitem}
\usepackage{epsf,epsfig}
\usepackage{amscd}
\usepackage{amsmath}
\usepackage{amssymb}
\usepackage{hyperref}

\usepackage{tensor}
\allowdisplaybreaks[1]

\hypersetup{colorlinks=true,linkcolor=blue,citecolor=red,urlcolor=blue}

\usepackage{feynmp}
\newcommand{\be}{\begin{equation}}
\newcommand{\ee}{\end{equation}}
\newcommand{\bea}{\begin{eqnarray}}
\newcommand{\eea}{\end{eqnarray}}

\newcommand{\nn}{\nonumber\\}

\def\la{\langle}
\def\ra{\rangle}

\def\vec#1{{\bm{#1}}}

\def\CE{\mathcal{E}}

\def\CI{\mathcal{I}}
\def\CJ{\mathcal{J}}

\def\CN{\mathcal{N}}
\def\CO{\mathcal{O}}
\def\CP{\mathcal{P}}
\def\CS{\mathcal{S}}
\def\CT{\mathcal{T}}
\def\CV{\mathcal{V}}

\def\qfr{\mathfrak{q}}
\def\wfr{\mathfrak{w}}

\def\delp{{\nabla_{\perp}}}
\def\parp{{\partial_{\perp}}}
\def\tp{\tau_{\Pi}}

\begin{document}

\title{Constructing higher-order hydrodynamics: The third order}
\author{Sa\v{s}o Grozdanov}
\email{grozdanov@lorentz.leidenuniv.nl}
\author{Nikolaos Kaplis}
\email{kaplis@lorentz.leidenuniv.nl}
\affiliation{Instituut-Lorentz for Theoretical Physics, Leiden University,\\Niels Bohrweg 2, Leiden 2333 CA, The Netherlands }

\begin{abstract}
Hydrodynamics can be formulated as the gradient expansion of conserved currents in terms of the fundamental fields describing the near-equilibrium fluid flow. In the relativistic case, the Navier-Stokes equations follow from the conservation of the stress-energy tensor to first order in derivatives. In this paper, we go beyond the presently understood second-order hydrodynamics and discuss the systematisation of obtaining the hydrodynamic expansion to an arbitrarily high order. As an example of the algorithm that we present, we fully classify the gradient expansion at third order for neutral fluids in four dimensions, thus finding the most general next-to-leading-order corrections to the relativistic Navier-Stokes equations in curved space-time. In doing so, we list $20$ new transport coefficient candidates in the conformal and $68$ in the non-conformal case. As we do not consider any constraints that could potentially arise from the local entropy current analysis, this is the maximal possible set of neutral third-order transport coefficients. To investigate the physical implications of these new transport coefficients, we obtain the third-order corrections to the linear dispersion relations that describe the propagation of diffusion and sound waves in relativistic fluids. We also compute the corrections to the scalar (spin-$2$) two-point correlation function of the third-order stress-energy tensor. Furthermore, as an example of a non-linear hydrodynamic flow, we calculate the third-order corrections to the energy density of a boost-invariant Bjorken flow. Finally, we apply our field theoretic results to the $\mathcal{N}=4$ supersymmetric Yang-Mills fluid at infinite 't Hooft coupling and infinite number of colours to find the values of five new linear combinations of the conformal transport coefficients. 
\end{abstract}

\maketitle

\newpage
\begingroup
\hypersetup{linkcolor=black}
\tableofcontents
\endgroup

\newpage
\section{Introduction}

Despite the fact that the history of research into phenomena involving the behaviour of fluids dates back millennia, hydrodynamics remains a subject of intense research both in mathematics and physics. This is a result of its extremely wide applicability across length scales: from the dynamics of quark-gluon plasma, to the evolution of the universe as a whole. In modern field theoretic language, the theory of hydrodynamics can be understood as the long-range, infrared effective theory that can be expressed in terms of the gradient expansion of the relevant fields. Recently, the formulation of hydrodynamics in the language of an effective field theory has been explored in several works, among them in \cite{Dubovsky:2005xd,Dubovsky:2011sj,Nicolis:2013lma,Endlich:2012vt,Grozdanov:2013dba,Bhattacharya:2012zx,Kovtun:2014hpa,Harder:2015nxa,Haehl:2015pja,Grozdanov:2015nea,Haehl:2013hoa,Burch:2015mea,Crossley:2015tka,deBoer:2015ija}.

In the non-relativistic limit, hydrodynamics is described by the Navier-Stokes equations, i.e. the equations of motion of energy and momentum transport,
\begin{align}
&\partial_0 \rho + \nabla \cdot \left(\rho \bm{v} \right) = 0, \label{ContEq}\\
&\rho\left(\partial_0 + \bm{v}\cdot \nabla \right) \bm{v}   = - \nabla P + \eta \nabla^2 \bm{v} + \left(\zeta + \eta/3 \right) \nabla \left(\nabla\cdot \bm{v}\right). \label{NavStok}
\end{align}
Here, $\rho$ denotes the sum of the fluid's energy density $\epsilon$ and pressure $P$. The two viscosities, the shear $\eta$ and the bulk $\zeta$, encode the microscopic properties of the fluid. In view of the gradient expansion, these famous equations, that is the continuity equation \eqref{ContEq} and the Navier-Stokes equation \eqref{NavStok}, can only sufficiently describe fluids at low energies. More precisely, the Navier-Stokes equations are sufficient to first order in the small parameter that controls the hydrodynamic approximation: $k \ell_{mfp}$, where $k$ is the momentum scale and $\ell_{mfp}$ the mean-free-path of the underlying microscopic processes.

To describe fluids at higher energies, like the quark-gluon plasma being tested at the LHC and RHIC, we must work with the relativistic version of hydrodynamics. Furthermore, it is important that the theory permits for microscopic processes at higher momentum scales. Hence, it is natural to expect that higher-order corrections in $k \ell_{mfp}$ will play an important role in the gradient expansion series, relevant for such high-energy fluid flows. This is the {\em main motivation} for this work. 

Second-order terms, which enter as the leading-order corrections to the Navier-Stokes equations, were first considered by Burnett \cite{BurnettD}. In a more modern language, they were studied by M\"{u}ller, Israel and Stewart \cite{Muller:1967zza,Israel:1976tn,Israel1976213,Israel:1979wp}, whose work was initially motivated by the well-known problem that first-order hydrodynamic solutions of the Navier-Stokes equations suffer from acausal propagation. They showed that the presence of second-order terms could cure these issues. In particular, the M\"{u}ller-Israel-Stewart theory is of phenomenological nature \cite{Baier:2007ix} where the variables are the energy density $\varepsilon$, velocity $u^\mu$ and the viscous part of the momentum flow $\Pi^{ab}$, which are governed by the conservation laws $\nabla_a T^{ab}=0$ along with the phenomenological equation $\tau_\Pi D \Pi^{ab} = - \Pi^{ab} - \eta \sigma^{ab}$. The coefficient $\tau_\Pi$ is known as the {\em relaxation time}, $D$ represents the longitudinal derivative taken in the direction of the fluid flow, $D = u^a \nabla_a$, and $\sigma^{ab}$ is the relativistic shear tensor, to be defined later. These equations are hyperbolic and encode non-instantaneous relaxation of momentum. Given that this is a phenomenological equation, one can naturally extend it beyond just linear order on the right-hand side. In fact, not only is this extension natural but it becomes necessary once conformal fluids are considered, in which case non-linear terms like $\Pi^{\mu\nu}(\nabla\cdot u)$ are present in the stress-energy tensor.

While second-order hydrodynamics has been important for stabilising hydrodynamic simulations, to our knowledge, no second-order transport coefficient has yet been measured.

The M\"{u}ller-Israel-Stewart theory did not include all possible terms consistent with the symmetries of hydrodynamics -- a systematic procedure that is normally dictated within the effective field theory approach. Their work was extended in \cite{Baier:2007ix,Bhattacharyya:2008jc}, where the full second-order conformal hydrodynamic stress-energy tensor was constructed. In \cite{Baier:2007ix,Bhattacharyya:2008jc}, it was shown that such an extension required five new transport coefficients, $\tau_{\Pi}$, $\kappa$, $\lambda_1$, $\lambda_2$ and $\lambda_3$. In the non-conformal case, all possible structures were found by Romatschke in \cite{Romatschke:2009kr}, resulting in fifteen transport coefficients.

Construction of hydrodynamics normally asserts the existence of a local thermal equilibrium and an entropy current. Consistency with the second law of thermodynamics then requires {\em global} entropy production to be non-negative. It is well known that this condition imposes constraints on the values of the hydrodynamic transport coefficients: In first-order Navier-Stokes hydrodynamics, one finds that shear and bulk viscosities must be non-negative, $\eta \geq 0$, $\zeta \geq 0$. For a pedagogical discussion, see \cite{Kovtun:2012rj}. In higher-order hydrodynamics, entropy production is more subtle. Many recent works, among them \cite{Loganayagam:2008is,Romatschke:2009kr,Haack:2008xx,Kovtun:2012rj,Bhattacharyya:2012nq,Jensen:2012jh,Moore:2012tc,Banerjee:2012iz}, have investigated this issue. To find constraints on second-order transport coefficients, they imposed a much stricter condition - namely that each linearly-independent (tensorial) term in the divergence of the {\em local} entropy current must be non-negative, even those formally sub-leading in the gradient expansion. It was shown by Bhattacharyya \cite{Bhattacharyya:2012nq} that under these condition, the number of independent, non-conformal second-order transport coefficients is reduced from fifteen to ten. The five conformal second-order transport coefficients remain un-constrained. While such a construction certainly suffices in ensuring non-negative entropy production, it is not completely clear whether it is necessary in all physical hydrodynamic systems.      

The question of entropy production in higher-order hydrodynamics become even more mysterious if one takes into account the results of recent numerical holographic simulations combined with experimental observations of the dynamics of strongly coupled quark-gluon plasma. A robust and remarkable property that they display is that the hydrodynamic description of time-evolution becomes extremely accurate very shortly after the collision of two heavy ion nuclei.\footnote{For a recent review, see \cite{jorge-book}.} Typically, this {\em hydrodynamisation time} $\tau_H$ is of the order of $1\, \text{fm}/\text{c}$. In many cases, the hydrodynamisation time is shorter than the thermalisation time, which means that hydrodynamic evolution may become applicable even before local thermal equilibrium is established in the plasma. What does hydrodynamic evolution in absence of a local thermal equilibrium imply for the existence of an entropy current? Furthermore, is there a notion of hydrodynamics at zero temperature? 

In this work, we will leave these important questions unanswered. As we will only study the un-constrained hydrodynamic gradient expansion, the new transport coefficients that we will find will correspond to the maximal possible set of transport coefficients in conformal and non-conformal fluids.

While exactly conformal high-energy fluids are not known in nature, the beta function in asymptotically free theories, such as QCD, becomes very small at high temperatures. In such regimes, conformal hydrodynamics can accurately approximate the behaviour of fluid-like states. Furthermore, conformal hydrodynamics is extremely important for understanding the AdS/CFT correspondence, within which the connection between gravity duals and hydrodynamics was established in \cite{Policastro:2002se,Policastro:2002tn}. As was clearly demonstrated by the duality, the IR limit of various large-$N$ quantum field theories behaves in accordance with the hydrodynamical gradient expansion. This is apparent for example from the dispersion relations of the extreme IR modes in the spectrum \cite{Kovtun:2005ev,Kovtun:2006pf}, which can be expressed in a series expansion,
\begin{align}\label{DispRelInt}
\omega = \sum_{n=0}^{\CO_H} \alpha_n k^{n+1},
\end{align}
where $\alpha_n$ can depend on any of the $m$-th-order hydrodynamic transport coefficients, with $m \leq n$. The order of the hydrodynamic expansion $\CO_H$ to which it is sensible to expand the series in Eq. \eqref{DispRelInt} is presently unknown. The reason for this is that adding the terms above some order $\CO_H$ may cause the sum to rapidly grow and spoil the series approximation to the mode's physical behaviour. Leaving these issues aside for the moment, it is important to note that the most powerful feature of the gravitational dual of some field theory is that it encodes full information about the entire hydrodynamic series -- the structure of the gradient expansion and the values of its microscopically determined transport coefficients. Beyond our desire to understand the general structure of relativistic hydrodynamics suitable for the description of high-energy fluid-like states of matter, AdS/CFT should therefore be in itself a very important motivation for studying higher-order hydrodynamics.\footnote{For works on second-order hydrodynamics in AdS/CFT, see \cite{Baier:2007ix,Bhattacharyya:2008jc,Banerjee:2010zd,Moore:2010bu,Moore:2012tc,Shaverin:2012kv,Haack:2008xx,Romatschke:2009im,Grozdanov:2015asa,Grozdanov:2014kva} and references therein.}

The problem that we will study in this work is precisely how hydrodynamics can be systematically generated at {\em all} orders in the gradient expansion, i.e. higher powers in $k \ell_{mfp}$. What is clear is that the complexity of such an expansion makes it rather forbidding for the higher-derivative terms to be analysed in the usual way. For this purpose, we will formulate a computational algorithm that can generate relativistic hydrodynamic gradient expansion at any order. To demonstrate its power, we will directly extend the works of \cite{Baier:2007ix,Romatschke:2009kr} and classify both conformal and non-conformal uncharged hydrodynamics at third order in four space-time dimensions. We will show that third-order hydrodynamic stress-energy tensor requires us to introduce $20$ new tensorial structures (and transport coefficients $\lambda_n^{(3)}$ with $n=1,\,2,\,\ldots,\,20$) for conformal fluids and $68$ tensors in the non-conformal case. Thus, we will find the most general next-to-leading-order corrections to the relativistic uncharged Navier-Stokes equations, with terms up to and including $\CO\left(\partial^4\right)$ in Eqs. \eqref{ContEq} and \eqref{NavStok}.

Next, we will use the AdS/CFT correspondence to compute some of the conformal, third-order transport coefficients, $\lambda^{(3)}_n$, in the supersymmetric $\CN=4$ Yang-Mills theory with an infinite number of colours $N_c$ and at infinite 't Hooft coupling $\lambda$. Using the linear-response dispersion relations for shear and sound modes, a two-point function of the stress-energy tensor and the non-linear boost-invariant Bjorken flow, we will derive the values for five linear combinations of these coefficients, namely 
\begin{align}
\lambda^{(3)}_1 + \lambda^{(3)}_2 + \lambda^{(3)}_4 \equiv - \theta_1 &= - \frac{N_c^2 T}{32 \pi} , \\
\lambda^{(3)}_3 +  \lambda^{(3)}_5 + \lambda^{(3)}_6 \equiv - \theta_2 &= \frac{N_c^2 T}{48 \pi} \left( \frac{\pi^2}{12} + 4 \ln 2 - \ln^2 2 - \frac{9}{2} \right), \\ 
\lambda^{(3)}_{1} - \lambda^{(3)}_{16} &= \frac{N_c^2 T}{16 \pi} \left(\frac{\pi^2}{12} + 4 \ln 2 - \ln^2 2 \right),\\  
\lambda^{(3)}_{17} &= \frac{N_c^2 T}{16 \pi} \left(\frac{\pi^2}{12} + 2 \ln 2 - \ln^2 2 \right) , 
\end{align}
and
\begin{align}
&\frac{\lambda^{(3)}_1}{6 } + \frac{4 \lambda^{(3)}_2}{3 } + \frac{4 \lambda^{(3)}_3}{3 } + \frac{5 \lambda^{(3)}_4}{6}+ \frac{5 \lambda^{(3)}_5}{6 } + \frac{4 \lambda^{(3)}_6}{3 }   -  \frac{\lambda^{(3)}_7}{2 }  \nn
& + \frac{3\lambda^{(3)}_8}{2 } + \frac{\lambda^{(3)}_9}{2 }  - \frac{2\lambda^{(3)}_{10}}{3 } - \frac{11\lambda^{(3)}_{11}}{6 } - \frac{\lambda^{(3)}_{12}}{3 } + \frac{\lambda^{(3)}_{13}}{6} - \lambda^{(3)}_{15}  = \frac{N_c^2 T}{648 \pi} \left( 15 - 2 \pi^2 - 45 \ln 2 + 24 \ln^2 2 \right).
\end{align}
From the point of view of phenomenological hydrodynamics, there is no clear way to define the basis of linearly-independent tensor structures at a given order in the gradient expansion. As a result, any linear combination of the transport coefficients can be declared as a set of {\em independent} transport coefficients. 

Beyond hydrodynamics to third order, recent works \cite{Bu:2014sia,Bu:2014ena,Bu:2015ika,Bu:2015bwa} have studied the linearised sub-sector of the hydrodynamic expansion to all orders. In this work, however, we will be interested in the full non-linear extension of the relativistic Navier-Stokes equations in all of the relevant fields.

There are several other corrections to hydrodynamics that could in principle invalidate the gradient expansion at some order. Firstly, it has long been known that the correlation functions of conserved operators, such as $\langle T_{ab} T_{cd}\rangle$, exhibit non-analytic behaviour beyond first-order hydrodynamics. This effect is known as the long-time tails and was recently studied and reviewed in \cite{Kovtun:2003vj,CaronHuot:2009iq,Kovtun:2012rj}. However, because these non-analyticities arise from loop corrections, they are suppressed by inverse factors of the number of colours in non-Abelian gauge theories ($1/N_c$) \cite{Kovtun:2003vj}. Thus, in large-$N_c$ field theories, such as those most readily described by the gauge-gravity duality, long-time tails are sub-leading compared to the gradient expansion series. Furthermore, the interplay between the non-analyticities and the higher-order expansion in $k$ has not yet been fully analysed and understood.

Secondly, as already mentioned above, an important question regarding the hydrodynamic expansion is the convergence of the series (cf. Eq. \eqref{DispRelInt}). It is believed that the hydrodynamic series is an asymptotic one, analogous to the perturbative series in QFT. Recent works \cite{Heller:2013fn,Heller:2015dha} have studied the behaviour of the hydrodynamic expansion from this point of view and \cite{Heller:2013fn} showed that in the holographic Bjorken flow \cite{Bjorken:1982qr,Janik:2005zt,Janik:2006ft,Heller:2007qt}, as expected from a divergent asymptotic series with a zero radius of convergence, the series indeed breaks down at some order. However, the order at which the series may break down in general is unknown and therefore it is important to study the expansion at higher orders. Similarly, in QED, the results of the renormalised perturbation theory are expected to be divergent \cite{PhysRev.85.631}. However, higher-order loop computations in QED have yielded some of the most successful high-energy theoretical predictions, consistent with experiments.

This paper is structured in the following way: In Sec. \ref{Sec:Syst}, we will proceed with our analysis of the hydrodynamic gradient expansion by first constructing a systematic algorithm that can be used for its classification at any order. We will then use the outlined procedure to write down the stress-energy tensor for conformal and non-conformal relativistic fluids at third order in Sec. \ref{Sec:3rdOrdHydro}. In Sec. \ref{Sec:DispRel}, we will look for the linear dispersion relations and find corrections to diffusion and the propagation of sound. Moreover, we will compute the scalar (spin-$2$) correlation function of the stress-energy tensor and the energy density corresponding to the Bjorken flow. These results will then be used in Sec. \ref{Sec.N4} to compute some of the new transport coefficients in the $\CN=4$ supersymmetric Yang-Mills theory by using the AdS/CFT correspondence. The appendices are used to prove a statement about the required tensorial ingredients in the classification of hydrodynamics, simplifications that occur in tensors that do not include any curvature tensors and to list all third-order scalars, vectors and two-tensors that enter the gradient expansion.

\section{Systematics of the construction}\label{Sec:Syst}

\subsection{Hydrodynamic variables and the generalised Navier-Stokes equations}\label{Sec:Gen}

Let us begin our exploration of hydrodynamics by identifying the relevant hydrodynamic variables and the equations of motion that govern the behaviour of fluids. In this work, we will focus on the uncharged relativistic fluids, although the systematics we outline here can easily be applied to the classification of the gradient expanded Noether currents that would correspond to additional conserved charges in the system. The steps of the construction that we employ will closely follow those presented in \cite{Kovtun:2012rj}.

An uncharged fluid on a manifold with an arbitrary metric $g_{ab}$ can be described by two near-equilibrium functions: the velocity field $u^a(x)$ and the temperature field $T(x)$. In the presence of a Noether current $J^a$, one also needs to promote the chemical potential to a space-time dependent, near-equilibrium field, $\mu(x)$. For present purposes, we only assume the existence of a conserved stress-energy tensor, which we write in the gradient expansion of the velocity and temperature fields,
\begin{align}\label{DefTasExp}
T^{ab} = T^{ab}_{(0)} \left(u,T\right) + T^{ab}_{(1)} \left(\partial u,\partial T\right) + \ldots + T^{ab}_{(n)} \left(\partial^n u, \ldots ,\partial^n T\right) + \ldots \equiv  T^{ab}_{(0)} + \Pi^{ab}.
\end{align}
The conservation equation,
\begin{align}\label{ConservationOfT}
\nabla_a T^{ab} = 0,
\end{align}
then provides dynamical equations (of motion) for $u^a$ and $T$. It is worth noting that Eq. \eqref{ConservationOfT} gives us $d$ equations in $d$ space-time dimensions, which can be solved by the $d-1$ independent components of $u^a$ together with a single scalar field $T$ ($u^a$ is normalised to $u_a u^a = -1$). However, for conciseness, we will only work in $d=4$ dimensions in this paper, although our construction could be easily generalised to an arbitrary number of dimensions.

Following \cite{Kovtun:2012rj}, it is convenient to decompose $T^{ab}$ into
\begin{align}
T^{ab}  = \CE u^a u^b + \CP \Delta^{ab} + \left(q^a u^b + u^a q^b \right) + t^{ab},
\end{align}
where $\CE$ and $\CP$ are scalars, $q^a$ a transverse vector and $t^{ab}$ a transverse, symmetric and traceless (TST) tensor. Here we have introduced the projector,
\begin{align}
\Delta^{ab} \equiv u^a u^b + g^{ab}.
\end{align}
Each one of the operators $\CE$, $\CP$, $q^a$ and $t^{ab}$ is then gradient-expanded in $u^a$, $T$ and $g_{ab}$, to facilitate the expansion in Eq. \eqref{DefTasExp}.\footnote{Throughout this work, we will be using the $(-,+,+,+)$ Lorentzian signature for the metric tensor.}

In order to construct a physically sensible theory of hydrodynamics, it is not sufficient to only find all possible tensor structures for $\CE$, $\CP$, $q^a$ and $t^{ab}$. The first issue arises from the fact that we have not microscopically specified the meaning of the velocity and temperature fields. Because we are working with the gradient expansion, it should therefore be possible to transform the fields by adding to them terms sub-leading in the gradient expansion while keeping the physical predictions of hydrodynamics invariant,
\begin{align}
u^a \to u^a + f^a_{(1),u} \left( \partial T, \partial u \right) + f^a_{(2),u} \left( \partial^2 T, \partial T\partial u, \partial^2 u \right) + \ldots \, , \label{uFrame} \\
T \to T + f_{(1),T} \left( \partial T, \partial u \right) + f_{(2),T} \left( \partial^2 T, \partial T\partial u, \partial^2 u \right) + \ldots \, .\label{TFrame}
\end{align}
The hydrodynamical description of a system should thus remain invariant under such {\em frame} re-definitions. To study uncharged fluids, it is most convenient to work in the Landau frame with
\begin{align}
u_a \Pi^{ab} = 0,
\end{align}
which requires us to set all $\CE_{(n)} = 0$ and $q^a_{(n)} = 0$, for $n \geq 1$. The zeroth-order $q^a_{(0)} = 0$ because there are no transverse vectors without derivatives among our hydrodynamics variables. The remaining term in the expansion of $\CE$, i.e. $\CE = \CE_{(0)}$, is then identified as the energy density of the fluid, $\CE = \CE_{(0)} \equiv \varepsilon$. After such an adjustment of the arbitrary functions in Eqs. \eqref{uFrame} and \eqref{TFrame}, we are left with
\begin{align}
T^{ab}  = \varepsilon u^a u^b + \CP \Delta^{ab} + t^{ab},
\end{align}
where $\CP_{(0)} \equiv P$ is the pressure of the fluid. Hence, what remains to be gradient-expanded in the classification of uncharged fluids, are the scalar $\CP$ and the transverse, symmetric and traceless $t^{ab}$.

There is an additional source of restrictions that we must impose on the structure of the hydrodynamic tensors in order to obtain the set of all ``physically" independent structures: when working at the $n$-th order of the gradient expansion, we can always use the equations of motion (the conservation equation \eqref{ConservationOfT}) to eliminate some terms at the expense of introducing corrections at all sub-leading orders in the gradient expansion. More precisely, Eq. \eqref{ConservationOfT} gives
\begin{align}
&u_b \nabla_a T^{ab} = - D \varepsilon - \left(\varepsilon + P \right) \nabla \cdot u + \text{higher derivatives} = 0 \,,\label{EoMT1} \\
&\Delta^{a}_{b}\nabla_c T^{cb} = \left(\varepsilon + P \right) D u^a + \delp^a P + \text{higher derivatives} = 0 \,,\label{EoMT2}
\end{align}
where we have defined the longitudinal and the transverse derivatives as
\begin{align}
D = u^a \nabla_a \,, && \delp_a = \Delta_{ab} \nabla^b \,,
\end{align}
so that $\nabla_a = \delp_a - u_a D$. The most important property of the transverse derivatives is that when acting on an arbitrary tensor,
\begin{align}\label{DelpProperty}
u^c \delp_c V_{b_1b_2\ldots} = 0 \,.
\end{align}

Eqs. \eqref{EoMT1} and \eqref{EoMT2} are the generalisations of the relativistic Navier-Stokes equations for fluids with a gradient expanded stress-energy tensor to an arbitrary order. Following \cite{Romatschke:2009kr}, it is convenient to use the thermodynamic relation $\varepsilon + P = s T$, where $s$ is the entropy density, to re-write Eqs. \eqref{EoMT1} and \eqref{EoMT2} as
\begin{align}
&D \ln s = - \nabla \cdot u   + \text{higher derivatives} \, , \label{EoMs1} \\
&D u^a = - c_s^2 \delp^a \ln s  + \text{higher derivatives} \, . \label{EoMs2}
\end{align}
The constant $c_s$ is the speed of sound in the fluid. Note also that $\nabla\cdot u = \delp\cdot u$. As a result of the form of Eqs. \eqref{EoMs1} and \eqref{EoMs2}, we will use the scalar function $\ln s$ instead of the temperature field $T$ to construct the gradient expansion.

The key observation that follows from Eqs. \eqref{EoMs1} and \eqref{EoMs2} is that it is most convenient to only work with the transverse derivatives of $u^a$ and $\ln s$, and the Riemann tensor $R_{abcd}$. This is because all longitudinal derivatives of $u^a$ and $\ln s$ can always be written as purely transverse derivatives and commutators of the covariant derivatives, which can then be expressed in terms of various combinations of the Riemann tensor, its covariant derivatives and transverse derivatives of $u^a$ and $\ln s$. The proof of this statement is presented in Appendix \ref{sec:RiemProof}.

\subsection{Constructing the gradient expansion}\label{Sec:ConstGradExp}

We are now ready to begin classifying the hydrodynamic expansion. By choosing the frame for $T$, or the entropy density field $s$, and defining $\CP_{(0)}$, we have already fixed zeroth-order hydrodynamics, which describes the behaviour of ideal, non-dissipative fluids. There exist no transverse vectors, neither TST tensors that could be constructed out of only $u^a$ or $g_{ab}$.

We can thus turn our attention to the hydrodynamic tensors with non-trivial derivative structures. To construct the hydrodynamic expansion at higher orders, it is most efficient to carefully select the relevant ingredients that will build the irreducible tensor structures, i.e. those that cannot be eliminated by the equations of motion \eqref{EoMs1} and \eqref{EoMs2}. As discussed at the end of Sec. \ref{Sec:Gen} and Appendix \ref{sec:RiemProof}, all the longitudinal components of $\nabla_a u_b$ and $\nabla_a \ln s $ can be eliminated. Therefore, it suffices to use transverse derivatives of $u^a$ and $\ln s$, as well as the Riemann tensor $R_{abcd}$ and its covariant derivatives.

Let us begin the classification of hydrodynamics by considering first-order tensors. The only possible one-derivative ingredients that can be used to build $\CE$, $\CP$, $q^\mu$ and $t^{\mu\nu}$ are $\delp_a u_b$ and $\delp_a \ln s$, as there are no one-derivative tensorial structures that would include the metric tensor $g_{ab}$. We can therefore write the first-order set of tensorial ingredients as
\begin{align}\label{Ord1Ing}
\CI^{(1)} = \left\{ \delp_a u_b, \,\delp_a \ln s \right\}.
\end{align}
Similarly, we define $\CI^{(n)}$ to be the set of all $n$-derivative objects relevant at the $n$-th order of the gradient expansion. The two and one-index tensors in Eq. \eqref{Ord1Ing} must then be contracted in all possible ways with the zeroth-order structures, i.e. $u_a$ and $g_{ab}$,
\begin{align}\label{Ord0Ing}
\CI^{(0)} = \left\{ u_a, \, g_{ab} \right\}.
\end{align}

Imagine that we wish to form all two-tensors with single entries from $\CI^{(1)}$ and an arbitrary number of the $\CI^{(0)}$ structures. It is easiest to count such combinations by using covariant and contravariant indices on the two different sets, as it would be redundant to add additional strings of $\CI^{(0)}$ and then contract them amongst themselves. The total number of free indices from the $\CI^{(0)}$ tensors is $a_1 + 2 a_2$, which should be directly contracted with either a $2$- or a $1$-index object from $\CI^{(1)}$. Hence, $a_1 + 2 a_2 - 2$ or $a_1 + 2 a_2 - 1$ can equal either $+2$ or $-2$.

To generalise the preceding discussion to tensors of any rank, let us now denote by $\left[\CI^{(n)}_m\right] $ the number of un-contracted indices in the $m$-th entry of $\CI^{(n)}$. All possible contractions of $\CI^{(0)}$ with $\CI^{(1)}$ can thus be found by solving
\begin{align}\label{PointerEq}
\sum_{n=1}^2 a_n \left[\CI^{(0)}_n\right] - \left[\CI^{(1)}_m\right] = \pm N,
\end{align}
for each $m$, such that $a_n \in \mathbb{Z}^+ \cup \{0\}$. The three different values of $N = \{0,1,2\}$ correspond to scalars (terms in $\CE$ and $\CP$), vectors (terms in $q^a$) and tensors (terms in $t^{ab}$), respectively. Throughout this work, we will use the symbol $\CJ^{(n)}_N$ to denote the set of all possible tensors with $n$ derivatives that can be used to form the tensors of rank $N$.

Using Eq. \eqref{PointerEq}, we find that in first-order hydrodynamics,
\begin{align}
&m=1: && a_1 + 2 a_2 - 2 = \pm N ,   \\
&m=2: && a_1 + 2 a_2 - 1 = \pm N.
\end{align}
For scalars with $N=0$, the possible combinations are
\begin{align}
&m=1: && \left\{\left(a_1,a_2\right)\right\} = \left\{ (2,0), (0,1)\right\} ,  \\
&m=2: && \left\{\left(a_1,a_2\right)\right\} = \left\{ (1,0)\right\},
\end{align}
which give us the possible tensors,
\begin{align}
\CJ^{(1)}_0 = \left\{ u_{a} u_{b} \delp_{c} u_{d},  \,  g_{ab} \delp_{c} u_{d} , \, u_{a} \delp_{b} \ln s \right\} .
\end{align}
For a vector with $N=1$, we need to solve for the right-hand-side of \eqref{PointerEq} with $\pm N = \pm 1$, giving us a set of equations
\begin{align}
&m=1: && a_1 + 2 a_2  = 3  ~~~~\lor~~~~ a_1 + 2 a_2  = 1, \\
&m=2: && a_1 + 2 a_2 = 2 ~~~~\lor~~~~ a_1 + 2 a_2  = 0,
\end{align}
which have solutions
\begin{align}
&m=1: && \left\{\left(a_1,a_2\right)\right\} = \left\{ (3,0), (1,1), (1,0)\right\} ,  \\
&m=2: && \left\{\left(a_1,a_2\right)\right\} = \left\{ (2,0), (0,1), (0,0) \right\}.
\end{align}
The possible tensors from which the one-derivative vectors can be constructed are thus
\begin{align}
\CJ^{(1)}_1 = \left\{ u_{a} u_{b} u_{c} \delp_{d} u_{e}, u_{a} g_{bc} \delp_{d} u_{e} ,  u_{a} \delp_{b} u_{c} , u_{a} u_{b} \delp_{c} \ln s, g_{ab} \delp_{c} \ln s, \delp_{a} \ln s \right\}.
\end{align}
Similarly, the two-tensors in $t^{ab}$ can be constructed by solving
\begin{align}
&m=1: && a_1 + 2 a_2  = 4 ~~~~\lor~~~~a_1 + 2 a_2  = 0, \\
&m=2: && a_1 + 2 a_2 =  3~~~~\lor~~~~a_1 + 2 a_2  = -1,
\end{align}
which gives
\begin{align}
&m=1: && \left\{\left(a_1,a_2\right)\right\} = \left\{ (4,0), (2,1), (0,2), (0,0) \right\} ,  \\
&m=2: && \left\{\left(a_1,a_2\right)\right\} = \left\{ (3,0), (1,1)\right\}.
\end{align}
Hence,
\begin{align}
\CJ^{(1)}_2 =  \big\{ &  u_{a} u_{b} u_{c} u_d \delp_{e} u_{f} , \, u_{a} u_{b} g_{cd} \delp_{e} u_{f} ,\, g_{ab} g_{cd} \delp_{e} u_{f} ,\, \delp_{a} u_{b},  \nn
& u_{a} u_{b} u_{c} \delp_{d} \ln s , \, u_{a}g_{bc} \delp_{d} \ln s \big\}.
\end{align}

All the tensors in the sets $\CJ^{(1)}_0$, $\CJ^{(1)}_1$ and $\CJ^{(1)}_2$ must now be contracted in all possible ways with $g_{ab}$ to form scalars, vectors and two-tensors, respectively. The next step in the construction is the elimination of duplicates and simplifications that use the properties of the participating tensors. In particular, we can use the fact that
\begin{align}\label{UNorm}
u_a u^a = -1,
\end{align}
hence $u_a \delp_b u^a = 0$, and so on. Furthermore, the property of the transverse derivatives that was stated in Eq. \eqref{DelpProperty} can also be used to vastly simplify the resulting expressions. In fact, it should be clear that the combination of Eqs. \eqref{DelpProperty} and \eqref{UNorm} makes any contraction between $u^a$ in $\CI^{(0)}$ and the entries of $\CI^{(1)}$ vanish. However, this will not be the case for the curvature tensors in $\CI^{(n)}$

All scalars that are made from $\CJ^{(1)}_0$ can immediately be used in $\CE$ and $\CP$. On the other hand, the vector $q^a$ and tensor $t^{ab}$ require additional treatment. Since $q^a$ must be transverse, we may only keep the vectors that vanish after they are contracted with $u_a$. To construct all TST tensors relevant for $t^{ab}$, it is easiest to form manifestly TST structures out of all two-tensors by applying the operation
\begin{align}\label{TST}
A^{\langle ab \rangle} \equiv \frac{1}{2} \Delta^{ac} \Delta^{bd} \left(A_{cd} + A_{dc}\right) - \frac{1}{d-1} \Delta^{ab} \Delta^{cd} A_{cd} \,.
\end{align}
All linearly independent tensors that survive this operation can then be considered as terms in $t^{ab}$.

Although the above procedure yields the correct results, it is computationally more efficient to construct the tensors that do not include any Riemann tensors separately. As already noted below Eq. \eqref{UNorm}, this is because our choice of the tensorial ingredients (Appendix \ref{sec:RiemProof}) vastly simplifies the possible contractions between $\CI^{(0)}$ and the tensors excluding various curvature tensors in $\CI^{(n)}$ with $n\geq 1$ (curvature tensors can have covariant derivatives acting on them). In Appendix \ref{sec:BasisAppendix}, we prove the claim that when constructing scalars, transverse vectors and TST two-tensors with only $u^a$ and $\ln s$ at any order of the gradient expansion, we can only use the $g_{ab}$ component of $\CI^{(0)}$, i.e. set $a_1 = 0$ in Eq.  \eqref{PointerEq}. For tensors containing $R_{abcd}$, we have to allow for any $a_1$.

Once all the linearly independent tensor structures are found, we associate hydrodynamic {\it transport coefficients} with each one of the tensors and write them as a sum in the stress-energy tensor. The transport coefficients cannot be determined phenomenologically, but must be calculated from the underlying microscopic theory.\footnote{Such calculations can be rather involved and use a combination of linear response theory and kinetic theory techniques or lattice gauge computations. See for example \cite{York:2008rr,Moore:2012tc}.} To denote the transport coefficients at all orders, we will use the lower case Greek letter ``upsilon" with three indices,
\begin{align}
\upsilon^{(n,N)}_i,
\end{align}
where $n$ is the order in the gradient expansion and $N$ determines whether a particular transport coefficients comes from the scalar ($\CP$), vector ($q^a$) or two-tensor ($t^{ab}$) sector of $T^{ab}$, denoted by $N=0$, $N=1$ and $N=2$, respectively. In case one wanted to work in a frame with $\CE \neq \CE_{(0)}$, some additional notation should be introduced to distinguish between the two $\upsilon^{(n,0)}_i$ cases. Finally, $i$ will count the number of transport coefficients in each of the sectors, at each order of the gradient expansion.

Since we are working in the Landau frame and in the frame with $\CE = \CE_{(0)} = \varepsilon$, we only need to find $\CP_{(1)}$ and $t^{ab}_{(1)}$ at first order. What we find from the above considerations is
\begin{align}
\CP_{(1)} &= - \zeta\, \delp \cdot u ,\\
t^{ab}_{(1)} &= - \eta \,\sigma^{ab},
\end{align}
where $\upsilon^{(1,0)}_1 = \zeta$ and $\upsilon^{(1,2)}_1 = \eta$ are the two first-order transport coefficients: the bulk and shear viscosity, respectively. The only first-order TST tensor is defined as
\begin{align}\label{Sigma}
\sigma^{ab} = 2 {}^\langle \delp^{a}u^{b\rangle}.
\end{align}
For future use, we also introduce the most general decomposition,
\begin{align}
\delp_a u_b =\frac{1}{2} \sigma_{ab} + \Omega_{ab} + \frac{1}{3}  \Delta_{ab}\delp_c u^c ,
\end{align}
where the vorticity is
\begin{align}
\Omega_{ab} = \frac{1}{2} \left( \delp_a u_b - \delp_b u_a \right).
\end{align}

At second order in the gradient expansion, the non-linear tensorial ingredients are
\begin{align}\label{Ord2Ing}
\CI^{(2)} = \left\{ \delp_a \delp_b u_c, \delp_a u_b\, \delp_c u_d,  \delp_a u_b\, \delp_c \ln s, \delp_a \delp_b \ln s, \delp_a \ln s \,\delp_b \ln s, R_{abcd} \right\},
\end{align}
where $\CI^{(2)}$ includes the Riemann tensor that involves two-derivative metric structures. The full structure of second-order hydrodynamics was first found by Romatschke in \cite{Romatschke:2009kr} and is given by a sum of fifteen linearly independent tensors with fifteen transport coefficients,
\begin{align}
\CP_{(2)} =& ~ \zeta\tau_\pi D \left(\nabla\cdot u\right) + \xi_1 \sigma^{ab} \sigma_{ab} +\xi_2\left(\nabla\cdot u\right)^2 + \xi_3 \Omega^{ab}\Omega_{ab} +\xi_4 \delp_a\ln s \delp^a \ln s \nn
&+ \xi_5 R + \xi_6 u_a u_b R^{ab}   ,  \label{P2expansion} \\
t^{ab}_{(2)} =& ~ \eta \tp \left[ {}^{\langle}D\sigma^{ab\rangle} + \frac{1}{3} \sigma^{ab} \left(\nabla\cdot u\right) \right] + \kappa \left[ R^{\langle ab \rangle} - 2 u_c u_d R^{c \langle ab \rangle d}  \right] +  \frac{1}{3} \eta\tp^* \sigma^{ab} \left(\nabla\cdot u\right) \nn
& + 2 \kappa^* u_c R^{c \langle ab \rangle d} u_d  + \lambda_1 \sigma^{\langle a}_{~~c} \sigma^{b\rangle c}  +\lambda_2 \sigma^{\langle a}_{~~c} \Omega^{b\rangle c}  +\lambda_3 \Omega^{\langle a}_{~~c} \Omega^{b\rangle c}  + \lambda_4 \delp^{\la a} \ln s \delp^{b\ra} \ln s. \label{t2expansion}
\end{align}
For historical reasons, in the M\"{u}ller-Israel-Stewart theory \cite{Muller:1967zza,Israel:1976tn,Israel1976213,Israel:1979wp}, the transport coefficient $\eta\tau_\Pi$ is written as a product of the {\em shear viscosity} $\eta$ and the {\em relaxation time} $\tau_\Pi$. Furthermore, the physically motivated combinations of the tensors (shear tensor, vorticity) that appear in Eqs. \eqref{P2expansion} and \eqref{t2expansion} are not necessarily those that would directly follow from our systematics, but can easily be written in terms of our tensors by simple linear combinations. What is invariant, regardless of the choice of the linearly independent tensors, is their number.

For calculations that do not employ the Landau frame, it is useful to also write down the transverse vector $q^a$. At first order, the only term that enters into the expansion is
\begin{align}
q^a_{(1)} =  \upsilon^{(1,1)}_1 \delp^a \ln s,
\end{align}
while at second order, the five independent transverse vectors give
\begin{align}
q^a_{(2)} =& \,  \upsilon^{(2,1)}_1 \delp_b \sigma^{ba} + \upsilon^{(2,1)}_2 \delp_b \Omega^{ba} + \upsilon^{(2,1)}_3 \sigma^{ba} \delp_b \ln s \nn
& + \upsilon^{(2,1)}_4 \delp_b u^b \delp^a \ln s + \upsilon^{(2,1)}_5 \Delta^{ab} u^c R_{bc} .
\end{align}

Beyond the possibility to work in a different frame, the knowledge of the vectors is also required for building the gradient expansion of the entropy current $S^a$ \cite{Loganayagam:2008is,Romatschke:2009kr,Kovtun:2012rj,Bhattacharyya:2012nq}. However, in that case, we would not impose that $S^a$ be transverse. In this paper, we do not consider the entropy current analysis nor do we study potential constraints that may arise from demanding positive local entropy production \cite{Loganayagam:2008is,Romatschke:2009kr,Haack:2008xx,Kovtun:2012rj,Bhattacharyya:2012nq,Jensen:2012jh,Moore:2012tc,Banerjee:2012iz}. As discussed in the introduction, it is well known that in first-order hydrodynamics this analysis results in the condition that shear and bulk viscosities must be non-negative \cite{Kovtun:2012rj}. In second-order hydrodynamics, \cite{Bhattacharyya:2012nq} argued that the entropy current considerations give five constraints, resulting in the reduction of the $15$ second-order transport coefficients to $10$ independent ones.

In a different frame for the temperature field $T$, which would result in $\CE_{(1)} \neq \ldots \neq \CE_{(n)} \neq 0 $, the structure of the gradient expansion for $\CE_{(n)}$ would be exactly the same as for $\CP_{(n)}$. In such cases, we would therefore only need to associate a new transport coefficient with each term in $\CE_{(n)}$, independent from those in $\CP_{(n)}$.

Finally, we note that by following the above procedure, we could also include the fluctuating chemical potential $\mu(x)$, which would then allow us to construct the gradient expansion of the conserved Noether current, such as for example the baryon number current. We will not explore this option in this paper.

\subsection{Conformal hydrodynamics}\label{Sec:ConfHydro}

A very important sub-class of fluids are the {\em conformal fluids}, which have played a central role in applications of the gauge-gravity duality to hydrodynamics \cite{Policastro:2002se,Policastro:2002tn}. Fluids with conformal symmetry are characterised by a vanishing trace of their stress energy tensor,
\begin{align}\label{TTraceless}
T^a_{~a} = 0,
\end{align}
and the Weyl-covariance of $T^{ab}$, i.e. each term in the stress-energy tensor needs to transform homogeneously under the non-linear Weyl transformations \cite{Baier:2007ix}.

In quantum field theory, the perturbations of the metric tensor, $g_{ab} \to g_{ab} + h_{ab}$, can be used to source the stress-energy tensor correlation functions. To first order in the source $h_{ab}$, the generating functional includes the coupling
\begin{align}\label{GenFunTh}
\int d^4 x \sqrt{-g} \, T^{ab} h_{ab}.
\end{align}
Under the Weyl transformation, the metric tensor transforms with conformal weight of $\Delta_{g_{ab}} = -2$,
\begin{align}\label{WeylMetric}
g_{ab} \to e^{-2\omega} g_{ab} \,.
\end{align}
Eqs. \eqref{GenFunTh} and\eqref{WeylMetric} then imply that in order for the stress-energy tensor to scale homogeneously, it must have the conformal weight $\Delta_{T^{ab}} = 6$ in four dimensions,
\begin{align}\label{WeylT}
T^{ab} \to e^{6\omega} \,T^{ab}.
\end{align}
In general, a tensor $A^{a_1\ldots a_m}_{b_1 \ldots b_n} $ with the conformal weight $\Delta_A = [A] + m - n$ that transforms homogeneously under the Weyl transformation, transforms as
\begin{align}\label{WeylTensor}
A^{a_1\ldots a_m}_{b_1 \ldots b_n} = e^{\Delta_A \omega} A^{a_1\ldots a_m}_{b_1 \ldots b_n} .
\end{align}
In the expression for the conformal weight, $[A]$ denotes the mass dimension of the tensor operator and $m$ and $n$ count the number of contravariant and covariant indices, respectively. Hence, $u_a$, $\ln s$ and the transport coefficients transform as
\begin{align}
&u^a \to  e^{\omega} u^a ,\\
&\ln s \to \ln \left( e^{3\omega} s \right), \\
&\upsilon^{(n,N)}_i \to e^{(4-n)\omega } \upsilon^{(n,N)}_i.
\end{align}
For a detailed discussion of the Weyl transformations in hydrodynamics, we refer the reader to \cite{Baier:2007ix}.

The procedure for finding the most general conformal hydrodynamic stress-energy tensor is then the following: since we are working in the Landau frame with only $\CE_{(0)} \neq 0$, the traceless condition \eqref{TTraceless} implies that $\varepsilon = 3 P$. Furthermore, it implies that all $\CP_{(n)}$ with $n \geq 1$ must vanish. We must then use the full series of tensors inside the non-conformal expression for the traceless $t^{ab}$, discussed in Section \ref{Sec:ConstGradExp}, and find how each one of them transforms under the Weyl transformation \eqref{WeylTensor}. Since \eqref{WeylT} must be ensured, those tensors that transform homogeneously can be immediately used in the conformal $T^{ab}$. Those that do not, must be combined by linear relations into the maximal possible set of homogeneously transforming tensors. Each of the remaining linearly independent homogeneously transforming TST two-tensors can then be assigned an independent transport coefficient and the full series constitutes the stress-energy tensor $T^{ab}$ of a conformal fluid.

Up to second order and in the Landau frame, the above procedure gives us the conformal stress-energy tensor \cite{Baier:2007ix,Bhattacharyya:2008jc},
\begin{align}\label{tab2ndOrd}
\Pi^{ab} = t^{ab}_{(1)} + t^{ab}_{(2)} =&\, -\eta \sigma^{ab} + \eta \tp \left[ {}^{\langle}D\sigma^{ab\rangle} + \frac{1}{3} \sigma^{ab} \left(\nabla \cdot u \right) \right] + \kappa \left[ R^{\langle ab \rangle} - 2 u_c R^{c \langle ab \rangle d} u_d  \right]  \nn
&+\lambda_1 \sigma^{\langle a}_{~~c} \sigma^{b\rangle c}  +\lambda_2 \sigma^{\langle a}_{~~c} \Omega^{b\rangle c}  +\lambda_3 \Omega^{\langle a}_{~~c} \Omega^{b\rangle c} ,
\end{align}
where $\Pi^{ab}$ was defined in Eq. \eqref{DefTasExp}. What remains from the list of all non-conformal transport coefficients is one first-order transport coefficient $\eta$ (shear viscosity) and five second-order transport coefficients $\tau_\Pi$, $\kappa$, $\lambda_1$, $\lambda_2$ and $\lambda_3$.

Finally, we should note that we can completely ignore the effects of the conformal (Weyl) anomaly on the hydrodynamic transport in this work, as we are only considering third-order hydrodynamics in $d=4$ dimensions. It is well known that the Weyl anomaly in four dimensions is given by
\begin{align}
\left\langle T^a_{~a} \right\rangle &= - \frac{a}{16\pi^2}\left( R_{abcd} R^{abcd} - 4 R_{ab} R^{ab} + R^2 \right) + \frac{c}{16\pi^2} \left(R_{abcd}R^{abcd} - 2 R_{ab} R^{ab} + \frac{1}{3} R^2 \right)    \label{TraceAnomaly} \\
& \sim \left( \partial g_{ab} \right)^4.
\end{align}
Hence, this Weyl trace anomaly could only affect fourth-order hydrodynamics \cite{Baier:2007ix}. In that case, the use of only $t^{ab}_{(4)}$ in the expansion of $T^{ab}_{(4)}$ would be insufficient, as $t^{ab}_{(4)}$ is manifestly traceless. In fact, Eq. \eqref{TraceAnomaly} could simply be incorporated into $T^{ab}$, in the Landau frame, by writing
\begin{align}
\CP_{(4)} =\frac{c - a}{48 \pi^2} R_{abcd} R^{abcd} + \frac{2 a-c}{24 \pi^2} R_{ab} R^{ab} + \frac{c - 3 a}{144 \pi^2}  R^2 + \ldots \, .
\end{align}
The three new transport coefficients,
\begin{align}
&\upsilon^{(4,0)}_1 = \frac{c - a}{48 \pi^2}  , \\
&\upsilon^{(4,0)}_2 = \frac{2 a-c}{24 \pi^2} , \\
&\upsilon^{(4,0)}_3 = \frac{c - 3 a}{144 \pi^2} ,
\end{align}
are thus completely determined by the central charges $a$ and $c$ of the underlying conformal field theory. We defer a more detailed study of fourth-order hydrodynamics to future work.

\section{Third-order hydrodynamics}\label{Sec:3rdOrdHydro}

We are now ready to construct the hydrodynamic stress-energy tensor at third order in the gradient expansion. To use the systematics described in Sec. \ref{Sec:Syst}, we first need the set of third-order tensorial ingredients,
\begin{align}\label{Ord3Ing}
\CI^{(3)} = \big\{ & \delp_a \delp_b \delp_c u_d, \delp_a \delp_b u_c\, \delp_d u_e,  \delp_a u_b\, \delp_c u_d \, \delp_e u_f,   \nn
& \delp_a \delp_b u_c \, \delp_d \ln s , \delp_a u_b \, \delp_c  \delp_d \ln s , \delp_a \delp_b\delp_c \ln s,  \nn
&  \delp_a \delp_b \ln s \, \delp_c \ln s, \delp_a \ln s \, \delp_b \ln s \, \delp_c \ln s , \delp_a u_b \, R_{cdef}, \nn
&\delp_a \ln s \, R_{bcde}, \nabla_a R_{bcde}  \big\} .
\end{align}
We can then show that the third-order non-conformal stress-energy tensor, in the Landau frame, takes the form
\begin{align}\label{T3ordFull}
T^{ab}_{(3)} = \CP_{(3)} \Delta^{ab} + t_{(3)}^{ab},
\end{align}
where $\CP_{(3)}$ is a sum of $23$ terms and $t^{ab}_{(3)}$ a sum of $45$ TST two-tensors, each with a new transport coefficient,
\begin{align}
&\CP_{(3)} = \sum_{i=1}^{23} \upsilon^{(3,0)}_i \CS_i,
&t_{(3)}^{ab} = \sum_{i=1}^{45} \upsilon^{(3,2)}_i \CT^{ab}_i.
\end{align}
The scalars and the two-tensors are listed in Appendices \ref{Sec:AppScalars} and \ref{AppTensors}, respectively. In total, these give $68$ new transport coefficients. If we were to work in a different frame, it is also useful to list all possible hydrodynamic vectors that may enter the expansion of $q^a_{(3)}$. The $28$ transverse third-order vectors are listed in Appendix \ref{Sec:AppVectors}. 

Out of the two-tensors, we can now construct conformal third-order hydrodynamics, as described in Sec. \ref{Sec:ConfHydro}. The expansion of the Weyl-covariant conformal stress-energy tensor takes the form
\begin{align}\label{T3ordConfFull}
T^{ab}_{(3)} = \sum_{i=1}^{20} \lambda^{(3)}_i \CO_i^{ab},
\end{align}
in which there are $20$ new Weyl-covariant tensors that require us to introduce $20$ new transport coefficients. In terms of $\CT^{ab}_i$, the conformal tensors are
\begin{small}
\begin{align}
\mathcal{O}_{1}=&\,\mathcal{T}_1+\frac{2 \mathcal{T}_{11}}{3}-\frac{\mathcal{T}_{12}}{3}+\frac{2 \mathcal{T}_{13}}{3}-\frac{2 \mathcal{T}_{14}}{3}-\frac{2 \mathcal{T}_{16}}{3}+\frac{2 \mathcal{T}_{17}}{9}-\frac{\mathcal{T}_{18}}{3}-\frac{\mathcal{T}_{19}}{3}-\frac{\mathcal{T}_{20}}{6}-2 \mathcal{T}_{27}+\frac{\mathcal{T}_{28}}{2}-\frac{\mathcal{T}_{40}}{2}+\frac{3 \mathcal{T}_{41}}{2}\nonumber\\
&\,+\mathcal{T}_{42}+\frac{3 \mathcal{T}_{43}}{2}+\frac{3 \mathcal{T}_{44}}{2}+\frac{\mathcal{T}_{45}}{2}  ,\\
\mathcal{O}_{2}=&\,\mathcal{T}_2 + \frac{2 \mathcal{T}_{11}}{3}+\frac{\mathcal{T}_{13}}{3}-\frac{\mathcal{T}_{15}}{3}-\mathcal{T}_{16}+\frac{2 \mathcal{T}_{17}}{9}-\frac{2 \mathcal{T}_{18}}{9}-\frac{4 \mathcal{T}_{19}}{9}-\frac{\mathcal{T}_{20}}{6}+\mathcal{T}_{27}-\frac{\mathcal{T}_{28}}{2}-\frac{3 \mathcal{T}_{29}}{2}-\frac{3 \mathcal{T}_{30}}{2}+\frac{\mathcal{T}_{33}}{2}\nonumber\\
&\,-\frac{\mathcal{T}_{35}}{2}+\frac{\mathcal{T}_{41}}{2}+\mathcal{T}_{42}-\frac{\mathcal{T}_{43}}{2}+\frac{5 \mathcal{T}_{44}}{2}+\frac{\mathcal{T}_{45}}{2},     \\
\mathcal{O}_{3}=&\,\mathcal{T}_3 + \frac{2 \mathcal{T}_{11}}{3}+\frac{\mathcal{T}_{13}}{3}-\mathcal{T}_{14}-\frac{\mathcal{T}_{15}}{3}+\frac{2 \mathcal{T}_{17}}{9}-\frac{2 \mathcal{T}_{18}}{9}-\frac{4 \mathcal{T}_{19}}{9}-\frac{\mathcal{T}_{20}}{6}-\frac{\mathcal{T}_{28}}{2}+\frac{\mathcal{T}_{29}}{2}-\frac{5 \mathcal{T}_{30}}{2}+\frac{\mathcal{T}_{33}}{2}-\frac{\mathcal{T}_{35}}{2}\nonumber\\
&\,+\frac{\mathcal{T}_{41}}{2}+\mathcal{T}_{42}+\frac{\mathcal{T}_{43}}{2}+\frac{3 \mathcal{T}_{44}}{2}+\frac{\mathcal{T}_{45}}{2}, \\
\mathcal{O}_{4}=&\,\mathcal{T}_4 + \frac{2 \mathcal{T}_{11}}{3}+\frac{\mathcal{T}_{13}}{3}-\frac{\mathcal{T}_{15}}{3}-\mathcal{T}_{16}+\frac{2 \mathcal{T}_{17}}{9}-\frac{2 \mathcal{T}_{18}}{9}-\frac{4 \mathcal{T}_{19}}{9}-\frac{\mathcal{T}_{20}}{6}-\frac{\mathcal{T}_{28}}{2}-\mathcal{T}_{30}+\frac{\mathcal{T}_{33}}{2}-\frac{\mathcal{T}_{35}}{2}+\mathcal{T}_{40}-\frac{5 \mathcal{T}_{41}}{2}\nonumber\\
&\,-\frac{3 \mathcal{T}_{43}}{2}+\frac{\mathcal{T}_{44}}{2}+\frac{\mathcal{T}_{45}}{2} , \\
\mathcal{O}_{5}=&\, \mathcal{T}_5 + \frac{2 \mathcal{T}_{11}}{3}+\frac{\mathcal{T}_{13}}{3}-\mathcal{T}_{14}-\frac{\mathcal{T}_{15}}{3}+\frac{2 \mathcal{T}_{17}}{9}-\frac{2 \mathcal{T}_{18}}{9}-\frac{4 \mathcal{T}_{19}}{9}-\frac{\mathcal{T}_{20}}{6}-\frac{\mathcal{T}_{28}}{2}-\mathcal{T}_{30}+\frac{\mathcal{T}_{33}}{2}-\frac{\mathcal{T}_{35}}{2}+\mathcal{T}_{40}-\frac{5 \mathcal{T}_{41}}{2}\nonumber\\
&\,-\frac{3 \mathcal{T}_{43}}{2}+\frac{\mathcal{T}_{44}}{2}+\frac{\mathcal{T}_{45}}{2},\\
\mathcal{O}_{6}=&\,\mathcal{T}_6 + \mathcal{T}_{11}-\frac{4 \mathcal{T}_{15}}{3}+\frac{2 \mathcal{T}_{17}}{9}-\frac{2 \mathcal{T}_{19}}{3}-\frac{\mathcal{T}_{20}}{6}+\frac{\mathcal{T}_{28}}{2}+\frac{\mathcal{T}_{29}}{2}-\frac{3 \mathcal{T}_{30}}{2}+\frac{3 \mathcal{T}_{33}}{2}-\frac{3 \mathcal{T}_{35}}{2}+2 \mathcal{T}_{40}-\frac{9 \mathcal{T}_{41}}{2}-\frac{9 \mathcal{T}_{43}}{2}\nonumber\\
&\,+\frac{3 \mathcal{T}_{44}}{2}+\frac{\mathcal{T}_{45}}{2},\\
\mathcal{O}_{7}=&\,\mathcal{T}_7-3 \mathcal{T}_9 -\frac{\mathcal{T}_{17}}{3}+\mathcal{T}_{19}-\frac{\mathcal{T}_{20}}{2}+\frac{3 \mathcal{T}_{28}}{2}-3 \mathcal{T}_{29}+\frac{3 \mathcal{T}_{40}}{2}-\frac{9 \mathcal{T}_{41}}{2}-\frac{9 \mathcal{T}_{43}}{2}+\frac{9 \mathcal{T}_{44}}{2}+\frac{3 \mathcal{T}_{45}}{2},\\
\mathcal{O}_{8}=&\,\mathcal{T}_8-\mathcal{T}_9-\frac{\mathcal{T}_{18}}{3}+\frac{\mathcal{T}_{19}}{3}+3 \mathcal{T}_{27}-\frac{3 \mathcal{T}_{29}}{2}-\frac{3 \mathcal{T}_{30}}{2}-3 \mathcal{T}_{43}+3 \mathcal{T}_{44} , \\
\mathcal{O}_{9}=&\,\mathcal{T}_{10}+\frac{\mathcal{T}_{20}}{6}+\frac{3 \mathcal{T}_{28}}{2}-\frac{3 \mathcal{T}_{40}}{2}+\frac{9 \mathcal{T}_{41}}{2}+3 \mathcal{T}_{42}+\frac{9 \mathcal{T}_{43}}{2}+\frac{9 \mathcal{T}_{44}}{2}+\frac{3 \mathcal{T}_{45}}{2},\\
\mathcal{O}_{10}=&\,\mathcal{T}_{21}-3 \mathcal{T}_{28},\\
\mathcal{O}_{11}=&\,\mathcal{T}_{22}-3 \mathcal{T}_{27}-\mathcal{T}_{28}+\frac{\mathcal{T}_{29}}{2}+\frac{3 \mathcal{T}_{30}}{2},\\
\mathcal{O}_{12}=&\,\mathcal{T}_{23}-\mathcal{T}_{28}-\mathcal{T}_{29},\\
\mathcal{O}_{13}=&\,\mathcal{T}_{24}-\mathcal{T}_{28}+\frac{\mathcal{T}_{29}}{2}-\frac{3 \mathcal{T}_{30}}{2},\\
\mathcal{O}_{14}=&\,\mathcal{T}_{25}-\mathcal{T}_{28},\\
\mathcal{O}_{15}=&\,\mathcal{T}_{26}-2 \mathcal{T}_{27}+\mathcal{T}_{30},\\
\mathcal{O}_{16}=&\,\mathcal{T}_{31}+\frac{\mathcal{T}_{33}}{2}+\mathcal{T}_{34}-\frac{\mathcal{T}_{35}}{6}-\frac{\mathcal{T}_{36}}{3}+\frac{\mathcal{T}_{40}}{2}+\mathcal{T}_{45},\\
\mathcal{O}_{17}=&\,\mathcal{T}_{32}+2 \mathcal{T}_{34}-\frac{2 \mathcal{T}_{35}}{3}-\frac{4 \mathcal{T}_{36}}{3}+\frac{2 \mathcal{T}_{40}}{3}+\frac{4 \mathcal{T}_{45}}{3},\\
\mathcal{O}_{18}=&\,\mathcal{T}_{37}-2 \mathcal{T}_{40}+6 \mathcal{T}_{41}+6 \mathcal{T}_{42}+6 \mathcal{T}_{43}+6 \mathcal{T}_{44}+2 \mathcal{T}_{45},\\
\mathcal{O}_{19}=&\,\mathcal{T}_{38}-\mathcal{T}_{40}+2 \mathcal{T}_{41}+\mathcal{T}_{42}+2 \mathcal{T}_{44},\\
\mathcal{O}_{20}=&\, \mathcal{T}_{39}-\mathcal{T}_{40}+2 \mathcal{T}_{41}+\mathcal{T}_{42}+2 \mathcal{T}_{43}.
\end{align}
\end{small}
It should be understood that all $\CO$ and $\CT$ tensors carry two indices, $a$ and $b$. Clearly, the choice of $\CO$ is not unique and we could have chosen a different linear combination. What is important is the number of linearly independent combinations that gives us the number of corresponding transport coefficients.

Furthermore, it is interesting to note that in flat space, only the first $15$ Weyl-covariant tensors, $\CO^{ab}_1$ -- $\CO^{ab}_{15}$, remain non-zero. This is consistent with the results from kinetic theory, within the context of which \cite{Jaiswal:2013vta,Jaiswal:2014raa,Chattopadhyay:2014lya} found $14$ transport coefficients in third-order hydrodynamics. Similarly, in second-order hydrodynamics, kinetic theory gave only $3$ instead of $4$ conformal transport coefficients in flat space. There, $\lambda_3 = 0$, which is the transport coefficient corresponding to the term consisting only of vorticity in the gradient expansion, i.e. $\Omega^{\langle a}_{~~c} \Omega^{b\rangle c}$. It is therefore plausible that the tensor missing from the kinetic theory analysis is the one consisting solely of vorticity tensors, i.e. $\sim \Omega^3$.\footnote{We thank Amaresh Jaiswal for bringing the results of kinetic theory to our attention.}

Finally, we stress that the number of new independent transport coefficients should be understood as an upper bound on their actual ``physical" number (a maximal possible set), since there could be further constraints coming from imposing non-negativity of the entropy current divergence \cite{Loganayagam:2008is,Romatschke:2009kr,Haack:2008xx,Kovtun:2012rj,Bhattacharyya:2012nq,Jensen:2012jh,Moore:2012tc,Banerjee:2012iz}. As we have not yet performed the local entropy current analysis in third-order hydrodynamics, we cannot establish whether the non-negativity of entropy production imposes any constraints on the $68$ and $20$ un-constrained transport coefficients in non-conformal and conformal hydrodynamics, respectively. It is interesting to note that while such an analysis restricts the number of non-conformal coefficients, at least in second-order hydrodynamics, the number of conformal coefficients remains the same \cite{Bhattacharyya:2012nq}.

\section{Properties of linear and non-linear third-order hydrodynamic transport}

In this section, we analyse the effects of third-order hydrodynamics on linear transport as well as an example of non-linear hydrodynamics. First, we compute the dispersion relations for shear and sound modes to third order in the hydrodynamic expansion. Then, we compute the scalar channel stress-energy tensor two-point function. Since the dispersion relations and two-point functions follow from linear perturbations of $T^{ab}$ in the hydrodynamic variables, only the stress-energy tensor terms linear in $u^a$, $\ln s$ and $g_{ab}$ are relevant. Lastly, as an example of a hydrodynamic process that is sensitive to non-linear terms, we compute the energy density of the boost-invariant Bjorken flow to third order.

\subsection{Shear and sound dispersion relations}\label{Sec:DispRel}

To compute the shear and sound dispersion relations in flat space, we need the linear part of the non-conformal third-order stress-energy tensor in Eq. \eqref{T3ordFull} that includes derivatives of the velocity and entropy density fields. This is given by the following series of terms:
\begin{align}\label{T3lin}
T^{ab}_{(3),lin}= \sum_{i=1}^3 \upsilon^{(3,0)}_i \CS_i \, \Delta^{ab} + \sum_{i=1}^{6}\upsilon^{(3,2)}_i \CT^{ab}_i.
\end{align}
It is then easy to find the relevant dispersion relations by following e.g. \cite{Romatschke:2009im,Romatschke:2009kr}. What we need to do is perturb $u^a$ and $\ln s$ around the equilibrium values $u^a = \left(1,0,0,0\right)$ and $\ln s = \text{const.}$, and solve the generalised relativistic Navier-Stokes equations \eqref{EoMs1} and \eqref{EoMs2}, that result from $\nabla_a T^{ab} = 0$, for $\omega(k)$ in the small $k$ expansion.

In flat space, $\delp_a = u_a u^b \partial_b + \partial_a$, which evaluated on the background, $u^a = (1,0,0,0)$, gives $\delp_a = u_a \partial_0 + \partial_a$. Hence, to zeroth order in linear perturbations, $\delp_0 = 0$ and $\delp_i = \partial_i$. For the purposes of linearised hydrodynamics, therefore
\begin{align}\label{LinDelPerp}
\partial_{\perp a}  = (0, \partial_i) + \,\text{non-linear}\,,
\end{align}
which makes it clear that the linearised Navier-Stokes equations have no additional time derivatives acting on the hydrodynamic variables that come from higher-order hydrodynamics. Using the notation of Eq. \eqref{LinDelPerp}, the linearised stress-energy tensor \eqref{T3lin} in flat space becomes
\begin{align}\label{T3lin2}
T^{ab}_{(3),lin} =& \left( \upsilon^{(3,0)}_1 + \upsilon^{(3,0)}_2 + \upsilon^{(3,0)}_3 \right) \eta^{ab} \parp_d \parp^d \parp_c  u^c  + \left( \upsilon^{(3,2)}_1 + \upsilon^{(3,2)}_2 + \upsilon^{(3,2)}_4 \right) \parp_c \parp^c \parp^{\la a} u^{ b\ra } \nn
&+  \left( \upsilon^{(3,2)}_3 + \upsilon^{(3,2)}_5 + \upsilon^{(3,2)}_6 \right) \parp^{\la a} \parp^{ b\ra} \parp_c u^c .
\end{align}

In the two different channels, i.e. the transverse (shear) and the longitudinal (sound), we need to turn on the following fluctuations:
\begin{align}
&\text{shear:}    & \delta u^y = e^{-i\omega t + i k x} f^{t}_{\omega,k}&  & \delta\ln s = e^{-i\omega t + i k x} g_{\omega, k}, \\
&\text{sound:}   & \delta u^x = e^{-i\omega t + i k x} f^{l}_{\omega,k} & & \delta\ln s = e^{-i\omega t + i k x} g_{\omega, k}.
\end{align}
The two dispersion relations in third-order hydrodynamics then follow from solving the equations of motion \eqref{EoMs1} and \eqref{EoMs2} with the linearised stress-energy tensor \eqref{T3lin2}. We find that the dispersion relations, which are expressed in the form of Eq. \eqref{DispRelInt}, involve nine new third-order transport coefficients,
\begin{align}
&\text{shear:}   & & \omega = - i \frac{\eta}{\varepsilon + P} k^2 - i  \left[ \frac{\eta^2 \tp  }{\left(\varepsilon + P\right)^2}   +  \frac{1}{2} \frac{  \upsilon^{(3,2)}_1 + \upsilon^{(3,2)}_2 + \upsilon^{(3,2)}_4  }{ \varepsilon + P} \right]  k^4  + \CO\left(k^5\right),  \label{ShearDisp} \\
&\text{sound:}  & & \omega =  \pm c_s k - i \Gamma k^2 \mp \frac{1}{2 c_s} \left[ \Gamma^2 -2 c_s^2 \left(\frac{2}{3}\frac{\eta\tp}{\varepsilon+P} + \frac{1}{2}\frac{\zeta\tau_\pi}{\varepsilon+P}  \right) \right] k^3  \nn
& && ~~~~~\, - i \left[ 2 \Gamma  \left(\frac{2}{3}\frac{\eta\tp}{\varepsilon+P} + \frac{1}{2}\frac{\zeta\tau_\pi}{\varepsilon+P}  \right)   +  \frac{3 \sum_{i=1}^3 \upsilon^{(3,0)}_i + 2 \sum_{i=1}^6 \upsilon^{(3,2)}_i }{6 \left(\varepsilon + P\right)}  \right] k^4 + \CO\left(k^5\right) ,\label{SoundDisp}
\end{align}
where
\begin{align}
\Gamma = \left(\frac{2}{3}\frac{\eta}{\varepsilon+P} + \frac{1}{2}\frac{\zeta}{\varepsilon+P}  \right).
\end{align}
These relations should be understood as generalisations of diffusion and the sound propagation in relativistic fluids.

In conformal hydrodynamics, we find that six of the $\lambda^{(3)}_i$ coefficients give non-vanishing contributions to the linearised stress-energy tensor: from $\lambda^{(3)}_1$ to $\lambda^{(3)}_6$. These transport coefficients precisely correspond to the first six non-conformal $\upsilon^{(3,2)}_i$. The relevant part of the linearised conformal third-order stress-energy tensor \eqref{T3ordConfFull} is then
\begin{align}\label{T3linConf}
T^{ab}_{(3),con,lin} = \left( \lambda^{(3)}_1 + \lambda^{(3)}_2 +  \lambda^{(3)}_4 \right) \parp_c \parp^c \parp^{\la a} u^{ b\ra }+  \left( \lambda^{(3)}_3 +  \lambda^{(3)}_5 + \lambda^{(3)}_6 \right) \parp^{\la a} \parp^{ b\ra} \parp_c u^c .
\end{align}
As only two combinations of the six $\lambda^{(3)}_i$ enter the linearised stress-energy tensor in flat space, it is convenient to define the following two combinations of the six conformal transport coefficients:
\begin{align}
&\theta_1 \equiv - \left( \lambda^{(3)}_1 + \lambda^{(3)}_2 +  \lambda^{(3)}_4 \right) , \\
&\theta_2 \equiv  - \left( \lambda^{(3)}_3 +  \lambda^{(3)}_5 + \lambda^{(3)}_6 \right).
\end{align}
Hence, the conformal shear and sound dispersion relations become
\begin{align}
&\text{shear:}   & & \omega = - i \frac{\eta}{\varepsilon + P} k^2 -  i  \left[ \frac{\eta^2 \tp  }{\left(\varepsilon + P\right)^2}   -  \frac{1}{2} \frac{  \theta_1 }{ \varepsilon + P} \right]  k^4 + \CO\left(k^5\right),  \label{ShearDispConf} \\
&\text{sound:}  & & \omega =  \pm c_s k - i \Gamma_c k^2 \mp \frac{\Gamma_c}{2 c_s} \left( \Gamma_c -2 c_s^2 \tp \right) k^3 - i \left[ \frac{8}{9} \frac{\eta^2 \tp  }{\left(\varepsilon + P\right)^2}   -  \frac{1}{3} \frac{  \theta_1 + \theta_2 }{ \varepsilon + P} \right]  k^4 + \CO\left(k^5\right) ,\label{SoundDispConf}
\end{align}
where the conformal $\Gamma = \Gamma_c$, in the absence of bulk viscosity $\zeta$, is
\begin{align}
\Gamma_c = \frac{2}{3}\frac{\eta}{\varepsilon+P},
\end{align}
and the speed of sound $c_s$ is fixed to $c_s = 1/\sqrt{3}$ for conformal fluids in four dimensions.

\subsection{Stress-energy tensor two-point function}\label{sec:TxyxyFT}

A particularly simple and useful two-point function of the stress-energy tensor $T^{ab}$ can be obtained by perturbing the flat Minkowski metric tensor in the $g_{xy}$ component, with the momentum flowing in the $z$-direction. Perturbing $T^{xy}$ to first order in $\delta g_{xy}$ and taking a derivative with respect to the metric perturbation gives the scalar two-point function $\la T^{xy}, T^{xy}\ra$.\footnote{The name {\em scalar} is inspired by the AdS/CFT correspondence where the $\delta g_{xy} \sim \exp\left\{-it\omega + i k z\right\}$ fluctuation in the bulk behaves as a minimally-coupled massless scalar. Alternatively, this channel is also referred to as the {\em tensor} channel because the fluctuation $\delta g_{xy}$ transforms as the spin-$2$ tensor fluctuation.} In this channel, the velocity and entropy density fields can be kept constant as they decouple from $\delta g_{xy}$. For a detailed description of such a calculation, see e.g. \cite{Baier:2007ix,Moore:2010bu}.

Using the non-conformal third-order stress-energy tensor that includes Eq. \eqref{T3ordFull}, we can compute the two-point function. We find  
\begin{align}\label{TxyxyFTnonC}
\la T^{xy}, T^{xy}\ra \equiv G^{xy,xy} (\omega, k) =&~ P - i \eta \omega + \left( \eta \tp - \frac{\kappa}{2} +\kappa^*\right) \omega^2 - \frac{\kappa}{2}   k^2 \nn
& + \frac{i}{2}  \left( \upsilon^{(3,2)}_{31} + \upsilon^{(3,2)}_{32} - \upsilon^{(3,2)}_{34}  \right) \omega^3  + \frac{i}{2} \left(\upsilon^{(3,2)}_1 - \upsilon^{(3,2)}_{31} - \upsilon^{(3,2)}_{32}  \right) \omega k^2  .
\end{align}
In order to use our results for a calculation of transport coefficients in a theory with a gravitational dual, we need the correlator for the conformal sub-class of third-order fluids, which follows from using the stress-energy tensor \eqref{T3ordConfFull}. The result is
\begin{align}\label{TxyxyFT}
G^{xy,xy}(\omega, k) =&~ P - i \eta \omega + \left( \eta \tp - \frac{\kappa}{2} \right) \omega^2 - \frac{\kappa}{2}   k^2 \nn
& - \frac{i}{2}  \lambda^{(3)}_{17} \omega^3  + \frac{i}{2} \left(\lambda^{(3)}_1 - \lambda^{(3)}_{16} - \lambda^{(3)}_{17}\right) \omega k^2  .
\end{align}

\subsection{The Bjorken flow}
We now turn to an example of a non-linear hydrodynamic calculation in a conformally invariant theory: We will analyse the effects of third-order hydrodynamics on the boost-invariant Bjorken flow \cite{Bjorken:1982qr}, which is relevant to the study of relativistic heavy ion collisions. The configuration describes hydrodynamic propagation of a four-dimensional boost-invariant plasma along the $z$-axis with velocity $z/t$. The solution can be conveniently written down in terms of proper time $\tau = \sqrt{t^2 - z^2}$, the rapidity parameter $\xi = \text{arctanh} (z/t)$ and a two-dimensional Euclidean plane, perpendicular to the direction of the flow. In these (flat metric) coordinates, the fluid is at rest. Hence, the solution for the velocity field and the metric can be written as
\begin{align}
&u^a = \left(u^\tau, u^\xi, \vec{u}^\perp \right) = \left(1,0,0,0 \right), \\
&g_{ab} = - d\tau^2 + \tau^2 d\xi^2 + d \vec{x}_\perp^2.
\end{align}
We then need to solve the hydrodynamic equations of motion for the remaining scalar field. Following \cite{Baier:2007ix}, it is convenient to work with the energy density $\varepsilon(\tau)$ instead of the entropy density field. The function $\varepsilon(\tau)$ must solve Eq. \eqref{EoMT1}, i.e.
\begin{align}\label{Bjorken1}
D \varepsilon + \left(\varepsilon + P \right) \nabla_a u^a + \Pi^{ab} \nabla_a u_b = 0,
\end{align}
where $\varepsilon(\tau)$ is only a function of $\tau$. In the equation of motion, we used $\Pi^{ab}$ as defined in Eq. \eqref{DefTasExp}. Given that the $T^a_{~a} = 0$ condition for a CFT gives $P = \varepsilon / 3$ and that the only non-zero component of $\nabla_a u_b$ is $\nabla_\xi u_\xi = \delp_\xi u_\xi  =\tau$ \cite{Bjorken:1982qr}, in four space-time dimensions Eq. \eqref{Bjorken1} reduces to
\begin{align}\label{EpsBjorkenEq}
\partial_\tau \varepsilon + \frac{4}{3} \frac{\varepsilon}{\tau}  +  \tau \Pi^{\xi\xi} = 0.
\end{align}
In conformal hydrodynamics, we have $\Pi^{ab} = t^{ab}$. Using the stress-energy tensor to third order in the hydrodynamic expansion then gives 
\begin{align}
\Pi^{\xi\xi} = t^{\xi\xi} =& - \frac{4 \eta}{3} \frac{1}{\tau^3}  - \left[ \frac{8 \eta \tp}{9} - \frac{8 \lambda_1}{9}\right] \frac{1}{\tau^4}  - \left[ \frac{\lambda^{(3)}_1}{6 } + \frac{4 \lambda^{(3)}_2}{3 } + \frac{4 \lambda^{(3)}_3}{3 } + \frac{5 \lambda^{(3)}_4}{6} + \frac{5 \lambda^{(3)}_5}{6 } + \frac{4 \lambda^{(3)}_6}{3 } \right. \nn
&\left.  -  \frac{\lambda^{(3)}_7}{2 }  + \frac{3\lambda^{(3)}_8}{2 } + \frac{\lambda^{(3)}_9}{2 } - \frac{2\lambda^{(3)}_{10}}{3 } - \frac{11\lambda^{(3)}_{11}}{6 } - \frac{\lambda^{(3)}_{12}}{3 } + \frac{\lambda^{(3)}_{13}}{6} - \lambda^{(3)}_{15} \right] \frac{1}{\tau^5} + \CO\left(\tau^{-6}\right).
\end{align} 
We can see that $\lambda^{(3)}_{14}$ does not contribute to the energy density. Furthermore, as discussed in Section \ref{Sec:3rdOrdHydro}, the coefficients $\lambda^{(3)}_{16}$ -- $\lambda^{(3)}_{20}$ correspond to tensors consisting solely of derivatives of the metric tensor that vanish in flat space. As a result, the Bjorken flow solution does not depend on any of those transport coefficients.

The equation \eqref{EpsBjorkenEq} can now be solved by first noting that the transport coefficients are functions of thermodynamical variables and thus $\varepsilon$. Because we are working with a CFT, they scale as \cite{Baier:2007ix}
\begin{align}\label{BjorkenScaling}
\eta = C \bar{\eta} \left( \frac{\varepsilon}{C} \right)^{3/4} , && \eta \tp = C \bar\eta \bar{\tp} \left( \frac{\varepsilon}{C} \right)^{1/2}, &&  \lambda_1 = C \bar\lambda_{1} \left( \frac{\varepsilon}{C} \right)^{1/2},&& \lambda^{(3)}_n = C \bar\lambda^{(3)}_{n}  \left( \frac{\varepsilon}{C} \right)^{1/4},
\end{align}
where $C$, $\bar\eta$, $\bar{\tp}$ and $\bar\lambda^{(3)}_{n}$ are constants. The energy density function for the Bjorken flow in third-order hydrodynamics is then
\begin{align}\label{Bjorken3rdFT}
\frac{\varepsilon(\tau)}{C}  =&~ \frac{1}{\tau^{2 - \nu}} - 2 \bar\eta \frac{1}{\tau^2} + \left[ \frac{3 \bar\eta^2 }{2}- \frac{2 \bar\eta \bar{\tp}}{3} + \frac{2 \bar\lambda_{1}}{3} \right] \frac{1}{\tau^{2+\nu}} \nn
&- \left[ \frac{\bar\eta^3}{2} - \frac{7 \bar\eta ^2 \bar{\tp} }{9} + \frac{7 \bar\eta \bar\lambda_1  }{9}  +\frac{\bar\lambda^{(3)}_1}{12 } + \frac{2 \bar\lambda^{(3)}_2}{3 } + \frac{2 \bar\lambda^{(3)}_3}{3 } + \frac{5 \bar\lambda^{(3)}_4}{12} + \frac{5 \bar\lambda^{(3)}_5}{12 } + \frac{2 \bar\lambda^{(3)}_6}{3 } \right.    \nn
&\left.  -  \frac{\bar\lambda^{(3)}_7}{4 }  + \frac{3\bar\lambda^{(3)}_8}{4 } + \frac{\bar\lambda^{(3)}_9}{4 } - \frac{\bar\lambda^{(3)}_{10}}{3 } - \frac{11\bar\lambda^{(3)}_{11}}{12 } - \frac{\bar\lambda^{(3)}_{12}}{6 } + \frac{\bar\lambda^{(3)}_{13}}{12} - \frac{ \bar\lambda^{(3)}_{15}}{2}  \right] \frac{1}{\tau^{2+2\nu}} + \CO\left(\tau^{-2-3\nu}\right),
\end{align}
where $\nu = 2/3$.

\section{Conformal transport in the $\CN = 4$ supersymmetric Yang-Mills theory}\label{Sec.N4}

\subsection{Shear and sound dispersion relations} 

Using our classification of third-order hydrodynamics, we now turn to the AdS/CFT correspondence in order to compute the new transport coefficients $\theta_1$ and $\theta_2$ that entered into the linear dispersion relations \eqref{ShearDispConf} and \eqref{SoundDispConf} in Sec. \ref{Sec:DispRel}. We will rely on the analytic results for the shear and the sound dispersion relations in the $\CN=4$ supersymmetric Yang-Mills theory at infinite 't Hooft coupling and infinite number of colours, $N_c$. The shear dispersion, computed by using well-known AdS/CFT techniques \cite{Policastro:2002se,Policastro:2002tn,Kovtun:2005ev}, was obtained in \cite{Baier:2007ix}. To find the dispersion of the sound mode, we extend the calculation of \cite{Baier:2007ix} to one order higher and look for the small-$\omega$ and small-$k$ behaviour of the lowest (hydrodynamic) quasi-normal mode of the background geometry dual to the $\CN=4$ theory at finite temperature. The geometry is the five-dimensional, non-extremal asymptotically-AdS black brane solution of Einstein gravity with a negative cosmological constant: 
\begin{align}\label{BBsol}
ds^2 = \frac{\left(\pi T\right)^2}{u} \left( - f(u) dt^2 + d\vec{x}^2 \right) + \frac{du^2}{4 u^2 f(u)},
\end{align}
where the emblackening factor is $f(u) = 1 - u^2$, and we have set the AdS radius to one. The quasi-normal mode results that we find are consistent with \cite{Bu:2015bwa}. To quartic order in momentum $k$, the two dispersion relations are given by
\begin{align}
&\text{Shear:}   & & \omega = - \frac{i}{4 \pi T} k^2 - \frac{i \left(1-\ln 2\right)}{32 \pi^3 T^3}   k^4 , \\
&\text{Sound:}   & & \omega = \pm \frac{1}{\sqrt{3} } k - \frac{i}{6 \pi T} k^2 \pm \frac{3-2\ln 2}{24 \sqrt{3} \pi^2 T^2} k^3 - \frac{i \left(\pi^2 - 24 + 24 \ln 2 - 12 \ln^2 2  \right)}{864  \pi^3 T^3}   k^4 .
\end{align}

The $k^4$ term of the shear dispersion relation only depends on the third-order hydrodynamic transport coefficient $\theta_1$, as found in Eq. \eqref{ShearDispConf}, while the sound dispersion relation depends on both $\theta_1$ and $\theta_2$, as can be seen from Eq. \eqref{SoundDispConf}. Hence, we can immediately determine their values in the $\CN=4$ theory, by using the already known transport coefficients at first and second order \cite{Baier:2007ix,Bhattacharyya:2008jc}. To date, these coefficients are known up to leading-order corrections in the 't Hooft coupling, $\gamma = \lambda^{-3/2} \zeta(3)/8$, where $\lambda = g^2_{YM} N_c$ \cite{Buchel:2004di, Benincasa:2005qc, Buchel:2008sh, Buchel:2008ac,Buchel:2008bz,Buchel:2008kd,Saremi:2011nh,Grozdanov:2014kva}. The complete list was presented in \cite{Grozdanov:2014kva} by including the value of $\lambda_2$:
\begin{align}
\eta &= \frac{\pi}{8} N^2_c T^3\, \left[ 1 + 135 \gamma  + \CO\left(\gamma^2\right)   \right] , \label{tcsym1} \\
\tau_{\Pi}  &= \frac{ \left( 2 - \ln{2}\right)}{2\pi T}   + \frac{375 \gamma}{4 \pi T} \, + \CO\left(\gamma^2\right)  , \label{tcsym2} \\
\kappa &= \frac{N_c^2 T^2}{8}  \left[ 1 - 10 \gamma  + \CO\left(\gamma^2\right)  \right], \label{tcsym3} \\
\lambda_1 &=  \frac{N_c^2 T^2}{16}  \left[ 1 + 350 \gamma  + \CO\left(\gamma^2\right)   \right],  \label{tcsym4}  \\
\lambda_2 &=- \frac{N_c^2 T^2}{16} \left[ 2\ln{2}   + 5 \left(97+54 \ln{2}\right) \gamma + \CO\left(\gamma^2\right) \right], \label{tcsym5} \\
\lambda_3 &=  \frac{ 25 N_c^2 T^2}{2} \, \gamma +\CO\left(\gamma^2\right) .\label{tcsym6}
\end{align}
Using these results, along with $\eta / s = 1 / (4\pi) + \CO\left(\gamma\right)$ \cite{Kovtun:2003wp,Kovtun:2004de}, and $s=\pi^2 N_c^2 T^3 / 2$, we find the new third-order coefficients to be
\begin{align}
&\theta_1 = \frac{N_c^2 T}{32 \pi} + \CO\left(\gamma\right) , \\
&\theta_2 = \frac{N_c^2 T}{48 \pi} \left( \frac{9}{2} - \frac{\pi^2}{12} - 4 \ln 2 + \ln^2 2 \right) + \CO\left(\gamma\right).
\end{align}

\subsection{Stress-energy tensor two-point function}

In this section, we present the scalar channel, retarded two-point function of the stress energy tensor $G^R_{xy,xy}(\omega,k)$ to third order in the gradient expansion by following \cite{Son:2002sd,Policastro:2002se}. The correlation function can be computed by perturbing the $g_{xy}$ component of the background metric tensor \eqref{BBsol}. The fluctuation of $g_{xy}$, i.e. the $g_{xy} \to g_{xy} + h_{xy}$ must be propagating along the $z$ direction. We can then define $\phi = u h_{xy} / (\pi T)^2$ so that $\phi = h^x_{~y}$, giving us the momentum space equation of motion 
\begin{align}
\phi_k'' + \frac{1+u^2}{u f } \phi_k' + \frac{\wfr^2 - \qfr^2 f}{u f^2}\phi_k = 0,
\end{align}
where $\wfr \equiv \omega / (2\pi T)$ and $\qfr \equiv k / (2 \pi T)$ are the dimensionless energy-momentum variables. In order to recover the retarded two-point function, the solution must be in-falling at the horizon,
\begin{align}
\phi_k (u) = \left(1-u\right)^{-i \wfr/2 } F_k(u),
\end{align}
and $F_k (u)$ must be regular at the horizon ($u=1$). The expression for $F_k(u)$ to second order in the hydrodynamic expansion, i.e. $\CO\left(\wfr^2, \wfr\qfr,\qfr^2\right)$, and the prescription for computing the retarded two-point function can be found in \cite{Son:2002sd,Policastro:2002se}. For our purposes, we again need to extend this calculation to one order higher. Because the expression for $F_k(u)$ is rather lengthy, it will not be stated here. We only present the two-point function with the third-order hydrodynamic corrections:
\begin{align}
G_R^{xy,xy} (\wfr,\qfr)=&~ - \frac{N_c^2 \pi^2 T^4}{4} \left\{ - \frac{1}{2} + i \wfr  -  \left(1 - \ln 2  \right) \wfr^2 + \qfr^2 \right.\nn
&\left. +   \left( \frac{\pi^2}{12} +2 \ln 2 - \ln^2 2\right) i \wfr^3 - 2 \ln 2\,  i \wfr \qfr^2   \right\} . 
\end{align}
Using the field theoretic result from Eq. \eqref{TxyxyFT}, we can find two new (linear combinations of the) transport coefficients:
\begin{align}
\lambda^{(3)}_{1} - \lambda^{(3)}_{16} &= \frac{N_c^2 T}{16 \pi} \left(\frac{\pi^2}{12} + 4 \ln 2 - \ln^2 2 \right) + \CO\left(\gamma\right) ,\\  
\lambda^{(3)}_{17} &= \frac{N_c^2 T}{16 \pi} \left(\frac{\pi^2}{12} + 2 \ln 2 - \ln^2 2 \right) + \CO\left(\gamma\right) .
\end{align} 

\subsection{The holographic Bjorken flow}

For our final calculation, we consider the holographic dual of the boost-invariant Bjorken flow in the supersymmetric $\CN=4$ Yang-Mills theory, which was developed in \cite{Janik:2005zt,Janik:2006ft,Heller:2007qt,Heller:2008fg,Heller:2011ju}. The analytic result for the energy density written as a function of proper time, $\varepsilon(\tau)$, to third order in the gradient expansion was found in \cite{Booth:2009ct}. It is given by
\begin{align}
\varepsilon(\tau) = \frac{N_c^2 }{2\pi^2 } \left\{\frac{1}{\tau^{4/3}} - \frac{\sqrt{2}}{3^{3/4}} \frac{1}{\tau^2} + \left( \frac{1 + 2 \ln 2}{12 \sqrt{3} } \right) \frac{1}{\tau^{8/3}} - \left( \frac{3 - 2\pi^2 - 24 \ln 2 + 24 \ln^2 2}{324\sqrt{2} \cdot 3^{1/4}} \right) \frac{1}{\tau^{10/3}}  \right\}.
\end{align}
Using the field theoretic result from Eq. \eqref{Bjorken3rdFT} together with the knowledge of the standard $\CN=4$ results from second-order hydrodynamics:
\begin{align}
\varepsilon = \frac{3}{8} \pi^2 N_c^2 T^4 , && \eta = \frac{\pi}{8} N_c^2 T^3, && \tp = \frac{2-\ln 2}{2\pi T}, && \lambda_1 = \frac{\eta}{2\pi T},
\end{align}
we can first find
\begin{align}
C = \frac{N_c^2}{2\pi^2}, && \bar\eta = \frac{1}{\sqrt{2} \cdot 3^{3/4}}, &&\bar{\tp} = \frac{3^{1/4}\left( 2 - \ln 2\right) }{2 \sqrt{2}}, && \bar\lambda_1 = \frac{1}{4\sqrt{3}}.
\end{align}
These constants then allow us to solve for the third-order ones, giving us
\begin{align}
\frac{\bar\lambda^{(3)}_1}{6 } + \frac{4 \bar\lambda^{(3)}_2}{3 } + \frac{4 \bar\lambda^{(3)}_3}{3 } + \frac{5 \bar\lambda^{(3)}_4}{6} + \frac{5 \bar\lambda^{(3)}_5}{6 } + \frac{4 \bar\lambda^{(3)}_6}{3 }  -  \frac{\bar\lambda^{(3)}_7}{2 } & \nn
+ \frac{3\bar\lambda^{(3)}_8}{2 } + \frac{\bar\lambda^{(3)}_9}{2 } - \frac{2\bar\lambda^{(3)}_{10}}{3 } - \frac{11\bar\lambda^{(3)}_{11}}{6 } - \frac{\bar\lambda^{(3)}_{12}}{3 } + \frac{\bar\lambda^{(3)}_{13}}{6} - \bar\lambda^{(3)}_{15}  &= \frac{15 - 2 \pi^2 - 45 \ln 2 + 24 \ln^2 2}{162 \sqrt{2} \cdot 3^{1/4}}.
\end{align}
Using the scaling relations \eqref{BjorkenScaling} finally gives us a linear combination of $14$ third-order transport coefficients in the $\CN=4$ theory at infinite 't Hooft coupling:
\begin{align}
\frac{\lambda^{(3)}_1}{6 } + \frac{4 \lambda^{(3)}_2}{3 } + \frac{4 \lambda^{(3)}_3}{3 } + \frac{5 \lambda^{(3)}_4}{6} + \frac{5 \lambda^{(3)}_5}{6 } + \frac{4 \lambda^{(3)}_6}{3 }   -  \frac{\lambda^{(3)}_7}{2 } & \nn
+ \frac{3\lambda^{(3)}_8}{2 } + \frac{\lambda^{(3)}_9}{2 } - \frac{2\lambda^{(3)}_{10}}{3 } - \frac{11\lambda^{(3)}_{11}}{6 } - \frac{\lambda^{(3)}_{12}}{3 } + \frac{\lambda^{(3)}_{13}}{6} - \lambda^{(3)}_{15}  &=\frac{N_c^2 T}{648 \pi} \left( 15 - 2 \pi^2 - 45 \ln 2 + 24 \ln^2 2 \right) \nn
& ~~~+ \CO(\gamma).
\end{align}
The 't Hooft coupling corrections to all five linear combinations of the conformal third-order transport coefficients found in this section remain to be computed.  
 
\section{Discussion}

In this paper, we presented a systematic algorithm for constructing the tensors that can be used to build the hydrodynamic gradient expansion at any order. We then used it to classify third-order hydrodynamics of uncharged fluids, i.e. the stress-energy tensor in the absence of Noether currents. The conservation equation for this stress-energy tensor therefore represents the most general, non-linear extension of the relativistic Navier-Stokes equations to the next-to-leading-order in the small energy-momentum gradient expansion.

In the non-conformal case, we found that $23$ scalars and $45$ independent tensors could be included in the gradient expansion, thus giving us in total $68$ new transport coefficients. In terms of the first-order Navier-Stokes fluids, the transport coefficients from the scalar and the tensor channels can be thought of as higher-order corrections to the bulk and the shear viscosity terms, respectively. Together with the $2$ and the $15$ transport coefficients in first and second-order expansions, this implies that there are $85$ coefficients that are required to completely describe uncharged fluids to third order.

For the conformal sub-class of fluids, which have a traceless homogeneously transforming stress-energy tensor under the non-linear Weyl transformations, we further showed that $20$ linearly independent tensors could be included at third order in the gradient expansion, resulting in $20$ new conformal transport coefficients. Hence, by including the $1$ and $5$ conformal transport coefficients in first and second-order hydrodynamics, we conclude that conformal fluids are described by $26$ transport coefficients (and independent tensor structures) to third order.

Using our results, we computed the linear dispersion relations, $\omega(k)$, for the shear and sound mode, to fourth order in $k$. These corrections to diffusion and the sound propagation turned out to depend on $9$ and $6$ transport coefficients in the non-conformal and the conformal cases, respectively. By employing the AdS/CFT calculations of the dispersion relations for the $\CN=4$ supersymmetric Yang-Mills theory at infinite 't Hooft coupling and number of colours, we were then able to determine the values of two new third-order transport coefficients in this theory, which we named $\theta_1$ and $\theta_2$. Each of the two $\theta_i$ depend on three of the conformal transport coefficients, $\lambda^{(3)}_i$. In flat space, these are the only two independent transport coefficients (or their combinations) that enter the linearised stress-energy tensor. Within linear response, we further used the third-order stress-energy tensor to compute the scalar (spin-$2$) two-point function, which gave us another two linear combinations of the transport coefficients that could be computed from holography. Lastly, by computing the third-order correction to the energy density of the non-linear boost-invariant Bjorken flow, we were able to find an additional linear combination of the new transport coefficients in the $\CN=4$ theory. Altogether, these calculations provided {\em five} linear combinations of the third-order transport coefficients in the $\CN=4$ theory at infinite 't Hooft coupling. Another fifteen remain to be computed in the future; a calculation that can be done either by finding and using the Kubo formulae for four-point Green's functions \cite{Barnes:2010jp,Moore:2010bu,Arnold:2011ja,Arnold:2011hp,Saremi:2011nh,Grozdanov:2014kva} or employing the tensorial structure found in this paper for a calculation within the fluid/gravity correspondence \cite{Bhattacharyya:2008jc,Rangamani:2009xk}. 

Having classified third-order hydrodynamics, numerous questions regarding its details remain to be answered in the future. By following the work of \cite{Moore:2010bu} in second-order hydrodynamics, one important task is to determine the relevant Kubo formulae with four-point correlation functions of the stress-energy tensor that would allow for a computation of all the transport coefficients, beyond the ones that enter into the linearised dispersion relations. These relations could then be used to compute the transport coefficients in perturbative and lattice QFT, as well as in strongly coupled field theories by using the gauge-gravity techniques. A particularly interesting task would be to study the fluid/gravity duality \cite{Bhattacharyya:2008jc} to third order in the gradient expansion, which could not only give us the microscopic values of the transport coefficients in a field theory dual to a particular gravitational setup, but also confirm the counting of the linearly independent tensors identified in this work.

Perhaps the most important remaining task is the understanding of the entropy current in higher-order hydrodynamics \cite{DeGroot:1980dk,Loganayagam:2008is,Bhattacharyya:2008xc,Romatschke:2009kr,Bhattacharyya:2012nq,Banerjee:2012iz,Kovtun:2012rj}. Firstly, the precise form at second-order remains to be found, i.e. fully expressed in terms of the second-order transport coefficients in the stress-energy tensor. As argued in \cite{Romatschke:2009kr}, third-order hydrodynamics may be required to resolve this issue even at second order. The construction of the entropy current at third order also remains to be performed so that potential constraints on the new maximal set of the transport coefficients can be uncovered following from the non-negativity of local entropy production \cite{Loganayagam:2008is,Romatschke:2009kr,Haack:2008xx,Kovtun:2012rj,Bhattacharyya:2012nq,Jensen:2012jh,Moore:2012tc,Banerjee:2012iz}. As a check of these relations, it would also be interesting to use the gauge-gravity duality and compute {\em all} non-conformal second-order transport coefficients in a field theory with a holographic dual.

In gauge-gravity duality, it has been known that certain combinations of transport coefficients give universal values in large classes of gravitational setups. In first and second-order hydrodynamics, universality has been found for the values of $\eta/s = 1/(4\pi)$ and $2 \eta \tp - 4 \lambda_1 - \lambda_2 = 0$ \cite{Kovtun:2004de,Buchel:2003tz,Iqbal:2008by,Starinets:2008fb,Shaverin:2012kv,Grozdanov:2014kva}. It is therefore natural to ask whether there exists a similar universal relation between conformal (and non-conformal) third-order transport coefficients. Furthermore, as discussed in \cite{Haehl:2014zda,Grozdanov:2014kva}, these universal relations may play an important role in the minimisation of entropy production in strongly coupled fluids, appearing as the coefficients of different tensor structures in the entropy current.

Finally, it would be extremely interesting if some of the second and the new third-order transport coefficients could be measured in real fluids, either through more precise measurements of various dispersion relations, or by other means. In particular, the microscopic information about the fluid's constituents contained in these higher order transport coefficients may play an important role in the modelling and studying of high-energy fluids, such as the quark-gluon plasma.

\acknowledgments

We would like to thank Amaresh Jaiswal and Mukund Rangamani for helpful correspondence. We are particularly grateful to Andrei Starinets for numerous useful discussions on related topics and for his comments on the draft of the paper. S. G. is supported in part by a VICI grant of the Netherlands Organization for Scientific Research (NWO), and by the Netherlands Organization for Scientific Research/Ministry of Science and Education (NWO/OCW). N. K. is supported by a grant from the John Templeton foundation. The opinions expressed in this publication are those of the authors and do not necessarily reflect the views of the John Templeton foundation.

This work was carried out on the Dutch national e-infrastructure with the support of SURF Foundation.

\appendix

\section{The basis of tensors for the gradient expansion}\label{sec:RiemProof}

In this Appendix, we present the proof of the claim in Section \ref{Sec:Gen} that it is sufficient to use only transverse derivatives of $u^a$ and $\ln s$, and the covariant derivatives of the Riemann tensor to write down the entire hydrodynamic gradient expansion at all orders.

Let us begin by considering the action of covariant derivatives on the hydrodynamic variables,
\begin{align}
&\nabla_b u_a = \delp_b u_a - u_b \,D u_a  \,,\\
&\nabla_b \ln s = \delp_b \ln s - u_b \,D \ln s \, .
\end{align}
Using the equations of motion \eqref{EoMs1} and \eqref{EoMs2}, we can write
\begin{align}
&\nabla_b u_a = \delp_b u_a  + c_s^2 u_b \delp_a \ln s + \text{higher derivatives}\,,\\
&\nabla_b \ln s = \delp_b \ln s + u_b \delp \cdot u +  \text{higher derivatives}\,,
\end{align}
which shows that all one-derivative combinations can easily be expressed in terms of transverse derivatives $\delp_a$, at the expense of introducing higher-derivative corrections.

Consider now a two-derivative tensor $\nabla_{c} \delp_b u_a $, which can be manipulated to give
\begin{align}
\nabla_{c} \delp_b u_a &= \delp_c \delp_b u_a - u_c \, D\delp_b u_a = \delp_c \delp_b u_a - u_c \left[D , \delp_b \right] u_a + u_c \delp_b \,D u_a \nn
&=\delp_c \delp_b u_a - c_s^2 u_c \delp_b \delp_a \ln s - u_c \left[D , \delp_b \right] u_a +  \text{higher derivatives}\,.
\end{align}
The commutator appearing above can be expressed as
\begin{align}\label{CommDDp}
\left[D , \delp_b \right] u_a = \left[u_d \nabla^d , \Delta_{be} \nabla^e \right] u_a = D \Delta_{be} \nabla^e u_a - \delp_b u_d \nabla^d u_a + u_d \Delta_{be} \left[\nabla^d, \nabla^e \right]u_a \,.
\end{align}
The first two terms in \eqref{CommDDp} contain only single derivatives of $u_a$, so as before, the equations of motion allow us to eliminate the longitudinal derivatives. The commutator term can be written as a Riemann tensor, since for an arbitrary vector field $V_a$,
\begin{align}
\left[ \nabla_d , \nabla_c\right] V_b = R^a_{~bcd} V_a \,.
\end{align}
We find that
\begin{align}
\nabla_{c} \delp_b u_a = - u_c  \Delta_{b}^{~e} R_{faed} u^d u^f +  \text{ terms with } \left(\delp_c \delp_b u_a, \delp_c \delp_b \ln s \right) +  \text{higher derivatives}\,.
\end{align}
Therefore, as claimed, the most general second derivative of $u_a$ can be expressed in terms of only transverse derivatives, the Riemann tensor and sub-leading higher-derivative corrections, which are irrelevant at the order of the gradient expansion to which we are working.

It is easy to see that with the help of the identity
\begin{align}\label{RiemannHigherOrd}
\left[\nabla_d, \nabla_c \right] V_{b_1b_2\ldots b_n} = R^{s}_{~b_1cd} V_{sb_2\ldots b_n} + R^{s}_{~b_2cd} V_{b_1 s \ldots b_n} +  \ldots +  R^{s}_{~b_n cd} V_{b_1 b_2 \ldots s} \,,
\end{align}
this discussion can be iteratively generalised to tensors of an arbitrary order, such as
\begin{align}
\nabla_c \delp_{b_1} \delp_{b_2} \ldots u_a  \, .
\end{align}
Schematically, the proof proceed as follows: we begin by writing
\begin{align}
\nabla_c \delp_{b_1} \delp_{b_2} \ldots u_a = \delp_c \delp_{b_1} \delp_{b_2} \ldots u_a - u_c \,D  \delp_{b_1} \delp_{b_2} \ldots u_a \,.
\end{align}
We then commute the outer two derivatives ($D$ and $\delp_{b_1}$) in the second term, which gives us
\begin{align}\label{nthOrdCommofDDp}
\nabla_c \delp_{b_1} \delp_{b_2} \ldots u_a = \ldots - u_c \delp_{b_1} \,D \delp_{b_2} \ldots u_a \,.
\end{align}
The ellipsis ($\ldots$) in Eq. \eqref{nthOrdCommofDDp} stands for various terms composed of purely transverse derivatives, the Riemann tensors coming from Eq. \eqref{RiemannHigherOrd} and higher-derivative terms. We must then proceed by commuting $D$ all the way through the expression (to the right), until it acts only on $u_a$ and the equations of motion can be used for the last time to eliminate the remaining longitudinal derivative.

Hence, we have shown that all of the tensorial ingredients that need to be considered in the construction of the hydrodynamic gradient expansion are the transverse derivatives of $u^a$ and $\ln s$,
\begin{align}
\delp_{b_1} \delp_{b_2} \ldots \delp_{b_n} u_a \,, && \delp_{b_1} \delp_{b_2} \ldots \delp_{b_n} \ln s \, ,
\end{align}
their products at all possible orders, and the Riemann tensor with various metric contractions and arbitrary covariant derivatives acting on it,
\begin{align}
\nabla_{e_1} \nabla_{e_2} \ldots \nabla_{e_n} R_{abcd} \, .
\end{align}

Because transverse derivatives have the property stated in Eq. \eqref{DelpProperty}, the tensorial ingredients can be grouped into two sets: those with transverse derivatives and those with covariant derivatives, which always involve the Riemann tensor and its contractions.

\section{Construction of the hydrodynamics tensors that contain no Riemann tensor}\label{sec:BasisAppendix}

Here, we prove the claim that when constructing scalars, transverse vectors and TST two-tensors, we can omit $u^a$ from $\CI^{(0)}$ altogether and consider only the metric contractions of $\delp_a u_b$ and $\delp_a \ln s$ inside the hydrodynamic tensors, at all orders in the gradient expansion.

Let us consider what happens to tensors built solely out of $\delp_a u_b$ and $\delp_a \ln s$ that get contracted with $u^a$. It is clear from Eq. \eqref{DelpProperty} that for any tensor $\delp_a V_{b_1 b_2 \ldots}$,
\begin{align}
u^a \delp_a V_{b_1 b_2 \ldots} = 0\,,
\end{align}
where $V_{b_1 b_2 \ldots}$ can contain more transverse derivatives as well as combinations of $u^a$ and $\ln s$. Now, consider a term of the form $u^a \delp_{b_1} \delp_a V_{b_2 b_3 \ldots}$. With the help of Eq. \eqref{DelpProperty}, we can commute $u^a$ through the outer derivative to find
\begin{align}\label{EqCommPerpDer1}
u^a \delp_{b_1} \delp_a V_{b_2 b_3 \ldots} = - g^{ac} \delp_{b_1} u_a \delp_{c} V_{b_2 b_3 \ldots} \, .
\end{align}
The point is that the right-hand-side of Eq. \eqref{EqCommPerpDer1} could have been constructed out of $\delp_{b_1} u_a \delp_{c} V_{b_2 b_3 \ldots}$ in some set $\CI^{(n)}$ and $g_{ab}$ from $\CI^{(0)}$, without any use of $u^a$ from $\CI^{(0)}$. By employing the same logic, it is easy to see that any contraction of the form
\begin{align}
u^a \delp_{b_1} \delp_{b_2} \ldots \delp_a u_c V_{d_1 d_2 \ldots}\, ,
\end{align}
can also be expressed as a sum of metric contractions acting on various components of $\CI^{(n)}$. No tensor with an undifferentiated $u^a$ can ever be linearly independent from tensors that use only transverse derivatives acting on $u^a$.

By going through the same process of commuting $u^a$ through the expression (cf. Appendix \ref{sec:RiemProof}), we can also see that the same conclusion can be drawn regarding tensors of the form
\begin{align}
u^a \delp_{b_1} \delp_{b_2} \ldots u_a V_{c_1 c_2 \ldots} \, ,
\end{align}
where now we use Eq. \eqref{UNorm} and its derivatives, instead of \eqref{DelpProperty}.

The only other possibilities that remain to be discussed are potentially uncontracted factors of $u^a$ with no derivatives acting on them. Such cases can appear in the construction of the transverse vectors and the TST tensors. A possible vector could have the form
\begin{align}
\CV^a = u^a V,
\end{align}
where all indices inside $V$ are contracted. However, such a vector can never be transverse, i.e. $u_a \CV^a \neq 0$. On the other hand, a TST two-tensor with undifferentiated $u^a$ could have the form
\begin{align}
\CT^{ab} = u^{\la a} V^{b\ra}.
\end{align}
However, considering such tensors is redundant, as $\CT^{ab}$ of this form could only be identically zero, i.e. $\CT^{ab} = 0$, due to various contractions of $u^a$ with $\Delta_{ab}$ in \eqref{TST}.

We have therefore shown that as claimed, only $g_{ab}$ can be used from $\CI^{(0)}$ when constructing tensors that contain only transverse derivatives. As a result, finding tensors with $R_{abcd}$, or covariant derivatives $\nabla_a$ of its various contractions, still requires us to use the full procedure outlined in Sec. \ref{Sec:ConstGradExp}.

\section{Tensors in non-conformal third-order hydrodynamics}\label{AppTensors}
In this Appendix, we list the transverse, symmetric and traceless two-tensors that participate in the third-order gradient expansion of the non-conformal stress-energy tensor:
\newline
\[
\begin{array}{llll}
&\mathcal{T}_{1}^{ab}=\tensor{\nabla}{_\perp_c}\tensor{\nabla}{_\perp^c}\tensor{\nabla}{_\perp^{\la a}}\tensor{u}{^{ b \ra}}
&\mathcal{T}_{2}^{ab}=\tensor{\nabla}{_\perp_c}\tensor{\nabla}{_\perp^{\la a}}\tensor{\nabla}{_\perp^c}\tensor{u}{^{ b \ra}}
&\mathcal{T}_{3}^{ab}=\tensor{\nabla}{_\perp_c}\tensor{\nabla}{_\perp^{\la a}}\tensor{\nabla}{_\perp^{ b \ra}}\tensor{u}{^c}\\
&\mathcal{T}_{4}^{ab}=\tensor{\nabla}{_\perp^{\la a}}\tensor{\nabla}{_\perp_c}\tensor{\nabla}{_\perp^c}\tensor{u}{^{ b \ra}}
&\mathcal{T}_{5}^{ab}=\tensor{\nabla}{_\perp^{\la a}}\tensor{\nabla}{_\perp_c}\tensor{\nabla}{_\perp^{ b \ra}}\tensor{u}{^c}
&\mathcal{T}_{6}^{ab}=\tensor{\nabla}{_\perp^{\la a}}\tensor{\nabla}{_\perp^{ b \ra}}\tensor{\nabla}{_\perp_c}\tensor{u}{^c}\\
&\mathcal{T}_{7}^{ab}=\tensor{\nabla}{_\perp_c}\tensor{u}{^c}\tensor{\nabla}{_\perp^{\la a}}\tensor{\nabla}{_\perp^{ b \ra}}\ln s
&\mathcal{T}_{8}^{ab}=\tensor{\nabla}{_\perp^c}\tensor{u}{^{\la a}}\tensor{\nabla}{_\perp_c}\tensor{\nabla}{_\perp^{ b \ra}}\ln s
&\mathcal{T}_{9}^{ab}=\tensor{\nabla}{_\perp^{\la a}}\tensor{u}{^c}\tensor{\nabla}{_\perp_c}\tensor{\nabla}{_\perp^{ b \ra}}\ln s\\
&\mathcal{T}_{10}^{ab}=\tensor{\nabla}{_\perp_c}\tensor{\nabla}{_\perp^c}\ln s\tensor{\nabla}{_\perp^{\la a}}\tensor{u}{^{ b \ra}}
&\mathcal{T}_{11}^{ab}=\tensor{\nabla}{_\perp^c}\ln s\tensor{\nabla}{_\perp^{\la a}}\tensor{\nabla}{_\perp^{ b \ra}}\tensor{u}{_c}
&\mathcal{T}_{12}^{ab}=\tensor{\nabla}{_\perp^c}\ln s\tensor{\nabla}{_\perp_c}\tensor{\nabla}{_\perp^{\la a}}\tensor{u}{^{ b \ra}}\\
&\mathcal{T}_{13}^{ab}=\tensor{\nabla}{_\perp^c}\ln s\tensor{\nabla}{_\perp^{\la a}}\tensor{\nabla}{_\perp_c}\tensor{u}{^{ b \ra}}
&\mathcal{T}_{14}^{ab}=\tensor{\nabla}{_\perp^{\la a}}\ln s\tensor{\nabla}{_\perp_c}\tensor{\nabla}{_\perp^{ b \ra}}\tensor{u}{^c}
&\mathcal{T}_{15}^{ab}=\tensor{\nabla}{_\perp^{\la a}}\ln s\tensor{\nabla}{_\perp^{ b \ra}}\tensor{\nabla}{_\perp_c}\tensor{u}{^c}\\
&\mathcal{T}_{16}^{ab}=\tensor{\nabla}{_\perp^{\la a}}\ln s\tensor{\nabla}{_\perp_c}\tensor{\nabla}{_\perp^c}\tensor{u}{^{ b \ra}}
&\mathcal{T}_{17}^{ab}=\tensor{\nabla}{_\perp^{\la a}}\ln s\tensor{\nabla}{_\perp^{ b \ra}}\ln s\tensor{\nabla}{_\perp_c}\tensor{u}{^c}
&\mathcal{T}_{18}^{ab}=\tensor{\nabla}{_\perp^{\la a}}\ln s\tensor{\nabla}{_\perp^c}\ln s\tensor{\nabla}{_\perp_c}\tensor{u}{^{ b \ra}}\\
&\mathcal{T}_{19}^{ab}=\tensor{\nabla}{_\perp^{\la a}}\ln s\tensor{\nabla}{_\perp^c}\ln s\tensor{\nabla}{_\perp^{ b \ra}}\tensor{u}{_c}
&\mathcal{T}_{20}^{ab}=\tensor{\nabla}{_\perp_c}\ln s\tensor{\nabla}{_\perp^c}\ln s\tensor{\nabla}{_\perp^{\la a}}\tensor{u}{^{ b \ra}}
&\mathcal{T}_{21}^{ab}=\tensor{\nabla}{_\perp_c}\tensor{u}{^c}\tensor{\nabla}{_\perp_d}\tensor{u}{^d}\tensor{\nabla}{_\perp^{\la a}}\tensor{u}{^{ b \ra}}\\
&\mathcal{T}_{22}^{ab}=\tensor{\nabla}{_\perp_d}\tensor{u}{^d}\tensor{\nabla}{_\perp^c}\tensor{u}{^{\la a}}\tensor{\nabla}{_\perp_c}\tensor{u}{^{ b \ra}}
&\mathcal{T}_{23}^{ab}=\tensor{\nabla}{_\perp_d}\tensor{u}{^d}\tensor{\nabla}{_\perp^c}\tensor{u}{^{\la a}}\tensor{\nabla}{_\perp^{ b \ra}}\tensor{u}{_c}
&\mathcal{T}_{24}^{ab}=\tensor{\nabla}{_\perp_d}\tensor{u}{^d}\tensor{\nabla}{_\perp^{\la a}}\tensor{u}{^c}\tensor{\nabla}{_\perp^{ b \ra}}\tensor{u}{_c}\\
&\mathcal{T}_{25}^{ab}=\tensor{\nabla}{_\perp^{\la a}}\tensor{u}{^{ b \ra}}\tensor{\nabla}{_\perp_d}\tensor{u}{_c}\tensor{\nabla}{_\perp^d}\tensor{u}{^c}
&\mathcal{T}_{26}^{ab}=\tensor{\nabla}{_\perp^c}\tensor{u}{^{\la a}}\tensor{\nabla}{_\perp^d}\tensor{u}{^{ b \ra}}\tensor{\nabla}{_\perp_c}\tensor{u}{_d}
&\mathcal{T}_{27}^{ab}=\tensor{\nabla}{_\perp^c}\tensor{u}{^{\la a}}\tensor{\nabla}{_\perp^{ b \ra}}\tensor{u}{^d}\tensor{\nabla}{_\perp_c}\tensor{u}{_d}\\
&\mathcal{T}_{28}^{ab}=\tensor{\nabla}{_\perp^{\la a}}\tensor{u}{^{ b \ra}}\tensor{\nabla}{_\perp_c}\tensor{u}{_d}\tensor{\nabla}{_\perp^d}\tensor{u}{^c}
&\mathcal{T}_{29}^{ab}=\tensor{\nabla}{_\perp^{\la a}}\tensor{u}{^d}\tensor{\nabla}{_\perp^c}\tensor{u}{^{ b \ra}}\tensor{\nabla}{_\perp_d}\tensor{u}{_c}
&\mathcal{T}_{30}^{ab}=\tensor{\nabla}{_\perp^{\la a}}\tensor{u}{^c}\tensor{\nabla}{_\perp^{ b \ra}}\tensor{u}{^d}\tensor{\nabla}{_\perp_c}\tensor{u}{_d}\\
&\mathcal{T}_{31}^{ab}=\tensor{u}{^c}\tensor{\nabla}{_d}\tensor{R}{^{\la a}_c^{ b \ra}^d}
&\mathcal{T}_{32}^{ab}=\tensor{u}{^c}\tensor{\nabla}{_c}\tensor{R}{^{\la a}^{ b \ra}}
&\mathcal{T}_{33}^{ab}=\tensor{u}{^c}\tensor{\nabla}{^{\la a}}\tensor{R}{^{ b \ra}_c}\\
&\mathcal{T}_{34}^{ab}=\tensor{u}{^c}\tensor{u}{^d}\tensor{u}{^e}\tensor{\nabla}{_e}\tensor{R}{^{\la a}_c^{ b \ra}_d}
&\mathcal{T}_{35}^{ab}=\tensor{u}{^c}\tensor{R}{^{\la a}_c}\tensor{\nabla}{_\perp^{ b \ra}}\ln s
&\mathcal{T}_{36}^{ab}=\tensor{u}{^c}\tensor{\nabla}{_\perp^d}\ln s\tensor{R}{^{\la a}_c^{ b \ra}_d}\\
&\mathcal{T}_{37}^{ab}=R\tensor{\nabla}{_\perp^{\la a}}\tensor{u}{^{ b \ra}}
&\mathcal{T}_{38}^{ab}=\tensor{R}{^{\la a}_c}\tensor{\nabla}{_\perp^c}\tensor{u}{^{ b \ra}}
&\mathcal{T}_{39}^{ab}=\tensor{R}{^{\la a}_c}\tensor{\nabla}{_\perp^{ b \ra}}\tensor{u}{^c}\\
&\mathcal{T}_{40}^{ab}=\tensor{R}{^{\la a}^{ b \ra}}\tensor{\nabla}{_\perp_c}\tensor{u}{^c}
&\mathcal{T}_{41}^{ab}=\tensor{\nabla}{_\perp^d}\tensor{u}{^c}\tensor{R}{^{\la a}_d^{ b \ra}_c}
&\mathcal{T}_{42}^{ab}=\tensor{u}{^c}\tensor{u}{^d}\tensor{R}{_c_d}\tensor{\nabla}{_\perp^{\la a}}\tensor{u}{^{ b \ra}}\\
&\mathcal{T}_{43}^{ab}=\tensor{u}{^c}\tensor{u}{^d}\tensor{\nabla}{_\perp^e}\tensor{u}{^{\la a}}\tensor{R}{^{ b \ra}_c_d_e}
&\mathcal{T}_{44}^{ab}=\tensor{u}{^c}\tensor{u}{^d}\tensor{\nabla}{_\perp^{\la a}}\tensor{u}{^e}\tensor{R}{^{ b \ra}_c_d_e}
&\mathcal{T}_{45}^{ab}=\tensor{u}{^c}\tensor{u}{^d}\tensor{\nabla}{_\perp_e}\tensor{u}{^e}\tensor{R}{^{\la a}_c^{ b \ra}_d}\\
\end{array}
\]

\newpage
\section{Scalars in non-conformal third-order hydrodynamics}\label{Sec:AppScalars}
The following is the list of scalars that participate in the third-order gradient expansion of the non-conformal stress-energy tensor:
\begin{table}[h]
\begin{center}
\[
\arraycolsep=20pt
\begin{array}{lll}
&\mathcal{S}_{1}=\tensor{\nabla}{_\perp_b}\tensor{\nabla}{_\perp^b}\tensor{\nabla}{_\perp_a}\tensor{u}{^a}
&\mathcal{S}_{2}=\tensor{\nabla}{_\perp_b}\tensor{\nabla}{_\perp_a}\tensor{\nabla}{_\perp^b}\tensor{u}{^a}\\
&\mathcal{S}_{3}=\tensor{\nabla}{_\perp_a}\tensor{\nabla}{_\perp_b}\tensor{\nabla}{_\perp^b}\tensor{u}{^a}
&\mathcal{S}_{4}=\tensor{\nabla}{_\perp_a}\tensor{u}{^a}\tensor{\nabla}{_\perp_b}\tensor{\nabla}{_\perp^b}\ln s\\
&\mathcal{S}_{5}=\tensor{\nabla}{_\perp_b}\tensor{\nabla}{_\perp_a}\ln s\tensor{\nabla}{_\perp^b}\tensor{u}{^a}
&\mathcal{S}_{6}=\tensor{\nabla}{_\perp^a}\ln s\tensor{\nabla}{_\perp_b}\tensor{\nabla}{_\perp^b}\tensor{u}{_a}\\
&\mathcal{S}_{7}=\tensor{\nabla}{_\perp^a}\ln s\tensor{\nabla}{_\perp_a}\tensor{\nabla}{_\perp_b}\tensor{u}{^b}
&\mathcal{S}_{8}=\tensor{\nabla}{_\perp^a}\ln s\tensor{\nabla}{_\perp_b}\tensor{\nabla}{_\perp_a}\tensor{u}{^b}\\
&\mathcal{S}_{9}=\tensor{\nabla}{_\perp_a}\ln s\tensor{\nabla}{_\perp^a}\ln s\tensor{\nabla}{_\perp_b}\tensor{u}{^b}
&\mathcal{S}_{10}=\tensor{\nabla}{_\perp^a}\ln s\tensor{\nabla}{_\perp^b}\ln s\tensor{\nabla}{_\perp_b}\tensor{u}{_a}\\
&\mathcal{S}_{11}=\tensor{\nabla}{_\perp_a}\tensor{u}{^a}\tensor{\nabla}{_\perp_b}\tensor{u}{^b}\tensor{\nabla}{_\perp_c}\tensor{u}{^c}
&\mathcal{S}_{12}=\tensor{\nabla}{_\perp_a}\tensor{u}{^a}\tensor{\nabla}{_\perp_c}\tensor{u}{_b}\tensor{\nabla}{_\perp^c}\tensor{u}{^b}\\
&\mathcal{S}_{13}=\tensor{\nabla}{_\perp_a}\tensor{u}{^a}\tensor{\nabla}{_\perp_b}\tensor{u}{_c}\tensor{\nabla}{_\perp^c}\tensor{u}{^b}
&\mathcal{S}_{14}=\tensor{\nabla}{_\perp^b}\tensor{u}{^a}\tensor{\nabla}{_\perp^c}\tensor{u}{_a}\tensor{\nabla}{_\perp_c}\tensor{u}{_b}\\
&\mathcal{S}_{15}=\tensor{\nabla}{_\perp^b}\tensor{u}{^a}\tensor{\nabla}{_\perp_a}\tensor{u}{_c}\tensor{\nabla}{_\perp^c}\tensor{u}{_b}
&\mathcal{S}_{16}=\tensor{u}{^a}\tensor{\nabla}{_b}\tensor{R}{_a^b}\\
&\mathcal{S}_{17}=\tensor{u}{^a}\tensor{\nabla}{_a}R
&\mathcal{S}_{18}=\tensor{u}{^a}\tensor{u}{^b}\tensor{u}{^c}\tensor{\nabla}{_c}\tensor{R}{_a_b}\\
&\mathcal{S}_{19}=\tensor{u}{^a}\tensor{R}{_a_b}\tensor{\nabla}{_\perp^b}\ln s
&\mathcal{S}_{20}=R\tensor{\nabla}{_\perp_a}\tensor{u}{^a}\\
&\mathcal{S}_{21}=\tensor{R}{_a_b}\tensor{\nabla}{_\perp^b}\tensor{u}{^a}
&\mathcal{S}_{22}=\tensor{u}{^a}\tensor{u}{^b}\tensor{R}{_a_b}\tensor{\nabla}{_\perp_c}\tensor{u}{^c}\\
&\mathcal{S}_{23}=\tensor{u}{^a}\tensor{u}{^b}\tensor{\nabla}{_\perp^d}\tensor{u}{^c}\tensor{R}{_a_c_b_d}
\end{array}
\]
\end{center}
\label{table:transverse.scalars}
\end{table}

\section{Vectors in non-conformal third-order hydrodynamics}\label{Sec:AppVectors}
Here, we write down the list of vectors that participate in the third-order gradient expansion of the non-conformal stress-energy tensor:
\[
\arraycolsep=20pt
\begin{array}{lll}
&\mathcal{V}_{1}^{a}=\tensor{\nabla}{_\perp^a}\tensor{\nabla}{_\perp_b}\tensor{\nabla}{_\perp^b}\ln s
&\mathcal{V}_{2}^{a}=\tensor{\nabla}{_\perp^a}\ln s\tensor{\nabla}{_\perp_b}\tensor{\nabla}{_\perp^b}\ln s\\
&\mathcal{V}_{3}^{a}=\tensor{\nabla}{_\perp_b}\tensor{u}{^b}\tensor{\nabla}{_\perp^a}\tensor{\nabla}{_\perp_c}\tensor{u}{^c}
&\mathcal{V}_{4}^{a}=\tensor{\nabla}{_\perp^b}\tensor{u}{^a}\tensor{\nabla}{_\perp_b}\tensor{\nabla}{_\perp_c}\tensor{u}{^c}\\
&\mathcal{V}_{5}^{a}=\tensor{\nabla}{_\perp^c}\tensor{u}{^b}\tensor{\nabla}{_\perp^a}\tensor{\nabla}{_\perp_c}\tensor{u}{_b}
&\mathcal{V}_{6}^{a}=\tensor{\nabla}{_\perp^b}\tensor{u}{^a}\tensor{\nabla}{_\perp_c}\tensor{\nabla}{_\perp_b}\tensor{u}{^c}\\
&\mathcal{V}_{7}^{a}=\tensor{\nabla}{_\perp^c}\tensor{u}{^b}\tensor{\nabla}{_\perp^a}\tensor{\nabla}{_\perp_b}\tensor{u}{_c}
&\mathcal{V}_{8}^{a}=\tensor{\nabla}{_\perp^b}\tensor{u}{^a}\tensor{\nabla}{_\perp_c}\tensor{\nabla}{_\perp^c}\tensor{u}{_b}\\
&\mathcal{V}_{9}^{a}=\tensor{\nabla}{_\perp^a}\tensor{u}{^b}\tensor{\nabla}{_\perp_b}\tensor{\nabla}{_\perp_c}\tensor{u}{^c}
&\mathcal{V}_{10}^{a}=\tensor{\nabla}{_\perp^a}\tensor{u}{^b}\tensor{\nabla}{_\perp_c}\tensor{\nabla}{_\perp_b}\tensor{u}{^c}\\
&\mathcal{V}_{11}^{a}=\tensor{\nabla}{_\perp^a}\tensor{u}{^b}\tensor{\nabla}{_\perp_c}\tensor{\nabla}{_\perp^c}\tensor{u}{_b}
&\mathcal{V}_{12}^{a}=\tensor{\nabla}{_\perp^a}\ln s\tensor{\nabla}{_\perp_b}\ln s\tensor{\nabla}{_\perp^b}\ln s\\
&\mathcal{V}_{13}^{a}=\tensor{\nabla}{_\perp^a}\ln s\tensor{\nabla}{_\perp_b}\tensor{u}{^b}\tensor{\nabla}{_\perp_c}\tensor{u}{^c}
&\mathcal{V}_{14}^{a}=\tensor{\nabla}{_\perp^b}\ln s\tensor{\nabla}{_\perp_c}\tensor{u}{^c}\tensor{\nabla}{_\perp_b}\tensor{u}{^a}\\
&\mathcal{V}_{15}^{a}=\tensor{\nabla}{_\perp^b}\ln s\tensor{\nabla}{_\perp_c}\tensor{u}{^c}\tensor{\nabla}{_\perp^a}\tensor{u}{_b}
&\mathcal{V}_{16}^{a}=\tensor{\nabla}{_\perp^a}\ln s\tensor{\nabla}{_\perp_c}\tensor{u}{_b}\tensor{\nabla}{_\perp^c}\tensor{u}{^b}\\
&\mathcal{V}_{17}^{a}=\tensor{\nabla}{_\perp^b}\ln s\tensor{\nabla}{_\perp^c}\tensor{u}{^a}\tensor{\nabla}{_\perp_c}\tensor{u}{_b}
&\mathcal{V}_{18}^{a}=\tensor{\nabla}{_\perp^a}\ln s\tensor{\nabla}{_\perp_b}\tensor{u}{_c}\tensor{\nabla}{_\perp^c}\tensor{u}{^b}\\
&\mathcal{V}_{19}^{a}=\tensor{\nabla}{_\perp^b}\ln s\tensor{\nabla}{_\perp^a}\tensor{u}{^c}\tensor{\nabla}{_\perp_c}\tensor{u}{_b}
&\mathcal{V}_{20}^{a}=\tensor{\nabla}{_\perp^b}\ln s\tensor{\nabla}{_\perp^c}\tensor{u}{^a}\tensor{\nabla}{_\perp_b}\tensor{u}{_c}\\
&\mathcal{V}_{21}^{a}=\tensor{\nabla}{_\perp^b}\ln s\tensor{\nabla}{_\perp^a}\tensor{u}{^c}\tensor{\nabla}{_\perp_b}\tensor{u}{_c}
&\mathcal{V}_{22}^{a}=\tensor{u}{^b}\tensor{u}{^c}\tensor{\nabla}{_d}\tensor{R}{^a_b_c^d}\\
&\mathcal{V}_{23}^{a}=R\tensor{\nabla}{_\perp^a}\ln s
&\mathcal{V}_{24}^{a}=\tensor{u}{^b}\tensor{u}{^c}\tensor{R}{_b_c}\tensor{\nabla}{_\perp^a}\ln s\\
&\mathcal{V}_{25}^{a}=\tensor{u}{^b}\tensor{u}{^c}\tensor{\nabla}{_\perp^d}\ln s\tensor{R}{^a_b_c_d}
&\mathcal{V}_{26}^{a}=\tensor{u}{^b}\tensor{R}{_b_c}\tensor{\nabla}{_\perp^c}\tensor{u}{^a}\\
&\mathcal{V}_{27}^{a}=\tensor{u}{^b}\tensor{R}{_b_c}\tensor{\nabla}{_\perp^a}\tensor{u}{^c}
&\mathcal{V}_{28}^{a}=\tensor{u}{^b}\tensor{\nabla}{_\perp^d}\tensor{u}{^c}\tensor{R}{^a_b_c_d}
\end{array}
\]


\newpage
\bibliography{hydro}

\begin{thebibliography}{80}%
\makeatletter
\providecommand \@ifxundefined [1]{%
 \@ifx{#1\undefined}
}%
\providecommand \@ifnum [1]{%
 \ifnum #1\expandafter \@firstoftwo
 \else \expandafter \@secondoftwo
 \fi
}%
\providecommand \@ifx [1]{%
 \ifx #1\expandafter \@firstoftwo
 \else \expandafter \@secondoftwo
 \fi
}%
\providecommand \natexlab [1]{#1}%
\providecommand \enquote  [1]{``#1''}%
\providecommand \bibnamefont  [1]{#1}%
\providecommand \bibfnamefont [1]{#1}%
\providecommand \citenamefont [1]{#1}%
\providecommand \href@noop [0]{\@secondoftwo}%
\providecommand \href [0]{\begingroup \@sanitize@url \@href}%
\providecommand \@href[1]{\@@startlink{#1}\@@href}%
\providecommand \@@href[1]{\endgroup#1\@@endlink}%
\providecommand \@sanitize@url [0]{\catcode `\\12\catcode `\$12\catcode
  `\&12\catcode `\#12\catcode `\^12\catcode `\_12\catcode `\%12\relax}%
\providecommand \@@startlink[1]{}%
\providecommand \@@endlink[0]{}%
\providecommand \url  [0]{\begingroup\@sanitize@url \@url }%
\providecommand \@url [1]{\endgroup\@href {#1}{\urlprefix }}%
\providecommand \urlprefix  [0]{URL }%
\providecommand \Eprint [0]{\href }%
\providecommand \doibase [0]{http://dx.doi.org/}%
\providecommand \selectlanguage [0]{\@gobble}%
\providecommand \bibinfo  [0]{\@secondoftwo}%
\providecommand \bibfield  [0]{\@secondoftwo}%
\providecommand \translation [1]{[#1]}%
\providecommand \BibitemOpen [0]{}%
\providecommand \bibitemStop [0]{}%
\providecommand \bibitemNoStop [0]{.\EOS\space}%
\providecommand \EOS [0]{\spacefactor3000\relax}%
\providecommand \BibitemShut  [1]{\csname bibitem#1\endcsname}%
\let\auto@bib@innerbib\@empty
\bibitem [{\citenamefont {Dubovsky}\ \emph {et~al.}(2006)\citenamefont
  {Dubovsky}, \citenamefont {Gregoire}, \citenamefont {Nicolis},\ and\
  \citenamefont {Rattazzi}}]{Dubovsky:2005xd}%
  \BibitemOpen
  \bibfield  {author} {\bibinfo {author} {\bibfnamefont {S.}~\bibnamefont
  {Dubovsky}}, \bibinfo {author} {\bibfnamefont {T.}~\bibnamefont {Gregoire}},
  \bibinfo {author} {\bibfnamefont {A.}~\bibnamefont {Nicolis}}, \ and\
  \bibinfo {author} {\bibfnamefont {R.}~\bibnamefont {Rattazzi}},\ }\href
  {\doibase 10.1088/1126-6708/2006/03/025} {\bibfield  {journal} {\bibinfo
  {journal} {JHEP}\ }\textbf {\bibinfo {volume} {0603}},\ \bibinfo {pages}
  {025} (\bibinfo {year} {2006})},\ \Eprint
  {http://arxiv.org/abs/hep-th/0512260} {arXiv:hep-th/0512260 [hep-th]}
  \BibitemShut {NoStop}%
\bibitem [{\citenamefont {Dubovsky}\ \emph {et~al.}(2012)\citenamefont
  {Dubovsky}, \citenamefont {Hui}, \citenamefont {Nicolis},\ and\ \citenamefont
  {Son}}]{Dubovsky:2011sj}%
  \BibitemOpen
  \bibfield  {author} {\bibinfo {author} {\bibfnamefont {S.}~\bibnamefont
  {Dubovsky}}, \bibinfo {author} {\bibfnamefont {L.}~\bibnamefont {Hui}},
  \bibinfo {author} {\bibfnamefont {A.}~\bibnamefont {Nicolis}}, \ and\
  \bibinfo {author} {\bibfnamefont {D.~T.}\ \bibnamefont {Son}},\ }\href
  {\doibase 10.1103/PhysRevD.85.085029} {\bibfield  {journal} {\bibinfo
  {journal} {Phys.Rev.}\ }\textbf {\bibinfo {volume} {D85}},\ \bibinfo {pages}
  {085029} (\bibinfo {year} {2012})},\ \Eprint {http://arxiv.org/abs/1107.0731}
  {arXiv:1107.0731 [hep-th]} \BibitemShut {NoStop}%
\bibitem [{\citenamefont {Nicolis}\ \emph {et~al.}(2014)\citenamefont
  {Nicolis}, \citenamefont {Penco},\ and\ \citenamefont
  {Rosen}}]{Nicolis:2013lma}%
  \BibitemOpen
  \bibfield  {author} {\bibinfo {author} {\bibfnamefont {A.}~\bibnamefont
  {Nicolis}}, \bibinfo {author} {\bibfnamefont {R.}~\bibnamefont {Penco}}, \
  and\ \bibinfo {author} {\bibfnamefont {R.~A.}\ \bibnamefont {Rosen}},\ }\href
  {\doibase 10.1103/PhysRevD.89.045002} {\bibfield  {journal} {\bibinfo
  {journal} {Phys.Rev.}\ }\textbf {\bibinfo {volume} {D89}},\ \bibinfo {pages}
  {045002} (\bibinfo {year} {2014})},\ \Eprint {http://arxiv.org/abs/1307.0517}
  {arXiv:1307.0517 [hep-th]} \BibitemShut {NoStop}%
\bibitem [{\citenamefont {Endlich}\ \emph {et~al.}(2013)\citenamefont
  {Endlich}, \citenamefont {Nicolis}, \citenamefont {Porto},\ and\
  \citenamefont {Wang}}]{Endlich:2012vt}%
  \BibitemOpen
  \bibfield  {author} {\bibinfo {author} {\bibfnamefont {S.}~\bibnamefont
  {Endlich}}, \bibinfo {author} {\bibfnamefont {A.}~\bibnamefont {Nicolis}},
  \bibinfo {author} {\bibfnamefont {R.~A.}\ \bibnamefont {Porto}}, \ and\
  \bibinfo {author} {\bibfnamefont {J.}~\bibnamefont {Wang}},\ }\href {\doibase
  10.1103/PhysRevD.88.105001} {\bibfield  {journal} {\bibinfo  {journal}
  {Phys.Rev.}\ }\textbf {\bibinfo {volume} {D88}},\ \bibinfo {pages} {105001}
  (\bibinfo {year} {2013})},\ \Eprint {http://arxiv.org/abs/1211.6461}
  {arXiv:1211.6461 [hep-th]} \BibitemShut {NoStop}%
\bibitem [{\citenamefont {Grozdanov}\ and\ \citenamefont
  {Polonyi}(2015{\natexlab{a}})}]{Grozdanov:2013dba}%
  \BibitemOpen
  \bibfield  {author} {\bibinfo {author} {\bibfnamefont {S.}~\bibnamefont
  {Grozdanov}}\ and\ \bibinfo {author} {\bibfnamefont {J.}~\bibnamefont
  {Polonyi}},\ }\href {\doibase 10.1103/PhysRevD.91.105031} {\bibfield
  {journal} {\bibinfo  {journal} {Phys.Rev.}\ }\textbf {\bibinfo {volume}
  {D91}},\ \bibinfo {pages} {105031} (\bibinfo {year} {2015}{\natexlab{a}})},\
  \Eprint {http://arxiv.org/abs/1305.3670} {arXiv:1305.3670 [hep-th]}
  \BibitemShut {NoStop}%
\bibitem [{\citenamefont {Bhattacharya}\ \emph {et~al.}(2013)\citenamefont
  {Bhattacharya}, \citenamefont {Bhattacharyya},\ and\ \citenamefont
  {Rangamani}}]{Bhattacharya:2012zx}%
  \BibitemOpen
  \bibfield  {author} {\bibinfo {author} {\bibfnamefont {J.}~\bibnamefont
  {Bhattacharya}}, \bibinfo {author} {\bibfnamefont {S.}~\bibnamefont
  {Bhattacharyya}}, \ and\ \bibinfo {author} {\bibfnamefont {M.}~\bibnamefont
  {Rangamani}},\ }\href {\doibase 10.1007/JHEP02(2013)153} {\bibfield
  {journal} {\bibinfo  {journal} {JHEP}\ }\textbf {\bibinfo {volume} {1302}},\
  \bibinfo {pages} {153} (\bibinfo {year} {2013})},\ \Eprint
  {http://arxiv.org/abs/1211.1020} {arXiv:1211.1020 [hep-th]} \BibitemShut
  {NoStop}%
\bibitem [{\citenamefont {Kovtun}\ \emph {et~al.}(2014)\citenamefont {Kovtun},
  \citenamefont {Moore},\ and\ \citenamefont {Romatschke}}]{Kovtun:2014hpa}%
  \BibitemOpen
  \bibfield  {author} {\bibinfo {author} {\bibfnamefont {P.}~\bibnamefont
  {Kovtun}}, \bibinfo {author} {\bibfnamefont {G.~D.}\ \bibnamefont {Moore}}, \
  and\ \bibinfo {author} {\bibfnamefont {P.}~\bibnamefont {Romatschke}},\
  }\href@noop {} {\  (\bibinfo {year} {2014})},\ \Eprint
  {http://arxiv.org/abs/1405.3967} {arXiv:1405.3967 [hep-ph]} \BibitemShut
  {NoStop}%
\bibitem [{\citenamefont {Harder}\ \emph {et~al.}(2015)\citenamefont {Harder},
  \citenamefont {Kovtun},\ and\ \citenamefont {Ritz}}]{Harder:2015nxa}%
  \BibitemOpen
  \bibfield  {author} {\bibinfo {author} {\bibfnamefont {M.}~\bibnamefont
  {Harder}}, \bibinfo {author} {\bibfnamefont {P.}~\bibnamefont {Kovtun}}, \
  and\ \bibinfo {author} {\bibfnamefont {A.}~\bibnamefont {Ritz}},\ }\href@noop
  {} {\  (\bibinfo {year} {2015})},\ \Eprint {http://arxiv.org/abs/1502.03076}
  {arXiv:1502.03076 [hep-th]} \BibitemShut {NoStop}%
\bibitem [{\citenamefont {Haehl}\ \emph
  {et~al.}(2015{\natexlab{a}})\citenamefont {Haehl}, \citenamefont
  {Loganayagam},\ and\ \citenamefont {Rangamani}}]{Haehl:2015pja}%
  \BibitemOpen
  \bibfield  {author} {\bibinfo {author} {\bibfnamefont {F.~M.}\ \bibnamefont
  {Haehl}}, \bibinfo {author} {\bibfnamefont {R.}~\bibnamefont {Loganayagam}},
  \ and\ \bibinfo {author} {\bibfnamefont {M.}~\bibnamefont {Rangamani}},\
  }\href@noop {} {\  (\bibinfo {year} {2015}{\natexlab{a}})},\ \Eprint
  {http://arxiv.org/abs/1502.00636} {arXiv:1502.00636 [hep-th]} \BibitemShut
  {NoStop}%
\bibitem [{\citenamefont {Grozdanov}\ and\ \citenamefont
  {Polonyi}(2015{\natexlab{b}})}]{Grozdanov:2015nea}%
  \BibitemOpen
  \bibfield  {author} {\bibinfo {author} {\bibfnamefont {S.}~\bibnamefont
  {Grozdanov}}\ and\ \bibinfo {author} {\bibfnamefont {J.}~\bibnamefont
  {Polonyi}},\ }\href@noop {} {\  (\bibinfo {year} {2015}{\natexlab{b}})},\
  \Eprint {http://arxiv.org/abs/1501.06620} {arXiv:1501.06620 [hep-th]}
  \BibitemShut {NoStop}%
\bibitem [{\citenamefont {Haehl}\ \emph {et~al.}(2014)\citenamefont {Haehl},
  \citenamefont {Loganayagam},\ and\ \citenamefont
  {Rangamani}}]{Haehl:2013hoa}%
  \BibitemOpen
  \bibfield  {author} {\bibinfo {author} {\bibfnamefont {F.~M.}\ \bibnamefont
  {Haehl}}, \bibinfo {author} {\bibfnamefont {R.}~\bibnamefont {Loganayagam}},
  \ and\ \bibinfo {author} {\bibfnamefont {M.}~\bibnamefont {Rangamani}},\
  }\href {\doibase 10.1007/JHEP03(2014)034} {\bibfield  {journal} {\bibinfo
  {journal} {JHEP}\ }\textbf {\bibinfo {volume} {1403}},\ \bibinfo {pages}
  {034} (\bibinfo {year} {2014})},\ \Eprint {http://arxiv.org/abs/1312.0610}
  {arXiv:1312.0610 [hep-th]} \BibitemShut {NoStop}%
\bibitem [{\citenamefont {Burch}\ and\ \citenamefont
  {Torrieri}(2015)}]{Burch:2015mea}%
  \BibitemOpen
  \bibfield  {author} {\bibinfo {author} {\bibfnamefont {T.}~\bibnamefont
  {Burch}}\ and\ \bibinfo {author} {\bibfnamefont {G.}~\bibnamefont
  {Torrieri}},\ }\href@noop {} {\  (\bibinfo {year} {2015})},\ \Eprint
  {http://arxiv.org/abs/1502.05421} {arXiv:1502.05421 [hep-lat]} \BibitemShut
  {NoStop}%
\bibitem [{\citenamefont {Crossley}\ \emph {et~al.}(2015)\citenamefont
  {Crossley}, \citenamefont {Glorioso}, \citenamefont {Liu},\ and\
  \citenamefont {Wang}}]{Crossley:2015tka}%
  \BibitemOpen
  \bibfield  {author} {\bibinfo {author} {\bibfnamefont {M.}~\bibnamefont
  {Crossley}}, \bibinfo {author} {\bibfnamefont {P.}~\bibnamefont {Glorioso}},
  \bibinfo {author} {\bibfnamefont {H.}~\bibnamefont {Liu}}, \ and\ \bibinfo
  {author} {\bibfnamefont {Y.}~\bibnamefont {Wang}},\ }\href@noop {} {\
  (\bibinfo {year} {2015})},\ \Eprint {http://arxiv.org/abs/1504.07611}
  {arXiv:1504.07611 [hep-th]} \BibitemShut {NoStop}%
\bibitem [{\citenamefont {de~Boer}\ \emph {et~al.}(2015)\citenamefont
  {de~Boer}, \citenamefont {Heller},\ and\ \citenamefont
  {Pinzani-Fokeeva}}]{deBoer:2015ija}%
  \BibitemOpen
  \bibfield  {author} {\bibinfo {author} {\bibfnamefont {J.}~\bibnamefont
  {de~Boer}}, \bibinfo {author} {\bibfnamefont {M.~P.}\ \bibnamefont {Heller}},
  \ and\ \bibinfo {author} {\bibfnamefont {N.}~\bibnamefont
  {Pinzani-Fokeeva}},\ }\href@noop {} {\  (\bibinfo {year} {2015})},\ \Eprint
  {http://arxiv.org/abs/1504.07616} {arXiv:1504.07616 [hep-th]} \BibitemShut
  {NoStop}%
\bibitem [{\citenamefont {Burnett}(1936)}]{BurnettD}%
  \BibitemOpen
  \bibfield  {author} {\bibinfo {author} {\bibfnamefont {D.}~\bibnamefont
  {Burnett}},\ }\href {\doibase 10.1112/plms/s2-40.1.382} {\bibfield  {journal}
  {\bibinfo  {journal} {Proc. London Math. Soc.}\ }\textbf {\bibinfo {volume}
  {s2-40}},\ \bibinfo {pages} {382} (\bibinfo {year} {1936})}\BibitemShut
  {NoStop}%
\bibitem [{\citenamefont {Muller}(1967)}]{Muller:1967zza}%
  \BibitemOpen
  \bibfield  {author} {\bibinfo {author} {\bibfnamefont {I.}~\bibnamefont
  {Muller}},\ }\href {\doibase 10.1007/BF01326412} {\bibfield  {journal}
  {\bibinfo  {journal} {Z.Phys.}\ }\textbf {\bibinfo {volume} {198}},\ \bibinfo
  {pages} {329} (\bibinfo {year} {1967})}\BibitemShut {NoStop}%
\bibitem [{\citenamefont {Israel}(1976)}]{Israel:1976tn}%
  \BibitemOpen
  \bibfield  {author} {\bibinfo {author} {\bibfnamefont {W.}~\bibnamefont
  {Israel}},\ }\href {\doibase 10.1016/0003-4916(76)90064-6} {\bibfield
  {journal} {\bibinfo  {journal} {Annals Phys.}\ }\textbf {\bibinfo {volume}
  {100}},\ \bibinfo {pages} {310} (\bibinfo {year} {1976})}\BibitemShut
  {NoStop}%
\bibitem [{\citenamefont {Israel}\ and\ \citenamefont
  {Stewart}(1976)}]{Israel1976213}%
  \BibitemOpen
  \bibfield  {author} {\bibinfo {author} {\bibfnamefont {W.}~\bibnamefont
  {Israel}}\ and\ \bibinfo {author} {\bibfnamefont {J.}~\bibnamefont
  {Stewart}},\ }\href {\doibase http://dx.doi.org/10.1016/0375-9601(76)90075-X}
  {\bibfield  {journal} {\bibinfo  {journal} {Physics Letters A}\ }\textbf
  {\bibinfo {volume} {58}},\ \bibinfo {pages} {213 } (\bibinfo {year}
  {1976})}\BibitemShut {NoStop}%
\bibitem [{\citenamefont {Israel}\ and\ \citenamefont
  {Stewart}(1979)}]{Israel:1979wp}%
  \BibitemOpen
  \bibfield  {author} {\bibinfo {author} {\bibfnamefont {W.}~\bibnamefont
  {Israel}}\ and\ \bibinfo {author} {\bibfnamefont {J.}~\bibnamefont
  {Stewart}},\ }\href {\doibase 10.1016/0003-4916(79)90130-1} {\bibfield
  {journal} {\bibinfo  {journal} {Annals Phys.}\ }\textbf {\bibinfo {volume}
  {118}},\ \bibinfo {pages} {341} (\bibinfo {year} {1979})}\BibitemShut
  {NoStop}%
\bibitem [{\citenamefont {Baier}\ \emph {et~al.}(2008)\citenamefont {Baier},
  \citenamefont {Romatschke}, \citenamefont {Son}, \citenamefont {Starinets},\
  and\ \citenamefont {Stephanov}}]{Baier:2007ix}%
  \BibitemOpen
  \bibfield  {author} {\bibinfo {author} {\bibfnamefont {R.}~\bibnamefont
  {Baier}}, \bibinfo {author} {\bibfnamefont {P.}~\bibnamefont {Romatschke}},
  \bibinfo {author} {\bibfnamefont {D.~T.}\ \bibnamefont {Son}}, \bibinfo
  {author} {\bibfnamefont {A.~O.}\ \bibnamefont {Starinets}}, \ and\ \bibinfo
  {author} {\bibfnamefont {M.~A.}\ \bibnamefont {Stephanov}},\ }\href {\doibase
  10.1088/1126-6708/2008/04/100} {\bibfield  {journal} {\bibinfo  {journal}
  {JHEP}\ }\textbf {\bibinfo {volume} {0804}},\ \bibinfo {pages} {100}
  (\bibinfo {year} {2008})},\ \Eprint {http://arxiv.org/abs/0712.2451}
  {arXiv:0712.2451 [hep-th]} \BibitemShut {NoStop}%
\bibitem [{\citenamefont {Bhattacharyya}\ \emph
  {et~al.}(2008{\natexlab{a}})\citenamefont {Bhattacharyya}, \citenamefont
  {Hubeny}, \citenamefont {Minwalla},\ and\ \citenamefont
  {Rangamani}}]{Bhattacharyya:2008jc}%
  \BibitemOpen
  \bibfield  {author} {\bibinfo {author} {\bibfnamefont {S.}~\bibnamefont
  {Bhattacharyya}}, \bibinfo {author} {\bibfnamefont {V.~E.}\ \bibnamefont
  {Hubeny}}, \bibinfo {author} {\bibfnamefont {S.}~\bibnamefont {Minwalla}}, \
  and\ \bibinfo {author} {\bibfnamefont {M.}~\bibnamefont {Rangamani}},\ }\href
  {\doibase 10.1088/1126-6708/2008/02/045} {\bibfield  {journal} {\bibinfo
  {journal} {JHEP}\ }\textbf {\bibinfo {volume} {0802}},\ \bibinfo {pages}
  {045} (\bibinfo {year} {2008}{\natexlab{a}})},\ \Eprint
  {http://arxiv.org/abs/0712.2456} {arXiv:0712.2456 [hep-th]} \BibitemShut
  {NoStop}%
\bibitem [{\citenamefont {Romatschke}(2010{\natexlab{a}})}]{Romatschke:2009kr}%
  \BibitemOpen
  \bibfield  {author} {\bibinfo {author} {\bibfnamefont {P.}~\bibnamefont
  {Romatschke}},\ }\href {\doibase 10.1088/0264-9381/27/2/025006} {\bibfield
  {journal} {\bibinfo  {journal} {Class.Quant.Grav.}\ }\textbf {\bibinfo
  {volume} {27}},\ \bibinfo {pages} {025006} (\bibinfo {year}
  {2010}{\natexlab{a}})},\ \Eprint {http://arxiv.org/abs/0906.4787}
  {arXiv:0906.4787 [hep-th]} \BibitemShut {NoStop}%
\bibitem [{\citenamefont {Kovtun}(2012)}]{Kovtun:2012rj}%
  \BibitemOpen
  \bibfield  {author} {\bibinfo {author} {\bibfnamefont {P.}~\bibnamefont
  {Kovtun}},\ }\href {\doibase 10.1088/1751-8113/45/47/473001} {\bibfield
  {journal} {\bibinfo  {journal} {J.Phys.}\ }\textbf {\bibinfo {volume}
  {A45}},\ \bibinfo {pages} {473001} (\bibinfo {year} {2012})},\ \Eprint
  {http://arxiv.org/abs/1205.5040} {arXiv:1205.5040 [hep-th]} \BibitemShut
  {NoStop}%
\bibitem [{\citenamefont {Loganayagam}(2008)}]{Loganayagam:2008is}%
  \BibitemOpen
  \bibfield  {author} {\bibinfo {author} {\bibfnamefont {R.}~\bibnamefont
  {Loganayagam}},\ }\href {\doibase 10.1088/1126-6708/2008/05/087} {\bibfield
  {journal} {\bibinfo  {journal} {JHEP}\ }\textbf {\bibinfo {volume} {0805}},\
  \bibinfo {pages} {087} (\bibinfo {year} {2008})},\ \Eprint
  {http://arxiv.org/abs/0801.3701} {arXiv:0801.3701 [hep-th]} \BibitemShut
  {NoStop}%
\bibitem [{\citenamefont {Haack}\ and\ \citenamefont
  {Yarom}(2009)}]{Haack:2008xx}%
  \BibitemOpen
  \bibfield  {author} {\bibinfo {author} {\bibfnamefont {M.}~\bibnamefont
  {Haack}}\ and\ \bibinfo {author} {\bibfnamefont {A.}~\bibnamefont {Yarom}},\
  }\href {\doibase 10.1016/j.nuclphysb.2008.12.028} {\bibfield  {journal}
  {\bibinfo  {journal} {Nucl.Phys.}\ }\textbf {\bibinfo {volume} {B813}},\
  \bibinfo {pages} {140} (\bibinfo {year} {2009})},\ \Eprint
  {http://arxiv.org/abs/0811.1794} {arXiv:0811.1794 [hep-th]} \BibitemShut
  {NoStop}%
\bibitem [{\citenamefont {Bhattacharyya}(2012)}]{Bhattacharyya:2012nq}%
  \BibitemOpen
  \bibfield  {author} {\bibinfo {author} {\bibfnamefont {S.}~\bibnamefont
  {Bhattacharyya}},\ }\href {\doibase 10.1007/JHEP07(2012)104} {\bibfield
  {journal} {\bibinfo  {journal} {JHEP}\ }\textbf {\bibinfo {volume} {1207}},\
  \bibinfo {pages} {104} (\bibinfo {year} {2012})},\ \Eprint
  {http://arxiv.org/abs/1201.4654} {arXiv:1201.4654 [hep-th]} \BibitemShut
  {NoStop}%
\bibitem [{\citenamefont {Jensen}\ \emph {et~al.}(2012)\citenamefont {Jensen},
  \citenamefont {Kaminski}, \citenamefont {Kovtun}, \citenamefont {Meyer},
  \citenamefont {Ritz} \emph {et~al.}}]{Jensen:2012jh}%
  \BibitemOpen
  \bibfield  {author} {\bibinfo {author} {\bibfnamefont {K.}~\bibnamefont
  {Jensen}}, \bibinfo {author} {\bibfnamefont {M.}~\bibnamefont {Kaminski}},
  \bibinfo {author} {\bibfnamefont {P.}~\bibnamefont {Kovtun}}, \bibinfo
  {author} {\bibfnamefont {R.}~\bibnamefont {Meyer}}, \bibinfo {author}
  {\bibfnamefont {A.}~\bibnamefont {Ritz}},  \emph {et~al.},\ }\href {\doibase
  10.1103/PhysRevLett.109.101601} {\bibfield  {journal} {\bibinfo  {journal}
  {Phys.Rev.Lett.}\ }\textbf {\bibinfo {volume} {109}},\ \bibinfo {pages}
  {101601} (\bibinfo {year} {2012})},\ \Eprint {http://arxiv.org/abs/1203.3556}
  {arXiv:1203.3556 [hep-th]} \BibitemShut {NoStop}%
\bibitem [{\citenamefont {Moore}\ and\ \citenamefont
  {Sohrabi}(2012)}]{Moore:2012tc}%
  \BibitemOpen
  \bibfield  {author} {\bibinfo {author} {\bibfnamefont {G.~D.}\ \bibnamefont
  {Moore}}\ and\ \bibinfo {author} {\bibfnamefont {K.~A.}\ \bibnamefont
  {Sohrabi}},\ }\href {\doibase 10.1007/JHEP11(2012)148} {\bibfield  {journal}
  {\bibinfo  {journal} {JHEP}\ }\textbf {\bibinfo {volume} {1211}},\ \bibinfo
  {pages} {148} (\bibinfo {year} {2012})},\ \Eprint
  {http://arxiv.org/abs/1210.3340} {arXiv:1210.3340 [hep-ph]} \BibitemShut
  {NoStop}%
\bibitem [{\citenamefont {Banerjee}\ \emph {et~al.}(2012)\citenamefont
  {Banerjee}, \citenamefont {Bhattacharya}, \citenamefont {Bhattacharyya},
  \citenamefont {Jain}, \citenamefont {Minwalla},\ and\ \citenamefont
  {Sharma}}]{Banerjee:2012iz}%
  \BibitemOpen
  \bibfield  {author} {\bibinfo {author} {\bibfnamefont {N.}~\bibnamefont
  {Banerjee}}, \bibinfo {author} {\bibfnamefont {J.}~\bibnamefont
  {Bhattacharya}}, \bibinfo {author} {\bibfnamefont {S.}~\bibnamefont
  {Bhattacharyya}}, \bibinfo {author} {\bibfnamefont {S.}~\bibnamefont {Jain}},
  \bibinfo {author} {\bibfnamefont {S.}~\bibnamefont {Minwalla}}, \ and\
  \bibinfo {author} {\bibfnamefont {T.}~\bibnamefont {Sharma}},\ }\href
  {\doibase 10.1007/JHEP09(2012)046} {\bibfield  {journal} {\bibinfo  {journal}
  {JHEP}\ }\textbf {\bibinfo {volume} {09}},\ \bibinfo {pages} {046} (\bibinfo
  {year} {2012})},\ \Eprint {http://arxiv.org/abs/1203.3544} {arXiv:1203.3544
  [hep-th]} \BibitemShut {NoStop}%
\bibitem [{\citenamefont {Casalderrey-Solana}\ \emph
  {et~al.}(2014)\citenamefont {Casalderrey-Solana}, \citenamefont {Liu},
  \citenamefont {Mateos}, \citenamefont {Rajagopal},\ and\ \citenamefont
  {Wiedemann}}]{jorge-book}%
  \BibitemOpen
  \bibfield  {author} {\bibinfo {author} {\bibfnamefont {J.}~\bibnamefont
  {Casalderrey-Solana}}, \bibinfo {author} {\bibfnamefont {H.}~\bibnamefont
  {Liu}}, \bibinfo {author} {\bibfnamefont {D.}~\bibnamefont {Mateos}},
  \bibinfo {author} {\bibfnamefont {K.}~\bibnamefont {Rajagopal}}, \ and\
  \bibinfo {author} {\bibfnamefont {U.~A.}\ \bibnamefont {Wiedemann}},\
  }\href@noop {} {\emph {\bibinfo {title} {{Gauge/String Duality, Hot QCD and
  Heavy Ion Collisions}}}}\ (\bibinfo  {publisher} {Cambridge University
  Press},\ \bibinfo {address} {Cambridge, UK},\ \bibinfo {year}
  {2014})\BibitemShut {NoStop}%
\bibitem [{\citenamefont {Policastro}\ \emph
  {et~al.}(2002{\natexlab{a}})\citenamefont {Policastro}, \citenamefont {Son},\
  and\ \citenamefont {Starinets}}]{Policastro:2002se}%
  \BibitemOpen
  \bibfield  {author} {\bibinfo {author} {\bibfnamefont {G.}~\bibnamefont
  {Policastro}}, \bibinfo {author} {\bibfnamefont {D.~T.}\ \bibnamefont {Son}},
  \ and\ \bibinfo {author} {\bibfnamefont {A.~O.}\ \bibnamefont {Starinets}},\
  }\href@noop {} {\bibfield  {journal} {\bibinfo  {journal} {JHEP}\ }\textbf
  {\bibinfo {volume} {0209}},\ \bibinfo {pages} {043} (\bibinfo {year}
  {2002}{\natexlab{a}})},\ \Eprint {http://arxiv.org/abs/hep-th/0205052}
  {arXiv:hep-th/0205052 [hep-th]} \BibitemShut {NoStop}%
\bibitem [{\citenamefont {Policastro}\ \emph
  {et~al.}(2002{\natexlab{b}})\citenamefont {Policastro}, \citenamefont {Son},\
  and\ \citenamefont {Starinets}}]{Policastro:2002tn}%
  \BibitemOpen
  \bibfield  {author} {\bibinfo {author} {\bibfnamefont {G.}~\bibnamefont
  {Policastro}}, \bibinfo {author} {\bibfnamefont {D.~T.}\ \bibnamefont {Son}},
  \ and\ \bibinfo {author} {\bibfnamefont {A.~O.}\ \bibnamefont {Starinets}},\
  }\href@noop {} {\bibfield  {journal} {\bibinfo  {journal} {JHEP}\ }\textbf
  {\bibinfo {volume} {0212}},\ \bibinfo {pages} {054} (\bibinfo {year}
  {2002}{\natexlab{b}})},\ \Eprint {http://arxiv.org/abs/hep-th/0210220}
  {arXiv:hep-th/0210220 [hep-th]} \BibitemShut {NoStop}%
\bibitem [{\citenamefont {Kovtun}\ and\ \citenamefont
  {Starinets}(2005)}]{Kovtun:2005ev}%
  \BibitemOpen
  \bibfield  {author} {\bibinfo {author} {\bibfnamefont {P.~K.}\ \bibnamefont
  {Kovtun}}\ and\ \bibinfo {author} {\bibfnamefont {A.~O.}\ \bibnamefont
  {Starinets}},\ }\href {\doibase 10.1103/PhysRevD.72.086009} {\bibfield
  {journal} {\bibinfo  {journal} {Phys.Rev.}\ }\textbf {\bibinfo {volume}
  {D72}},\ \bibinfo {pages} {086009} (\bibinfo {year} {2005})},\ \Eprint
  {http://arxiv.org/abs/hep-th/0506184} {arXiv:hep-th/0506184 [hep-th]}
  \BibitemShut {NoStop}%
\bibitem [{\citenamefont {Kovtun}\ and\ \citenamefont
  {Starinets}(2006)}]{Kovtun:2006pf}%
  \BibitemOpen
  \bibfield  {author} {\bibinfo {author} {\bibfnamefont {P.}~\bibnamefont
  {Kovtun}}\ and\ \bibinfo {author} {\bibfnamefont {A.}~\bibnamefont
  {Starinets}},\ }\href {\doibase 10.1103/PhysRevLett.96.131601} {\bibfield
  {journal} {\bibinfo  {journal} {Phys.Rev.Lett.}\ }\textbf {\bibinfo {volume}
  {96}},\ \bibinfo {pages} {131601} (\bibinfo {year} {2006})},\ \Eprint
  {http://arxiv.org/abs/hep-th/0602059} {arXiv:hep-th/0602059 [hep-th]}
  \BibitemShut {NoStop}%
\bibitem [{\citenamefont {Banerjee}\ and\ \citenamefont
  {Dutta}(2010)}]{Banerjee:2010zd}%
  \BibitemOpen
  \bibfield  {author} {\bibinfo {author} {\bibfnamefont {N.}~\bibnamefont
  {Banerjee}}\ and\ \bibinfo {author} {\bibfnamefont {S.}~\bibnamefont
  {Dutta}},\ }\href {\doibase 10.1007/JHEP08(2010)041} {\bibfield  {journal}
  {\bibinfo  {journal} {JHEP}\ }\textbf {\bibinfo {volume} {1008}},\ \bibinfo
  {pages} {041} (\bibinfo {year} {2010})},\ \Eprint
  {http://arxiv.org/abs/1005.2367} {arXiv:1005.2367 [hep-th]} \BibitemShut
  {NoStop}%
\bibitem [{\citenamefont {Moore}\ and\ \citenamefont
  {Sohrabi}(2011)}]{Moore:2010bu}%
  \BibitemOpen
  \bibfield  {author} {\bibinfo {author} {\bibfnamefont {G.~D.}\ \bibnamefont
  {Moore}}\ and\ \bibinfo {author} {\bibfnamefont {K.~A.}\ \bibnamefont
  {Sohrabi}},\ }\href {\doibase 10.1103/PhysRevLett.106.122302} {\bibfield
  {journal} {\bibinfo  {journal} {Phys.Rev.Lett.}\ }\textbf {\bibinfo {volume}
  {106}},\ \bibinfo {pages} {122302} (\bibinfo {year} {2011})},\ \Eprint
  {http://arxiv.org/abs/1007.5333} {arXiv:1007.5333 [hep-ph]} \BibitemShut
  {NoStop}%
\bibitem [{\citenamefont {Shaverin}\ and\ \citenamefont
  {Yarom}(2013)}]{Shaverin:2012kv}%
  \BibitemOpen
  \bibfield  {author} {\bibinfo {author} {\bibfnamefont {E.}~\bibnamefont
  {Shaverin}}\ and\ \bibinfo {author} {\bibfnamefont {A.}~\bibnamefont
  {Yarom}},\ }\href {\doibase 10.1007/JHEP04(2013)013} {\bibfield  {journal}
  {\bibinfo  {journal} {JHEP}\ }\textbf {\bibinfo {volume} {1304}},\ \bibinfo
  {pages} {013} (\bibinfo {year} {2013})},\ \Eprint
  {http://arxiv.org/abs/1211.1979} {arXiv:1211.1979 [hep-th]} \BibitemShut
  {NoStop}%
\bibitem [{\citenamefont {Romatschke}(2010{\natexlab{b}})}]{Romatschke:2009im}%
  \BibitemOpen
  \bibfield  {author} {\bibinfo {author} {\bibfnamefont {P.}~\bibnamefont
  {Romatschke}},\ }\href {\doibase 10.1142/S0218301310014613} {\bibfield
  {journal} {\bibinfo  {journal} {Int.J.Mod.Phys.}\ }\textbf {\bibinfo {volume}
  {E19}},\ \bibinfo {pages} {1} (\bibinfo {year} {2010}{\natexlab{b}})},\
  \Eprint {http://arxiv.org/abs/0902.3663} {arXiv:0902.3663 [hep-ph]}
  \BibitemShut {NoStop}%
\bibitem [{\citenamefont {Grozdanov}\ and\ \citenamefont
  {Starinets}(2015{\natexlab{a}})}]{Grozdanov:2015asa}%
  \BibitemOpen
  \bibfield  {author} {\bibinfo {author} {\bibfnamefont {S.}~\bibnamefont
  {Grozdanov}}\ and\ \bibinfo {author} {\bibfnamefont {A.}~\bibnamefont
  {Starinets}},\ }\href {\doibase 10.1007/s11232-015-0245-7} {\bibfield
  {journal} {\bibinfo  {journal} {Theor.Math.Phys.}\ }\textbf {\bibinfo
  {volume} {182}},\ \bibinfo {pages} {61} (\bibinfo {year}
  {2015}{\natexlab{a}})}\BibitemShut {NoStop}%
\bibitem [{\citenamefont {Grozdanov}\ and\ \citenamefont
  {Starinets}(2015{\natexlab{b}})}]{Grozdanov:2014kva}%
  \BibitemOpen
  \bibfield  {author} {\bibinfo {author} {\bibfnamefont {S.}~\bibnamefont
  {Grozdanov}}\ and\ \bibinfo {author} {\bibfnamefont {A.~O.}\ \bibnamefont
  {Starinets}},\ }\href {\doibase 10.1007/JHEP03(2015)007} {\bibfield
  {journal} {\bibinfo  {journal} {JHEP}\ }\textbf {\bibinfo {volume} {1503}},\
  \bibinfo {pages} {007} (\bibinfo {year} {2015}{\natexlab{b}})},\ \Eprint
  {http://arxiv.org/abs/1412.5685} {arXiv:1412.5685 [hep-th]} \BibitemShut
  {NoStop}%
\bibitem [{\citenamefont {Bu}\ and\ \citenamefont
  {Lublinsky}(2014{\natexlab{a}})}]{Bu:2014sia}%
  \BibitemOpen
  \bibfield  {author} {\bibinfo {author} {\bibfnamefont {Y.}~\bibnamefont
  {Bu}}\ and\ \bibinfo {author} {\bibfnamefont {M.}~\bibnamefont {Lublinsky}},\
  }\href {\doibase 10.1103/PhysRevD.90.086003} {\bibfield  {journal} {\bibinfo
  {journal} {Phys.Rev.}\ }\textbf {\bibinfo {volume} {D90}},\ \bibinfo {pages}
  {086003} (\bibinfo {year} {2014}{\natexlab{a}})},\ \Eprint
  {http://arxiv.org/abs/1406.7222} {arXiv:1406.7222 [hep-th]} \BibitemShut
  {NoStop}%
\bibitem [{\citenamefont {Bu}\ and\ \citenamefont
  {Lublinsky}(2014{\natexlab{b}})}]{Bu:2014ena}%
  \BibitemOpen
  \bibfield  {author} {\bibinfo {author} {\bibfnamefont {Y.}~\bibnamefont
  {Bu}}\ and\ \bibinfo {author} {\bibfnamefont {M.}~\bibnamefont {Lublinsky}},\
  }\href {\doibase 10.1007/JHEP11(2014)064} {\bibfield  {journal} {\bibinfo
  {journal} {JHEP}\ }\textbf {\bibinfo {volume} {1411}},\ \bibinfo {pages}
  {064} (\bibinfo {year} {2014}{\natexlab{b}})},\ \Eprint
  {http://arxiv.org/abs/1409.3095} {arXiv:1409.3095 [hep-th]} \BibitemShut
  {NoStop}%
\bibitem [{\citenamefont {Bu}\ and\ \citenamefont
  {Lublinsky}(2015)}]{Bu:2015ika}%
  \BibitemOpen
  \bibfield  {author} {\bibinfo {author} {\bibfnamefont {Y.}~\bibnamefont
  {Bu}}\ and\ \bibinfo {author} {\bibfnamefont {M.}~\bibnamefont {Lublinsky}},\
  }\href {\doibase 10.1007/JHEP04(2015)136} {\bibfield  {journal} {\bibinfo
  {journal} {JHEP}\ }\textbf {\bibinfo {volume} {1504}},\ \bibinfo {pages}
  {136} (\bibinfo {year} {2015})},\ \Eprint {http://arxiv.org/abs/1502.08044}
  {arXiv:1502.08044 [hep-th]} \BibitemShut {NoStop}%
\bibitem [{\citenamefont {Bu}\ \emph {et~al.}(2015)\citenamefont {Bu},
  \citenamefont {Lublinsky},\ and\ \citenamefont {Sharon}}]{Bu:2015bwa}%
  \BibitemOpen
  \bibfield  {author} {\bibinfo {author} {\bibfnamefont {Y.}~\bibnamefont
  {Bu}}, \bibinfo {author} {\bibfnamefont {M.}~\bibnamefont {Lublinsky}}, \
  and\ \bibinfo {author} {\bibfnamefont {A.}~\bibnamefont {Sharon}},\
  }\href@noop {} {\  (\bibinfo {year} {2015})},\ \Eprint
  {http://arxiv.org/abs/1504.01370} {arXiv:1504.01370 [hep-th]} \BibitemShut
  {NoStop}%
\bibitem [{\citenamefont {Kovtun}\ and\ \citenamefont
  {Yaffe}(2003)}]{Kovtun:2003vj}%
  \BibitemOpen
  \bibfield  {author} {\bibinfo {author} {\bibfnamefont {P.}~\bibnamefont
  {Kovtun}}\ and\ \bibinfo {author} {\bibfnamefont {L.~G.}\ \bibnamefont
  {Yaffe}},\ }\href {\doibase 10.1103/PhysRevD.68.025007} {\bibfield  {journal}
  {\bibinfo  {journal} {Phys.Rev.}\ }\textbf {\bibinfo {volume} {D68}},\
  \bibinfo {pages} {025007} (\bibinfo {year} {2003})},\ \Eprint
  {http://arxiv.org/abs/hep-th/0303010} {arXiv:hep-th/0303010 [hep-th]}
  \BibitemShut {NoStop}%
\bibitem [{\citenamefont {Caron-Huot}\ and\ \citenamefont
  {Saremi}(2010)}]{CaronHuot:2009iq}%
  \BibitemOpen
  \bibfield  {author} {\bibinfo {author} {\bibfnamefont {S.}~\bibnamefont
  {Caron-Huot}}\ and\ \bibinfo {author} {\bibfnamefont {O.}~\bibnamefont
  {Saremi}},\ }\href {\doibase 10.1007/JHEP11(2010)013} {\bibfield  {journal}
  {\bibinfo  {journal} {JHEP}\ }\textbf {\bibinfo {volume} {1011}},\ \bibinfo
  {pages} {013} (\bibinfo {year} {2010})},\ \Eprint
  {http://arxiv.org/abs/0909.4525} {arXiv:0909.4525 [hep-th]} \BibitemShut
  {NoStop}%
\bibitem [{\citenamefont {Heller}\ \emph {et~al.}(2013)\citenamefont {Heller},
  \citenamefont {Janik},\ and\ \citenamefont {Witaszczyk}}]{Heller:2013fn}%
  \BibitemOpen
  \bibfield  {author} {\bibinfo {author} {\bibfnamefont {M.~P.}\ \bibnamefont
  {Heller}}, \bibinfo {author} {\bibfnamefont {R.~A.}\ \bibnamefont {Janik}}, \
  and\ \bibinfo {author} {\bibfnamefont {P.}~\bibnamefont {Witaszczyk}},\
  }\href {\doibase 10.1103/PhysRevLett.110.211602} {\bibfield  {journal}
  {\bibinfo  {journal} {Phys.Rev.Lett.}\ }\textbf {\bibinfo {volume} {110}},\
  \bibinfo {pages} {211602} (\bibinfo {year} {2013})},\ \Eprint
  {http://arxiv.org/abs/1302.0697} {arXiv:1302.0697 [hep-th]} \BibitemShut
  {NoStop}%
\bibitem [{\citenamefont {Heller}\ and\ \citenamefont
  {Spalinski}(2015)}]{Heller:2015dha}%
  \BibitemOpen
  \bibfield  {author} {\bibinfo {author} {\bibfnamefont {M.~P.}\ \bibnamefont
  {Heller}}\ and\ \bibinfo {author} {\bibfnamefont {M.}~\bibnamefont
  {Spalinski}},\ }\href@noop {} {\  (\bibinfo {year} {2015})},\ \Eprint
  {http://arxiv.org/abs/1503.07514} {arXiv:1503.07514 [hep-th]} \BibitemShut
  {NoStop}%
\bibitem [{\citenamefont {Bjorken}(1983)}]{Bjorken:1982qr}%
  \BibitemOpen
  \bibfield  {author} {\bibinfo {author} {\bibfnamefont {J.}~\bibnamefont
  {Bjorken}},\ }\href {\doibase 10.1103/PhysRevD.27.140} {\bibfield  {journal}
  {\bibinfo  {journal} {Phys.Rev.}\ }\textbf {\bibinfo {volume} {D27}},\
  \bibinfo {pages} {140} (\bibinfo {year} {1983})}\BibitemShut {NoStop}%
\bibitem [{\citenamefont {Janik}\ and\ \citenamefont
  {Peschanski}(2006)}]{Janik:2005zt}%
  \BibitemOpen
  \bibfield  {author} {\bibinfo {author} {\bibfnamefont {R.~A.}\ \bibnamefont
  {Janik}}\ and\ \bibinfo {author} {\bibfnamefont {R.~B.}\ \bibnamefont
  {Peschanski}},\ }\href {\doibase 10.1103/PhysRevD.73.045013} {\bibfield
  {journal} {\bibinfo  {journal} {Phys. Rev.}\ }\textbf {\bibinfo {volume}
  {D73}},\ \bibinfo {pages} {045013} (\bibinfo {year} {2006})},\ \Eprint
  {http://arxiv.org/abs/hep-th/0512162} {arXiv:hep-th/0512162 [hep-th]}
  \BibitemShut {NoStop}%
\bibitem [{\citenamefont {Janik}(2007)}]{Janik:2006ft}%
  \BibitemOpen
  \bibfield  {author} {\bibinfo {author} {\bibfnamefont {R.~A.}\ \bibnamefont
  {Janik}},\ }\href {\doibase 10.1103/PhysRevLett.98.022302} {\bibfield
  {journal} {\bibinfo  {journal} {Phys. Rev. Lett.}\ }\textbf {\bibinfo
  {volume} {98}},\ \bibinfo {pages} {022302} (\bibinfo {year} {2007})},\
  \Eprint {http://arxiv.org/abs/hep-th/0610144} {arXiv:hep-th/0610144 [hep-th]}
  \BibitemShut {NoStop}%
\bibitem [{\citenamefont {Heller}\ and\ \citenamefont
  {Janik}(2007)}]{Heller:2007qt}%
  \BibitemOpen
  \bibfield  {author} {\bibinfo {author} {\bibfnamefont {M.~P.}\ \bibnamefont
  {Heller}}\ and\ \bibinfo {author} {\bibfnamefont {R.~A.}\ \bibnamefont
  {Janik}},\ }\href {\doibase 10.1103/PhysRevD.76.025027} {\bibfield  {journal}
  {\bibinfo  {journal} {Phys.Rev.}\ }\textbf {\bibinfo {volume} {D76}},\
  \bibinfo {pages} {025027} (\bibinfo {year} {2007})},\ \Eprint
  {http://arxiv.org/abs/hep-th/0703243} {arXiv:hep-th/0703243 [HEP-TH]}
  \BibitemShut {NoStop}%
\bibitem [{\citenamefont {Dyson}(1952)}]{PhysRev.85.631}%
  \BibitemOpen
  \bibfield  {author} {\bibinfo {author} {\bibfnamefont {F.~J.}\ \bibnamefont
  {Dyson}},\ }\href {\doibase 10.1103/PhysRev.85.631} {\bibfield  {journal}
  {\bibinfo  {journal} {Phys. Rev.}\ }\textbf {\bibinfo {volume} {85}},\
  \bibinfo {pages} {631} (\bibinfo {year} {1952})}\BibitemShut {NoStop}%
\bibitem [{\citenamefont {York}\ and\ \citenamefont
  {Moore}(2009)}]{York:2008rr}%
  \BibitemOpen
  \bibfield  {author} {\bibinfo {author} {\bibfnamefont {M.~A.}\ \bibnamefont
  {York}}\ and\ \bibinfo {author} {\bibfnamefont {G.~D.}\ \bibnamefont
  {Moore}},\ }\href {\doibase 10.1103/PhysRevD.79.054011} {\bibfield  {journal}
  {\bibinfo  {journal} {Phys.Rev.}\ }\textbf {\bibinfo {volume} {D79}},\
  \bibinfo {pages} {054011} (\bibinfo {year} {2009})},\ \Eprint
  {http://arxiv.org/abs/0811.0729} {arXiv:0811.0729 [hep-ph]} \BibitemShut
  {NoStop}%
\bibitem [{\citenamefont {Jaiswal}(2013)}]{Jaiswal:2013vta}%
  \BibitemOpen
  \bibfield  {author} {\bibinfo {author} {\bibfnamefont {A.}~\bibnamefont
  {Jaiswal}},\ }\href {\doibase 10.1103/PhysRevC.88.021903} {\bibfield
  {journal} {\bibinfo  {journal} {Phys.Rev.}\ }\textbf {\bibinfo {volume}
  {C88}},\ \bibinfo {pages} {021903} (\bibinfo {year} {2013})},\ \Eprint
  {http://arxiv.org/abs/1305.3480} {arXiv:1305.3480 [nucl-th]} \BibitemShut
  {NoStop}%
\bibitem [{\citenamefont {Jaiswal}(2014)}]{Jaiswal:2014raa}%
  \BibitemOpen
  \bibfield  {author} {\bibinfo {author} {\bibfnamefont {A.}~\bibnamefont
  {Jaiswal}},\ }\href {\doibase 10.1016/j.nuclphysa.2014.08.035} {\bibfield
  {journal} {\bibinfo  {journal} {Nucl.Phys.}\ }\textbf {\bibinfo {volume}
  {A931}},\ \bibinfo {pages} {1205} (\bibinfo {year} {2014})},\ \Eprint
  {http://arxiv.org/abs/1407.0837} {arXiv:1407.0837 [nucl-th]} \BibitemShut
  {NoStop}%
\bibitem [{\citenamefont {Chattopadhyay}\ \emph {et~al.}(2015)\citenamefont
  {Chattopadhyay}, \citenamefont {Jaiswal}, \citenamefont {Pal},\ and\
  \citenamefont {Ryblewski}}]{Chattopadhyay:2014lya}%
  \BibitemOpen
  \bibfield  {author} {\bibinfo {author} {\bibfnamefont {C.}~\bibnamefont
  {Chattopadhyay}}, \bibinfo {author} {\bibfnamefont {A.}~\bibnamefont
  {Jaiswal}}, \bibinfo {author} {\bibfnamefont {S.}~\bibnamefont {Pal}}, \ and\
  \bibinfo {author} {\bibfnamefont {R.}~\bibnamefont {Ryblewski}},\ }\href
  {\doibase 10.1103/PhysRevC.91.024917} {\bibfield  {journal} {\bibinfo
  {journal} {Phys.Rev.}\ }\textbf {\bibinfo {volume} {C91}},\ \bibinfo {pages}
  {024917} (\bibinfo {year} {2015})},\ \Eprint {http://arxiv.org/abs/1411.2363}
  {arXiv:1411.2363 [nucl-th]} \BibitemShut {NoStop}%
\bibitem [{\citenamefont {Buchel}\ \emph {et~al.}(2005)\citenamefont {Buchel},
  \citenamefont {Liu},\ and\ \citenamefont {Starinets}}]{Buchel:2004di}%
  \BibitemOpen
  \bibfield  {author} {\bibinfo {author} {\bibfnamefont {A.}~\bibnamefont
  {Buchel}}, \bibinfo {author} {\bibfnamefont {J.~T.}\ \bibnamefont {Liu}}, \
  and\ \bibinfo {author} {\bibfnamefont {A.~O.}\ \bibnamefont {Starinets}},\
  }\href {\doibase 10.1016/j.nuclphysb.2004.11.055} {\bibfield  {journal}
  {\bibinfo  {journal} {Nucl.Phys.}\ }\textbf {\bibinfo {volume} {B707}},\
  \bibinfo {pages} {56} (\bibinfo {year} {2005})},\ \Eprint
  {http://arxiv.org/abs/hep-th/0406264} {arXiv:hep-th/0406264 [hep-th]}
  \BibitemShut {NoStop}%
\bibitem [{\citenamefont {Benincasa}\ and\ \citenamefont
  {Buchel}(2006)}]{Benincasa:2005qc}%
  \BibitemOpen
  \bibfield  {author} {\bibinfo {author} {\bibfnamefont {P.}~\bibnamefont
  {Benincasa}}\ and\ \bibinfo {author} {\bibfnamefont {A.}~\bibnamefont
  {Buchel}},\ }\href {\doibase 10.1088/1126-6708/2006/01/103} {\bibfield
  {journal} {\bibinfo  {journal} {JHEP}\ }\textbf {\bibinfo {volume} {0601}},\
  \bibinfo {pages} {103} (\bibinfo {year} {2006})},\ \Eprint
  {http://arxiv.org/abs/hep-th/0510041} {arXiv:hep-th/0510041 [hep-th]}
  \BibitemShut {NoStop}%
\bibitem [{\citenamefont {Buchel}(2008{\natexlab{a}})}]{Buchel:2008sh}%
  \BibitemOpen
  \bibfield  {author} {\bibinfo {author} {\bibfnamefont {A.}~\bibnamefont
  {Buchel}},\ }\href {\doibase 10.1016/j.nuclphysb.2008.05.024} {\bibfield
  {journal} {\bibinfo  {journal} {Nucl.Phys.}\ }\textbf {\bibinfo {volume}
  {B803}},\ \bibinfo {pages} {166} (\bibinfo {year} {2008}{\natexlab{a}})},\
  \Eprint {http://arxiv.org/abs/0805.2683} {arXiv:0805.2683 [hep-th]}
  \BibitemShut {NoStop}%
\bibitem [{\citenamefont {Buchel}(2008{\natexlab{b}})}]{Buchel:2008ac}%
  \BibitemOpen
  \bibfield  {author} {\bibinfo {author} {\bibfnamefont {A.}~\bibnamefont
  {Buchel}},\ }\href {\doibase 10.1016/j.nuclphysb.2008.03.009} {\bibfield
  {journal} {\bibinfo  {journal} {Nucl.Phys.}\ }\textbf {\bibinfo {volume}
  {B802}},\ \bibinfo {pages} {281} (\bibinfo {year} {2008}{\natexlab{b}})},\
  \Eprint {http://arxiv.org/abs/0801.4421} {arXiv:0801.4421 [hep-th]}
  \BibitemShut {NoStop}%
\bibitem [{\citenamefont {Buchel}\ and\ \citenamefont
  {Paulos}(2008)}]{Buchel:2008bz}%
  \BibitemOpen
  \bibfield  {author} {\bibinfo {author} {\bibfnamefont {A.}~\bibnamefont
  {Buchel}}\ and\ \bibinfo {author} {\bibfnamefont {M.}~\bibnamefont
  {Paulos}},\ }\href {\doibase 10.1016/j.nuclphysb.2008.07.002} {\bibfield
  {journal} {\bibinfo  {journal} {Nucl.Phys.}\ }\textbf {\bibinfo {volume}
  {B805}},\ \bibinfo {pages} {59} (\bibinfo {year} {2008})},\ \Eprint
  {http://arxiv.org/abs/0806.0788} {arXiv:0806.0788 [hep-th]} \BibitemShut
  {NoStop}%
\bibitem [{\citenamefont {Buchel}\ and\ \citenamefont
  {Paulos}(2009)}]{Buchel:2008kd}%
  \BibitemOpen
  \bibfield  {author} {\bibinfo {author} {\bibfnamefont {A.}~\bibnamefont
  {Buchel}}\ and\ \bibinfo {author} {\bibfnamefont {M.}~\bibnamefont
  {Paulos}},\ }\href {\doibase 10.1016/j.nuclphysb.2008.10.012} {\bibfield
  {journal} {\bibinfo  {journal} {Nucl.Phys.}\ }\textbf {\bibinfo {volume}
  {B810}},\ \bibinfo {pages} {40} (\bibinfo {year} {2009})},\ \Eprint
  {http://arxiv.org/abs/0808.1601} {arXiv:0808.1601 [hep-th]} \BibitemShut
  {NoStop}%
\bibitem [{\citenamefont {Saremi}\ and\ \citenamefont
  {Sohrabi}(2011)}]{Saremi:2011nh}%
  \BibitemOpen
  \bibfield  {author} {\bibinfo {author} {\bibfnamefont {O.}~\bibnamefont
  {Saremi}}\ and\ \bibinfo {author} {\bibfnamefont {K.~A.}\ \bibnamefont
  {Sohrabi}},\ }\href {\doibase 10.1007/JHEP11(2011)147} {\bibfield  {journal}
  {\bibinfo  {journal} {JHEP}\ }\textbf {\bibinfo {volume} {1111}},\ \bibinfo
  {pages} {147} (\bibinfo {year} {2011})},\ \Eprint
  {http://arxiv.org/abs/1105.4870} {arXiv:1105.4870 [hep-th]} \BibitemShut
  {NoStop}%
\bibitem [{\citenamefont {Kovtun}\ \emph {et~al.}(2003)\citenamefont {Kovtun},
  \citenamefont {Son},\ and\ \citenamefont {Starinets}}]{Kovtun:2003wp}%
  \BibitemOpen
  \bibfield  {author} {\bibinfo {author} {\bibfnamefont {P.}~\bibnamefont
  {Kovtun}}, \bibinfo {author} {\bibfnamefont {D.~T.}\ \bibnamefont {Son}}, \
  and\ \bibinfo {author} {\bibfnamefont {A.~O.}\ \bibnamefont {Starinets}},\
  }\href {\doibase 10.1088/1126-6708/2003/10/064} {\bibfield  {journal}
  {\bibinfo  {journal} {JHEP}\ }\textbf {\bibinfo {volume} {0310}},\ \bibinfo
  {pages} {064} (\bibinfo {year} {2003})},\ \Eprint
  {http://arxiv.org/abs/hep-th/0309213} {arXiv:hep-th/0309213 [hep-th]}
  \BibitemShut {NoStop}%
\bibitem [{\citenamefont {Kovtun}\ \emph {et~al.}(2005)\citenamefont {Kovtun},
  \citenamefont {Son},\ and\ \citenamefont {Starinets}}]{Kovtun:2004de}%
  \BibitemOpen
  \bibfield  {author} {\bibinfo {author} {\bibfnamefont {P.}~\bibnamefont
  {Kovtun}}, \bibinfo {author} {\bibfnamefont {D.~T.}\ \bibnamefont {Son}}, \
  and\ \bibinfo {author} {\bibfnamefont {A.~O.}\ \bibnamefont {Starinets}},\
  }\href {\doibase 10.1103/PhysRevLett.94.111601} {\bibfield  {journal}
  {\bibinfo  {journal} {Phys.Rev.Lett.}\ }\textbf {\bibinfo {volume} {94}},\
  \bibinfo {pages} {111601} (\bibinfo {year} {2005})},\ \Eprint
  {http://arxiv.org/abs/hep-th/0405231} {arXiv:hep-th/0405231 [hep-th]}
  \BibitemShut {NoStop}%
\bibitem [{\citenamefont {Son}\ and\ \citenamefont
  {Starinets}(2002)}]{Son:2002sd}%
  \BibitemOpen
  \bibfield  {author} {\bibinfo {author} {\bibfnamefont {D.~T.}\ \bibnamefont
  {Son}}\ and\ \bibinfo {author} {\bibfnamefont {A.~O.}\ \bibnamefont
  {Starinets}},\ }\href@noop {} {\bibfield  {journal} {\bibinfo  {journal}
  {JHEP}\ }\textbf {\bibinfo {volume} {0209}},\ \bibinfo {pages} {042}
  (\bibinfo {year} {2002})},\ \Eprint {http://arxiv.org/abs/hep-th/0205051}
  {arXiv:hep-th/0205051 [hep-th]} \BibitemShut {NoStop}%
\bibitem [{\citenamefont {Heller}\ \emph {et~al.}(2008)\citenamefont {Heller},
  \citenamefont {Janik},\ and\ \citenamefont {Peschanski}}]{Heller:2008fg}%
  \BibitemOpen
  \bibfield  {author} {\bibinfo {author} {\bibfnamefont {M.~P.}\ \bibnamefont
  {Heller}}, \bibinfo {author} {\bibfnamefont {R.~A.}\ \bibnamefont {Janik}}, \
  and\ \bibinfo {author} {\bibfnamefont {R.}~\bibnamefont {Peschanski}},\
  }\bibfield  {booktitle} {\emph {\bibinfo {booktitle} {{Aspects of duality.
  Proceedings, 48th Cracow School of Theoretical Physics, Zakopane, Poland,
  June 13-22, 2008}}},\ }\href@noop {} {\bibfield  {journal} {\bibinfo
  {journal} {Acta Phys. Polon.}\ }\textbf {\bibinfo {volume} {B39}},\ \bibinfo
  {pages} {3183} (\bibinfo {year} {2008})},\ \Eprint
  {http://arxiv.org/abs/0811.3113} {arXiv:0811.3113 [hep-th]} \BibitemShut
  {NoStop}%
\bibitem [{\citenamefont {Heller}\ \emph {et~al.}(2012)\citenamefont {Heller},
  \citenamefont {Janik},\ and\ \citenamefont {Witaszczyk}}]{Heller:2011ju}%
  \BibitemOpen
  \bibfield  {author} {\bibinfo {author} {\bibfnamefont {M.~P.}\ \bibnamefont
  {Heller}}, \bibinfo {author} {\bibfnamefont {R.~A.}\ \bibnamefont {Janik}}, \
  and\ \bibinfo {author} {\bibfnamefont {P.}~\bibnamefont {Witaszczyk}},\
  }\href {\doibase 10.1103/PhysRevLett.108.201602} {\bibfield  {journal}
  {\bibinfo  {journal} {Phys. Rev. Lett.}\ }\textbf {\bibinfo {volume} {108}},\
  \bibinfo {pages} {201602} (\bibinfo {year} {2012})},\ \Eprint
  {http://arxiv.org/abs/1103.3452} {arXiv:1103.3452 [hep-th]} \BibitemShut
  {NoStop}%
\bibitem [{\citenamefont {Booth}\ \emph {et~al.}(2009)\citenamefont {Booth},
  \citenamefont {Heller},\ and\ \citenamefont {Spalinski}}]{Booth:2009ct}%
  \BibitemOpen
  \bibfield  {author} {\bibinfo {author} {\bibfnamefont {I.}~\bibnamefont
  {Booth}}, \bibinfo {author} {\bibfnamefont {M.~P.}\ \bibnamefont {Heller}}, \
  and\ \bibinfo {author} {\bibfnamefont {M.}~\bibnamefont {Spalinski}},\ }\href
  {\doibase 10.1103/PhysRevD.80.126013} {\bibfield  {journal} {\bibinfo
  {journal} {Phys. Rev.}\ }\textbf {\bibinfo {volume} {D80}},\ \bibinfo {pages}
  {126013} (\bibinfo {year} {2009})},\ \Eprint {http://arxiv.org/abs/0910.0748}
  {arXiv:0910.0748 [hep-th]} \BibitemShut {NoStop}%
\bibitem [{\citenamefont {Barnes}\ \emph {et~al.}(2010)\citenamefont {Barnes},
  \citenamefont {Vaman}, \citenamefont {Wu},\ and\ \citenamefont
  {Arnold}}]{Barnes:2010jp}%
  \BibitemOpen
  \bibfield  {author} {\bibinfo {author} {\bibfnamefont {E.}~\bibnamefont
  {Barnes}}, \bibinfo {author} {\bibfnamefont {D.}~\bibnamefont {Vaman}},
  \bibinfo {author} {\bibfnamefont {C.}~\bibnamefont {Wu}}, \ and\ \bibinfo
  {author} {\bibfnamefont {P.}~\bibnamefont {Arnold}},\ }\href {\doibase
  10.1103/PhysRevD.82.025019} {\bibfield  {journal} {\bibinfo  {journal}
  {Phys.Rev.}\ }\textbf {\bibinfo {volume} {D82}},\ \bibinfo {pages} {025019}
  (\bibinfo {year} {2010})},\ \Eprint {http://arxiv.org/abs/1004.1179}
  {arXiv:1004.1179 [hep-th]} \BibitemShut {NoStop}%
\bibitem [{\citenamefont {Arnold}\ \emph {et~al.}(2011)\citenamefont {Arnold},
  \citenamefont {Vaman}, \citenamefont {Wu},\ and\ \citenamefont
  {Xiao}}]{Arnold:2011ja}%
  \BibitemOpen
  \bibfield  {author} {\bibinfo {author} {\bibfnamefont {P.}~\bibnamefont
  {Arnold}}, \bibinfo {author} {\bibfnamefont {D.}~\bibnamefont {Vaman}},
  \bibinfo {author} {\bibfnamefont {C.}~\bibnamefont {Wu}}, \ and\ \bibinfo
  {author} {\bibfnamefont {W.}~\bibnamefont {Xiao}},\ }\href {\doibase
  10.1007/JHEP10(2011)033} {\bibfield  {journal} {\bibinfo  {journal} {JHEP}\
  }\textbf {\bibinfo {volume} {1110}},\ \bibinfo {pages} {033} (\bibinfo {year}
  {2011})},\ \Eprint {http://arxiv.org/abs/1105.4645} {arXiv:1105.4645
  [hep-th]} \BibitemShut {NoStop}%
\bibitem [{\citenamefont {Arnold}\ and\ \citenamefont
  {Vaman}(2011)}]{Arnold:2011hp}%
  \BibitemOpen
  \bibfield  {author} {\bibinfo {author} {\bibfnamefont {P.}~\bibnamefont
  {Arnold}}\ and\ \bibinfo {author} {\bibfnamefont {D.}~\bibnamefont {Vaman}},\
  }\href {\doibase 10.1007/JHEP11(2011)033} {\bibfield  {journal} {\bibinfo
  {journal} {JHEP}\ }\textbf {\bibinfo {volume} {1111}},\ \bibinfo {pages}
  {033} (\bibinfo {year} {2011})},\ \Eprint {http://arxiv.org/abs/1109.0040}
  {arXiv:1109.0040 [hep-th]} \BibitemShut {NoStop}%
\bibitem [{\citenamefont {Rangamani}(2009)}]{Rangamani:2009xk}%
  \BibitemOpen
  \bibfield  {author} {\bibinfo {author} {\bibfnamefont {M.}~\bibnamefont
  {Rangamani}},\ }\href {\doibase 10.1088/0264-9381/26/22/224003} {\bibfield
  {journal} {\bibinfo  {journal} {Class.Quant.Grav.}\ }\textbf {\bibinfo
  {volume} {26}},\ \bibinfo {pages} {224003} (\bibinfo {year} {2009})},\
  \Eprint {http://arxiv.org/abs/0905.4352} {arXiv:0905.4352 [hep-th]}
  \BibitemShut {NoStop}%
\bibitem [{\citenamefont {De~Groot}\ \emph {et~al.}(1980)\citenamefont
  {De~Groot}, \citenamefont {Van~Leeuwen},\ and\ \citenamefont
  {Van~Weert}}]{DeGroot:1980dk}%
  \BibitemOpen
  \bibfield  {author} {\bibinfo {author} {\bibfnamefont {S.}~\bibnamefont
  {De~Groot}}, \bibinfo {author} {\bibfnamefont {W.}~\bibnamefont
  {Van~Leeuwen}}, \ and\ \bibinfo {author} {\bibfnamefont {C.}~\bibnamefont
  {Van~Weert}},\ }\href@noop {} {\emph {\bibinfo {title} {{Relativistic Kinetic
  Theory. Principles and Applications}}}}\ (\bibinfo {year} {1980})\BibitemShut
  {NoStop}%
\bibitem [{\citenamefont {Bhattacharyya}\ \emph
  {et~al.}(2008{\natexlab{b}})\citenamefont {Bhattacharyya}, \citenamefont
  {Hubeny}, \citenamefont {Loganayagam}, \citenamefont {Mandal}, \citenamefont
  {Minwalla} \emph {et~al.}}]{Bhattacharyya:2008xc}%
  \BibitemOpen
  \bibfield  {author} {\bibinfo {author} {\bibfnamefont {S.}~\bibnamefont
  {Bhattacharyya}}, \bibinfo {author} {\bibfnamefont {V.~E.}\ \bibnamefont
  {Hubeny}}, \bibinfo {author} {\bibfnamefont {R.}~\bibnamefont {Loganayagam}},
  \bibinfo {author} {\bibfnamefont {G.}~\bibnamefont {Mandal}}, \bibinfo
  {author} {\bibfnamefont {S.}~\bibnamefont {Minwalla}},  \emph {et~al.},\
  }\href {\doibase 10.1088/1126-6708/2008/06/055} {\bibfield  {journal}
  {\bibinfo  {journal} {JHEP}\ }\textbf {\bibinfo {volume} {0806}},\ \bibinfo
  {pages} {055} (\bibinfo {year} {2008}{\natexlab{b}})},\ \Eprint
  {http://arxiv.org/abs/0803.2526} {arXiv:0803.2526 [hep-th]} \BibitemShut
  {NoStop}%
\bibitem [{\citenamefont {Buchel}\ and\ \citenamefont
  {Liu}(2004)}]{Buchel:2003tz}%
  \BibitemOpen
  \bibfield  {author} {\bibinfo {author} {\bibfnamefont {A.}~\bibnamefont
  {Buchel}}\ and\ \bibinfo {author} {\bibfnamefont {J.~T.}\ \bibnamefont
  {Liu}},\ }\href {\doibase 10.1103/PhysRevLett.93.090602} {\bibfield
  {journal} {\bibinfo  {journal} {Phys.Rev.Lett.}\ }\textbf {\bibinfo {volume}
  {93}},\ \bibinfo {pages} {090602} (\bibinfo {year} {2004})},\ \Eprint
  {http://arxiv.org/abs/hep-th/0311175} {arXiv:hep-th/0311175 [hep-th]}
  \BibitemShut {NoStop}%
\bibitem [{\citenamefont {Iqbal}\ and\ \citenamefont
  {Liu}(2009)}]{Iqbal:2008by}%
  \BibitemOpen
  \bibfield  {author} {\bibinfo {author} {\bibfnamefont {N.}~\bibnamefont
  {Iqbal}}\ and\ \bibinfo {author} {\bibfnamefont {H.}~\bibnamefont {Liu}},\
  }\href {\doibase 10.1103/PhysRevD.79.025023} {\bibfield  {journal} {\bibinfo
  {journal} {Phys.Rev.}\ }\textbf {\bibinfo {volume} {D79}},\ \bibinfo {pages}
  {025023} (\bibinfo {year} {2009})},\ \Eprint {http://arxiv.org/abs/0809.3808}
  {arXiv:0809.3808 [hep-th]} \BibitemShut {NoStop}%
\bibitem [{\citenamefont {Starinets}(2009)}]{Starinets:2008fb}%
  \BibitemOpen
  \bibfield  {author} {\bibinfo {author} {\bibfnamefont {A.~O.}\ \bibnamefont
  {Starinets}},\ }\href {\doibase 10.1016/j.physletb.2008.11.028} {\bibfield
  {journal} {\bibinfo  {journal} {Phys.Lett.}\ }\textbf {\bibinfo {volume}
  {B670}},\ \bibinfo {pages} {442} (\bibinfo {year} {2009})},\ \Eprint
  {http://arxiv.org/abs/0806.3797} {arXiv:0806.3797 [hep-th]} \BibitemShut
  {NoStop}%
\bibitem [{\citenamefont {Haehl}\ \emph
  {et~al.}(2015{\natexlab{b}})\citenamefont {Haehl}, \citenamefont
  {Loganayagam},\ and\ \citenamefont {Rangamani}}]{Haehl:2014zda}%
  \BibitemOpen
  \bibfield  {author} {\bibinfo {author} {\bibfnamefont {F.~M.}\ \bibnamefont
  {Haehl}}, \bibinfo {author} {\bibfnamefont {R.}~\bibnamefont {Loganayagam}},
  \ and\ \bibinfo {author} {\bibfnamefont {M.}~\bibnamefont {Rangamani}},\
  }\href {\doibase 10.1103/PhysRevLett.114.201601} {\bibfield  {journal}
  {\bibinfo  {journal} {Phys.Rev.Lett.}\ }\textbf {\bibinfo {volume} {114}},\
  \bibinfo {pages} {201601} (\bibinfo {year} {2015}{\natexlab{b}})},\ \Eprint
  {http://arxiv.org/abs/1412.1090} {arXiv:1412.1090 [hep-th]} \BibitemShut
  {NoStop}%
\end{thebibliography}%

\end{document}